\documentclass{ws-ijmpa}
\usepackage[super,compress]{cite}
\usepackage{graphicx}
\usepackage{slashed,color}
\usepackage{amsmath,amsfonts,bm}
\usepackage{amssymb,physics}

\newcommand{\xB}{x_{\rm Bj}}

\newcommand{\nn}{\nonumber \\ }

\renewcommand{\theequation}{\arabic{section}.\arabic{equation}}

\def\trimmarks{}

\begin{document}

\markboth{A.~V.~Radyushkin}{Theory and applications of parton pseudodistributions}

%
\catchline{}{}{}{}{}
%

\title{Theory and applications of parton pseudodistributions }

\author{A.~V.~Radyushkin}  

\address{Physics Department, Old Dominion University, Norfolk,
             VA 23529, USA \\ 
Thomas Jefferson National Accelerator Facility,
              Newport News, VA 23606, USA \\
radyush@jlab.org}

\maketitle

\begin{history}
\end{history}

\begin{abstract}

We review the basic theory  of the parton pseudodistributions approach and
its applications to lattice extractions of parton distribution functions. 
The crucial idea  of the approach is the realization  that  the 
 correlator $M(z,p)$ of the parton fields  is  a function ${\cal M} (\nu, -z^2)$ 
of Lorentz invariants
$\nu =-(zp)$, the Ioffe time, and the invariant interval $z^2$. 
This observation allows to extract the Ioffe-time distribution 
${\cal M} (\nu, -z^2)$ from Euclidean separations $z$ accessible on the lattice.
Another basic  feature is the use of the ratio ${\mathfrak M} (\nu,-z^2) \equiv {\cal M} (\nu, -z^2)/{\cal M} (0, -z^2)$,
that allows to eliminate artificial ultraviolet divergence generated by
the gauge link for space-like intervals. 
The remaining $z^2$-dependence of the reduced Ioffe-time distribution
${\mathfrak M} (\nu,-z^2) $ corresponds to perturbative evolution,
and can be converted into the scale-dependence of parton distributions
$f(x,\mu^2)$ using matching relations. 
The $\nu$-dependence of ${\mathfrak M} (\nu,-z^2) $ governs 
the $x$-dependence of parton densities $f(x,\mu^2)$.
The perturbative evolution was successfully observed 
in exploratory quenched lattice calculation.
The analysis of its precise data provides a framework for 
extraction of parton densities using the pseudodistributions
approach. It was used in  the recently performed calculations of the nucleon 
and pion valence quark distributions. We also discuss matching conditions
for the pion distribution amplitude and generalized parton distributions,
the lattice studies  of which are  now in progress.

\end{abstract}

\ccode{PACS numbers:12.38.-t, 
      11.15.Ha,  
      12.38.Gc  
}


\section{Introduction:  Why pseudo-PDFs?}

Feynman's parton distribution functions\cite{Feynman:1973xc}   (PDFs)  $f(x)$ 
are  the basic  building blocks for  the description of hard inclusive processes 
in quantum chromodynamics (QCD).
Generically,  they are defined through matrix elements of  bilocal operators of the 
$\langle  p |  \phi (0 )  \phi (z) |p \rangle \equiv M(z,p)$  type 
taken on the light cone $z^2=0$. 

Since  PDFs 
accumulate nonperturbative  information about 
the hadron structure,  they   are a natural candidate  for 
a lattice study. 
 However, the  intervals,  that are strictly on the light cone, are   not  accessible on      Euclidean lattices.
Still, it is possible    to perform lattice simulations  for small space-like $z^2$, and  to 
arrange  then some method of reaching the $z^2=0$ limit.

The  starting  idea is to take the equal-time interval $z=\{0,0,0,z_3 \}$.
It was put forward  in Refs.  
 [\citen{Detmold:2005gg,Braun:2007wv}] and emphasized 
 by X. Ji  in the paper  [\citen{Ji:2013dva}]  that strongly stimulated   further development in the lattice  studies 
 of the  PDFs (see Ref. [\citen{Cichy:2018mum}]  for a recent review and references).
 Other objects for a  lattice investigation include   the 
 pion 
distribution amplitude\cite{Radyushkin:1977gp}  (DA),  
a function  playing a fundamental role in  perturbative QCD studies 
of  hard exclusive processes,  and  generalized parton distributions\cite{Mueller:1998fv,Ji:1996ek,Radyushkin:1997ki}.

 By Lorentz invariance, the matrix element $M(z,p)$  is a function of the 
 {\it Ioffe time}\cite{Ioffe:1969kf}  $(pz)\equiv -\nu$  and of the interval $z^2$.
 Writing it as a function of these invariants, 
 $M(z,p) \equiv {\cal M} (\nu, -z^2)$, one deals with the 
 Ioffe-time {\it pseudodistribution}\cite{Radyushkin:2017cyf}  ${\cal M} (\nu, -z^2)$, which is a generalization 
 of the light-cone Ioffe-time distribution\cite{Braun:1994jq}  (ITD) ${\cal I}(\nu, \mu^2)$ onto space-like intervals $z^2$. 
 By definition, ${\cal I}(\nu, \mu^2)$ is a Fourier transform of the light-cone PDF $f(x,\mu^2)$,
 with   the Ioffe time $\nu$ being the variable that is Fourier-conjugate 
 to the parton momentum fraction variable $x$. 
 
  Analogously, taking the Fourier transform in $\nu$ of the pseudo-ITD
 ${\cal M} (\nu, - z^2)$ gives the {\it pseudo-PDF}\cite{Radyushkin:2017cyf}   ${\cal P} (x, -z^2)$.
By construction, ${\cal P} (x, -z^2)$ is a Lorentz-covariant function, just like 
  ${\cal M} (\nu, - z^2)$.  This means that the ``$x$'' variable is also 
  Lorentz-invariant.   It does not depend on a specific frame choice.
  In particular, there  is no need to take an infinite momentum frame to define it.
 It can  be shown\cite{Radyushkin:1983wh,Radyushkin:2016hsy}   
  that, for any  contributing  Feynman diagram,   ${\cal P} (x, -z^2)$ has the same support $-1 \leq x \leq 1$
  as the light-cone PDFs $f(x,\mu^2)$ do, even though $z^2$ is space-like.
  
  Hence, the pseudo-PDF ${\cal P} (x, z_3^2)$ is the most  natural generalization 
  of the light-cone PDF $f(x,\mu^2)$ onto space-like intervals. 
 For  $f(x,\mu^2)$, the usual  interpretation of the scale $\mu$  is that $1/\mu$ {\it characterizes}  the distances 
  at which the hadron structure is probed. In this sense,  when one takes  $z=\{0,0,0,z_3 \}$, 
   the scale $z_3$ in ${\cal P} (x, z_3^2)$
\mbox{ \it is  literally} the distance at which the hadron structure is probed.

Thus, the $z_3$-dependence of the pseudo-ITD ${\cal M} (\nu=z_3 p_3, z_3^2)$ comes in  two
ways. 
First,  the  $z_3$-dependence may be  accompanied by the $p_3$-dependence:
it comes  through the product $z_3 p_3=\nu$. 
The $\nu$-dependence of  ${\cal M} (\nu, z_3^2)$ converts 
into the \mbox{$x$-dependence}  of  ${\cal P} (x, z_3^2)$ . 
The remaining 
$z_3^2$-dependence of ${\cal M} (\nu, z_3^2)$  specifies how the  $x$-shape of 
${\cal P} (x, z_3^2)$ changes with the change of  the probing distance $z_3$.

In fact, 	 the dependence  on the probing distance
may be interpreted in terms of  the distribution of the parton's transverse  momentum\cite{Radyushkin:2017cyf}.
Recall,  that  $\nu$ and $z^2$ are Lorentz invariants. Therefore, 
the pseudo-ITD ${\cal M} (\nu, -z^2)$ is the same universal function of them,
no matter how $\nu$ and $z^2$ were  obtained from specific  choices 
of $z$ and $p$.  In  particular, taking $z$ on the light front, $z=\{z_+=0, z_-, z_\perp \}$,
 and a longitudinal \mbox{$p=\{p_+, p_-,0_\perp\}$}    gives
${\cal M} (\nu, z_\perp^2)$, where $\nu = -p_+ z_-$. 
In this situation, the $\nu$-dependence of ${\cal M} (\nu, z_\perp^2)$ 
determines the \mbox{$x$-distribution}   of the
longitudinal 
``plus''-component of the parton momentum, while its 
$z_\perp^2$-dependence determines the distribution of its transverse momentum $k_\perp$.

Hence, the two arguments of the pseudo-ITD ${\cal M} (\nu, -z^2)$
correspond to two different physical phenomena. Its  dependence on the Ioffe time $\nu$ 
converts into  the $x$-dependence of the pseudo-PDF ${\cal P} (x, -z^2)$,
which  characterizes  how the  parton  momentum increases with the increase 
of the hadron momentum. 
On the other hand, the dependence on $z^2$ characterizes a 
distribution of that part of the  parton momentum  that does not depend on the hadron momentum,
so it may be connected with    a ``primordial'' 
parton momentum distribution in the hadron  rest frame\cite{Radyushkin:2017cyf}.

In Ref. [\citen{Ji:2013dva}], it was proposed to convert the matrix 
element $M(z_3,p)$ 
into the {\it quasi-PDF} 
$Q(y,p_3)$. This is achieved by taking the 
Fourier transform of ${\cal M} (z_3 p_3, z_3^2)$ with respect to 
$z_3$. The resulting function  $Q(y,p_3)$ characterizes 
the fraction $y$ of the 
 third component 
of the hadron momentum $p_3$ carried by the parton. 
This fraction may take any value, from $-\infty$ to $\infty$, there is no restriction on it.

Since $z_3$ enters both in $\nu$ and $z_3^2$,   the $y$-shape of $Q(y,p_3)$ is  governed 
both by  the \mbox{$\nu$-dependence} of the pseudo-ITD  ${\cal M} (\nu, z_3^2)$
and  by  its $z_3^2$-dependence. 
Thus, the two different physical phenomena reflected in the $\nu$- and $z_3^2$-dependences
of ${\cal M} (\nu, z_3^2)$ are mixed in $Q(y,p_3)$. 
Writing $z_3$ as $\nu/p_3$, one can convert the $z_3$-integral of ${\cal M} (z_3 p_3, z_3^2)$ 
into the $\nu$-integral of ${\cal M} (\nu, \nu^2/p_3^2)$. 
For large $p_3$, the second argument  tends to zero,
and one essentially deals with the $\nu$-integral of ${\cal M} (\nu, 0)$, which 
gives the light-cone PDF $f(x)$. 
In other words, the $y$-shape of $Q(y,p_3)$   depends on $p_3$,
and reaches the PDF limit $f(y)$ when $p_3 \to \infty$,
i.e. in the infinite momentum frame. Taking the large-$p_3$ limit  for $Q(y,p_3)$ is the 
main idea\cite{Ji:2014gla} of  the quasi-PDF approach\footnote{For a recent review on quasi-PDFs see Ref. [\citen{Zhao:2018fyu}]}.

 To have a large momentum  is always a challenge for 
 a lattice simulation. 
Thus, the question is to which extent the efforts 
 to get a  large $p_3$ are justified.
 If the reason is to get a small value for  the second argument
 of ${\cal M} (\nu, \nu^2/p_3^2)$,  then this can be achieved 
 by simply taking a small $z_3$. And one can take then 
 {\it any} value of $p_3$, from zero to
 the achievable maximum. For instance, 
  in the lattice study performed in
 Ref.  [\citen{Orginos:2017kos}], 
 there were 
 7 values of $p_3= p (2\pi/L)$, 
 with $0\leq p \leq 6$.     
 In other words, for each value of $z_3$,  there were 7  values 
 of the Ioffe-time parameter $\nu$, instead of just one 
 value of $\nu$  obtained  in a  measurement  for the largest achievable $p_3$.
 As we discussed, it is the $\nu$-dependence 
 of ${\cal M} (\nu, z_3^2)$ that determines the $x$-dependence of PDFs,
 and  the pseudo-PDF approach allows to get  a detailed  information about it. 
 
In this connection,  we want to mention  that  another      approach, {\it the good lattice cross-sections}, that was 
proposed and developed in Refs. [\citen{Ma:2014jla,Ma:2017pxb}], is also    based on the factorization 
 in  the coordinate   space and the  analysis of the Ioffe-time dependence. 

 In  the present  paper, we review the basic ideas of the pseudo-PDF approach 
 formulated in Refs. [\citen{Radyushkin:2017cyf,Radyushkin:2017sfi}], further developed in Refs.
  [\citen{Radyushkin:2017lvu,Radyushkin:2018cvn,Karpie:2018zaz,Radyushkin:2018nbf,Radyushkin:2019owq,Balitsky:2019krf}]
 and used in lattice analyses of Refs. [\citen{Orginos:2017kos,Joo:2019jct,Joo:2019bzr}]. 
 
 In Sec. 2,  we discuss  the   general  aspects of the PDF concept.
  We start by  illustrating  the parton idea  by using the simplest example of the handbag diagram for a scalar analog of 
  deep inelastic scattering, and continue by outlining modifications
  necessary in a  theory with  spin-1/2 quarks and gauge fields.
  We describe the Ioffe-time distributions and the ratio method that 
  is a very essential element of the pseudo-PDF approach. It  
  allows, in particular, to efficiently get rid of the link-related ultraviolet 
  divergences that are artifacts of using space-like field correlators.
  We also discuss in this section some general properties of the quasi-PDFs,  
  in particular, their relation to the transverse-momentum dependence of parton distributions.  
  We show that such a dependence  described a ``TMD''  ${\cal F} (x, k_perp^2)$ is  determined by the $z^2$-dependence  
  of the pseudo-ITD ${\cal M} (\nu, -z^2)$.  
  
  In Sec. 3, we investigate  the nonperturbative aspects of the $P$-dependence of quasi-PDFs 
  using some simple models for  
  the TMD ${\cal F} (x, k_perp^2)$. 
  We describe the main features of the nonperturbative evolution of the quasi-PDFs, 
  in particular, the rate of approach of quasi-PDFs $Q(y,P)$  to  their $P\to \infty$ limit that give the  light-cone PDFs. 
  We discuss the role of target-mass corrections  $\sim (M^2/P^2)^n$, and argue  that 
  they are actually much smaller than the $\sim (\langle k_\perp^2\rangle /P^2)^n$ 
  transverse-momentum corrections. 
  
In Sec. 4, we give a discussion of the perturbative structure of the ITDs at one-loop level,
concentrating both on the link-related ultraviolet divergences and on the infrared aspects
connected with the  perturbative evolution. A special attention is given to matching
relations that allow 
to convert the $z_3^2$-dependence of the reduced ITD into the $\mu^2$-dependence 
of the light-cone PDFs. 

In the next two sections, we discuss the results of  lattice calculations\cite{Orginos:2017kos,Joo:2019jct}  
guided by the pseudo-PDF approach ideas.  
We concentrate  on   applications of the pseudo-PDF approach concepts,  skipping  
the questions related to actual  lattice extraction of the data, the analysis of 
discretization errors, finite-volume effects, etc. 

The results of the exploratory lattice study\cite{Orginos:2017kos}  based on the pseudo-PDF approach 
are  described in Sec. 5.  The high statistical accuracy of its  data   allows 
to perform a lattice study of    perturbative evolution,  the  phenomenon that  no
other lattice simulations were able to detect yet. 
The  analysis of the quenched data forms a basis for future 
studies of the perturbative evolution
  within lattice setups that are closer to the real-world QCD.

The results of a recent calculation\cite{Joo:2019jct} with dynamical  
fermions are discussed in Sec. 6. The PDFs extracted in this study are in much better agreement
with phenomenological studies.
However,  larger statistical errors of the data do not allow to  detect 
perturbative evolution.
 
 In Sec. 7, we describe the derivation of matching relations for
 the pion distribution amplitude and generalized parton distributions
 that are necessary in the ongoing and future efforts for extraction of these 
 distributions from the lattice. 
 
Sec. 8 contains the summary of the paper.  

The derivation of the spectral property $|x|\leq 1$ for the pseudo-PDFs
is outlined in the Appendix.

  
   \setcounter{equation}{0}  
  
\section{Parton distributions}

  \subsection{Handbag diagram and   pseudo-PDFs}
  \label{HB}
  
        \begin{figure}[b]
    \centerline{\includegraphics[width=2in]{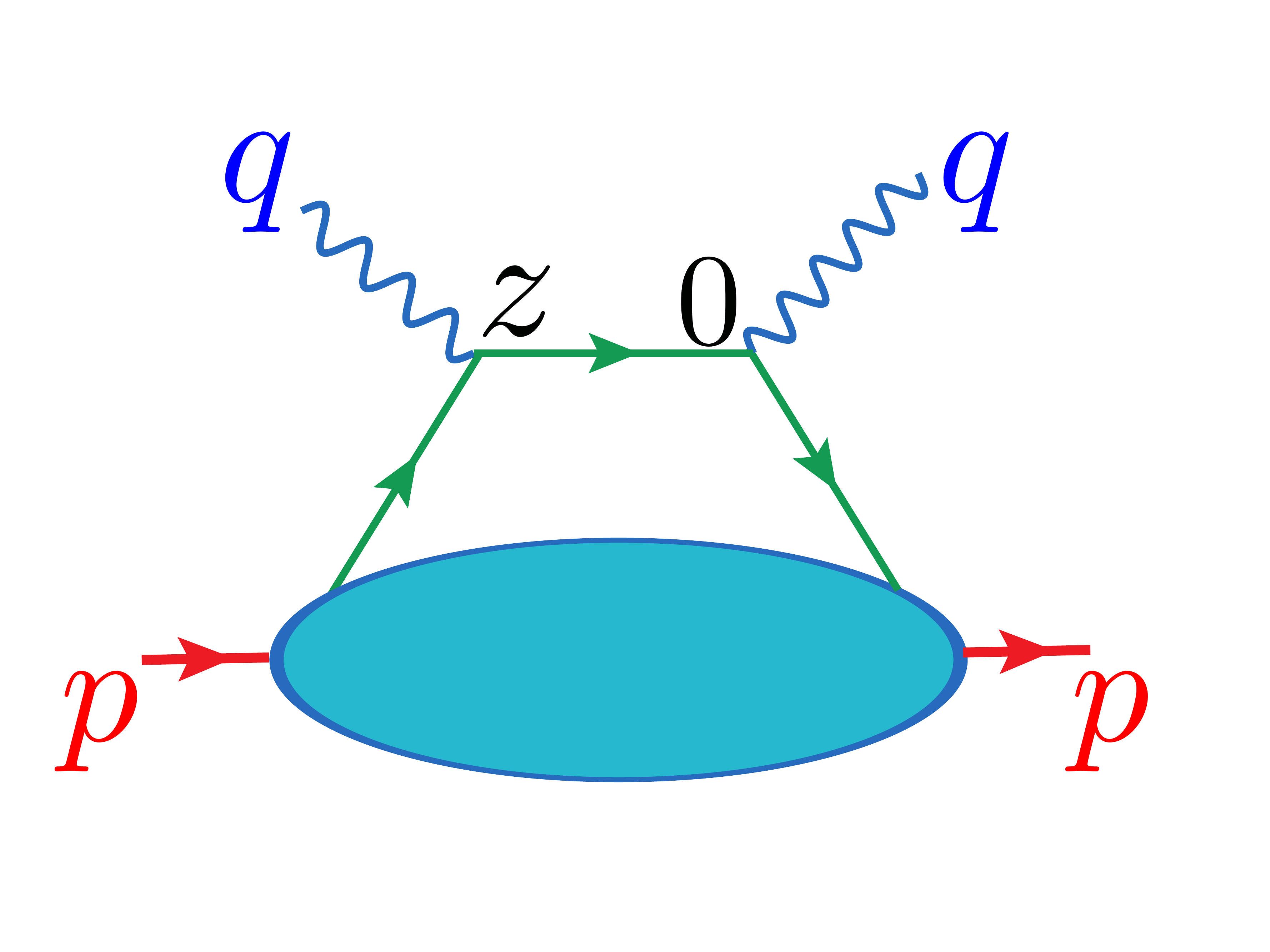}}
    \caption{General handbag diagram for  a virtual forward Compton amplitude
     in coordinate representation.
    \label{handbagZ}}
    \end{figure}

  Historically, parton distributions were  introduced
  to describe deep inelastic scattering (DIS). 
  The usual starting point of DIS analysis is the forward  Compton amplitude
  $T(q,p)$ 
  which,  
  in the lowest approximation,  is given by a handbag diagram (see Fig. \ref{handbagZ}).
  To skip inessential complications related to spin
  (they  do not affect  
  the very concept of parton distributions and may  be included when needed), we  start with 
  a simple example of a  scalar handbag diagram, and  write it in   
  the coordinate representation
  \begin{align}
T(q,p) = 
\int {\dd ^4z}\, e^{-i(qz)} \, D(z) \, 
\langle  p | \phi (0)  \phi(z)  | p \rangle  \  , 
\label{Tpqz}
\end{align}
where  $D(z) = -i/4\pi^2 z^2$ is  the scalar massless propagator, $p$ 
is the target  momentum and $q$ is the momentum of the hard probe.

 The matrix element $\langle  p | \phi (0)  \phi(z)  | p \rangle \equiv M(z,p)$
 accumulates
 information about the target.
 To proceed with the $\dd^4z$ integral, one needs to know something about 
 the dependence of $M(z,p)$ on the coordinate $z$.

 It can be shown \cite{Radyushkin:2016hsy,Radyushkin:1983wh}   
  that, for each of  contributing  Feynman diagrams,   
  $M(z,p)$ has the following representation  (see Appendix  A for some details)
     \begin{align}
 { M} (z,p) 
&   = 
 \int_{-1}^1 \dd x 
 \, e^{- i x (pz)  } \,  {\cal P} (x, -z^2)  \   ,
  \label{MPD}
\end{align}   
  where  ${\cal P} (x, -z^2)$    is  the 
   parton {\it pseudodistribution}  function or {\it pseudo-PDF},  introduced in Ref. 
  [\citen{Radyushkin:2017cyf}].  
   In the simplest case, when ${\cal P} (x, -z^2)$  has no $z^2$-dependence,
 so that  ${\cal P} (x, -z^2)=f(x)$,  the $\dd^4z$ integral becomes trivial, and we get 
     \begin{align}
T_0(q,p) = -
 \int_{-1}^1 \dd x \, \frac{f(x)}{(q+xp)^2+i \epsilon}  \ ,
\label{T0pqz}
\end{align}
which is the well-known parton-model expression for the 
forward Compton amplitude, with $f(x)$ being the {\it parton distribution function}
(PDF).

  Note that  Eq. (\ref{MPD})   introduces  the momentum fraction 
    variable $x$ in an absolutely  covariant way.  
 One has  no need to assume that 
$z^2=0$  or $p^2=0$ or  to  take an infinite momentum frame, etc., to define $x$. 
Of course, since  the representation  (\ref{MPD}) works in general case,
it also works if we take $z$ on the light cone. 
In particular,  taking  $z$ that has the  light-cone ``minus'' component $z_-$ only, gives the
  representation  
  \begin{align}
  \langle  p |\phi (0) \phi(z_-)  
| p \rangle =
   \int_{-1}^1 \dd x \, {\cal P} (x, 0) \, 
e^{-ixp_+ z_-} \,  \  , 
 \label{twist2par0}
\end{align}
which  has the standard interpretation that $x$ is the 
 fraction of  the  light-cone ``plus''  component $p_+$
 of the   target momentum 
carried by the parton.

However, we want to emphasize  that $p$ in Eq. (\ref{T0pqz}) is the  {\it actual}  hadron momentum satisfying $p^2=M^2$. 
 In particular, writing    
  \begin{align}   
  (q+xp)^2= 2 (qp)(x- \xB)+x^2 M^2
  \end{align} 
 (with $\xB = -q^2/2 (pq)$ is the Bjorken variable), and taking the  imaginary part, we 
 get  the \mbox{$\xi$-scaling}  expression \cite{Nachtmann:1973mr,Georgi:1976ve}  
 for the relevant structure function 
 \begin{align} 
W_0  (q,p)   
= &
   \int_{0}^1 dx 
     \, f (x) \ \delta [2 (qp)(x- \xB)+x^2 M^2]  
     = 
     \frac{f(\xi_{\rm N})}{ 2(pq)\sqrt{1+ \frac{4x_{\rm Bj}^2 M^2}{Q^2}}} \ , 
     \label{sDelta}
\end{align} 
where $Q^2=-q^2$ and $\xi_{ N} $   
 is the Nachtmann  variable\cite{Nachtmann:1973mr} 
  \begin{align}  
& 
\xi_{ N} = \frac{2 x_{\rm Bj}}{1+ \sqrt{1+ {4x_{\rm Bj}^2 M^2}/{Q^2}}} \ . 
\label{xiNa0}
 \end{align} 
Thus, Eq. (\ref{sDelta})  allows to calculate  target-mass corrections. 

On dimensional grounds, one may expect that further terms  
in the formal \mbox{$z^2$-expansion}  ${\cal P} (x, -z^2) =f(x) + z^2 f_4 (x) \ + \ldots + (z^2)^n  f_{2(n+1)} (x)  + \ldots $ 
will be accompanied by extra $(1/Q^2)^n$ factors, i.e. that such   ``higher twist'' contributions to $T(q,p)$ 
are suppressed by powers of $1/Q^2$ for large $Q^2$. 
However, the light-cone singularity of the massless scalar propagator (which is $D(z) \sim 1/ z^2$) 
is canceled by the $z^2$ factors, resulting in contributions containing $ \delta^4 (q+xp)$  and its derivatives. 
In DIS, $q$ is not proportional to $p$, and such contributions are treated as zero. 

Thus, if ${\cal P} (x, -z^2)$ is analytic on the light cone, the scalar handbag diagram 
is given by the twist-2 part alone. For spin-1/2 partons, the propagator \mbox{$S^c (z) \sim \slashed z /(z^2)^2$}  
is more singular, and the handbag diagram contains twist-2 and twist-4 terms. 
The twist-2 part is given by $\xi$-scaling expressions \cite{Nachtmann:1973mr,Georgi:1976ve}  
involving the twist-2 PDF $f(x)$, while the twist-4 part requires an independent function  
related to $\bar \psi D^2 \psi$-type operators.  


 \subsection{Light-cone singularities and factorization}

 One cannot use a   formal  \mbox{$z^2$-expansion}  if   ${\cal P} (x, -z^2)$ is singular for $z^2=0$.   
  In QCD and other renormalizable theories,   
  $ {\cal P} (x, -z^2)  $
 has   $\sim \ln (-z^2) $  terms.  
 These singularities are perfectly integrable when embedded in the 
 expression (\ref{Tpqz}) for 
$T(q,p)$:  they  just produce logarithmic $\ln (-q^2)$ contributions that violate
 a strict dimensional scaling present in  $T_0(q,p)$. 
 
 On the other hand, taking    $z^2 =0$  in the pseudo-PDFs  themselves  produces ultraviolet 
divergences in the perturbative expressions for  matrix elements 
of $ \phi (0)  \phi(z) $ operators. 
Introducing some UV cut-off $\Lambda$ converts 
 $\ln (-1/ z^2)$ into $ \ln \Lambda^2$, and the resulting PDFs   depend on
 the cut-off scale, $f(x) \to   f(x, \Lambda^2)$. 
The usual procedure is to use the dimensional regularization (DR) 
for momentum integrals 
$\dd^4 k \to (\mu^2)^\epsilon \dd^{4- 2 \epsilon} k $.
After subtraction of  the $1/\epsilon$ poles, one gets PDFs   
 depending  on the DR renormalization scale $\mu$. 
For the minimal $\overline{\rm MS}$ subtraction,
one obtains the standard  $\overline{\rm MS}$ parton densities
$f(x) \to   f(x, \mu^2)$. 

 It should be emphasized  that,  if one   keeps $z^2$  spacelike, then 
 ${\cal P} (x, -z^2)$ is finite, and  no  regularization  for  the $\ln (- z^2)$ terms  is   needed.
 In this sense,  the interval $z^2$ serves as an  UV cut-off, 
 and one may treat 
   ${\cal P} (x, -z^2)$ as just another type of a PDF,  that is defined in a 
   peculiar  ``$z^2$''-scheme rather in the $\overline{\rm MS}$ scheme. 
 In fact, the PDFs of this $z^2$-scheme are  more physical 
 than the $\overline{\rm MS}$ ones. One may say that they  literally measure  the hadron structure 
 at distances $d= \sqrt{-z^2}$. 

However, the established  standard is to use the $\overline{\rm MS}$-scheme
PDFs  $f(x, \mu^2)$. 
In the expression for  $T(q,p)$, written in terms of the momentum invariants
$q^2=-Q^2$ and \mbox{$x_{\rm B} =Q^2/2(pq)$,} they appear 
through the factorization formula
     \begin{align}
T( x_{\rm B}, Q^2) 
&   = 
 \int_{-1}^1 \frac{\dd x}{|x|}  \, 
 t(x_{\rm B}/x, Q^2/\mu^2) \, f(x,\mu^2) \ + \ {\rm higher \  twists} 
 \,  \    ,
  \label{Tqx}
\end{align} 
in which the scaling-violating $\ln Q^2$ terms are split into the ``short-distance'' part $\ln Q^2/\mu^2$
present in the coefficient function $ t(x_{\rm B}/x, Q^2/\mu^2)$ and the evolution 
logarithms $\ln \mu^2$ present in the scale-dependent PDF $f(x,\mu^2)$. 
This formula is obtained  by applying  the operator product expansion (OPE) 
 to $T(q,p)$ written  as  
  \begin{align}
T(q,p) = 
\int {\dd ^4z}\, e^{-i(qz)} \, 
\langle  p | j (0)  j(z)  | p \rangle  \  , 
\label{Tjj}
\end{align}
i.e., in terms of the probing currents $j(0)$, $j(z)$.   Similarly,
one can apply the OPE to the product of fields $\phi (0)  \phi(z) $ defining the pseudo-PDF. 
In non-gauge theories,
\begin{align} 
{\cal P } (x, -z^2 ) =&  \int_{-1}^1 \frac{\dd w}{w} \, C (w, z^2 \mu ^2) \,  f (x/w,\mu^2) 
 + {\cal O} (z^2) 
  \  .
\label{OPE}
\end{align}
In this expression,  the $\ln (-z^2)$   terms are split 
between the  coefficient function  $C (w, z^2 \mu ^2)$ and the PDF $ f (x/w,\mu^2) $. 
Here we write the  factorization relation in the  form 
following from 
 the  nonlocal  light-cone OPE \cite{Anikin:1978tj,Balitsky:1987bk}
 (see also  Ref. [\citen{Radyushkin:2017lvu}]).   
 
 \subsection{Gauge theories}

In QCD, the  quarks have spin 1/2, and  the 
handbag  diagram for the Compton  amplitude 
 is  given by 
 \begin{align}
 T^{\mu \nu} (q,p) = 
\int {d^4z}\, e^{-i(qz)} 
  \langle p |   \bar \psi(0)  & \gamma^\nu \, S^c (-z) \, \gamma^\mu \,  \psi (z)|p \rangle 
\ , 
\label{fermiC}
\end{align}
where $S^c (z) = \slashed z/ 2\pi^2 (z^2)^2$ 
is the  propagator for a massless fermion.
Writing  $\gamma^\nu \, \slashed z \, \gamma^\mu $
as $\left [ g^{\nu \beta} g^{\mu \alpha} + g^{\nu\beta}  g^{\mu \alpha} - g^{\mu \nu} g^{\alpha \beta} + i \epsilon^{\mu \nu \alpha \beta} \gamma_5  \right ] z_\beta\gamma_\alpha$
we 
get  matrix elements  $ \langle p |    \bar \psi(0)   \,  \gamma_\alpha   \,  \psi (z)|p \rangle  $ and 
 $\langle p |    \bar \psi(0)   \,\gamma_5 \gamma_\alpha   \,  \psi (z)|p \rangle  $ 
 corresponding to unpolarized and polarized PDFs, respectively.
 
  \begin{figure}[h]
  \centerline{\includegraphics[height=4cm]{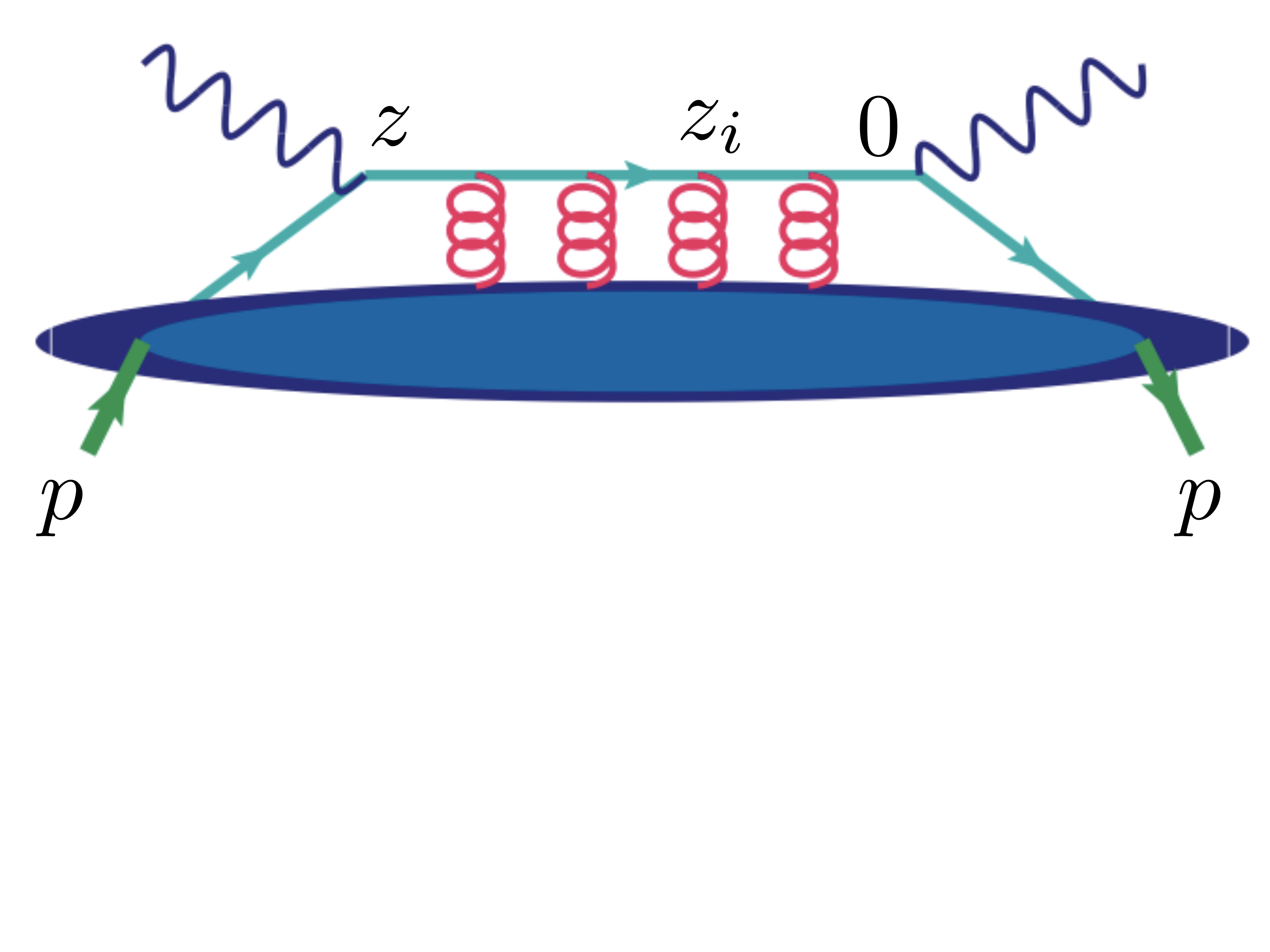} 
   }
   \vspace{-1.5cm}
  \caption{Structure of QCD factorization for DIS in covariant gauges.}
  \label{dis_fac.pdf}
\end{figure}
   
Furthermore, in gauge theories, the handbag  contribution in 
 covariant gauges should be complemented 
by   diagrams corresponding to operators 
$\bar \psi (0)  \ldots \slashed A(z_i) \ldots  \psi (z)$
containing twist-0 gluonic field $A_{\mu_i}  (z_i)$ inserted 
into the fermion line between the points \mbox{$z$ and $0$}
(see Fig.\,\ref{dis_fac.pdf}).
The sum of gluon  insertions is equivalent
to substituting the free propagator $S^c (z_1-z_2)$ by
a propagator ${\cal S}^c (z_1, z_2;  A)$ of a  quark 
in an external gluonic  field $A$. 
  This propagator 
 satisfies the Dirac equation 
\begin{align}
i \left [ \frac{\slashed \partial}{\partial z_1 } - ig \slashed A (z_1)  \right ]{\cal S}^c (z_1,z_2;  A) = - \delta^4 (z_1-z_2)  \ .
\label{Dirac}
\end{align}
The  solution of this equation  may be written in the form
\begin{align}
{\cal S}^c (z_1,z_2;  A) =  E (z_1,z_2; A)  {\mathfrak S}^c_{\rm FS }
(z_1,z_2;A) 
\label{FSProp}
\end{align}
involving the straight-line exponential
  \begin{align}
{ E}(z_1,z_2; A) \equiv P \exp{ \left [ ig \,   \int_0^1dt \,  (z_2^\alpha-z_1^\alpha)\, A_\alpha ((1-t)z_1+t z_2) 
 \right ] }  \  .
 \label{straightE}
\end{align}

In its turn, the factor $ {\mathfrak S}^c_{\rm FS }$   satisfies the Dirac 
equation (\ref{Dirac})  
with the general  vector potential  $A^\mu(z) $ substituted \cite{Efremov:1978fi,Efremov:1978xm,Efremov:1980ub} 
by the vector potential $ {\mathfrak A}^\mu  (z; z_1) $ in the Fock-Schwinger (FS) 
gauge \cite{Fock:1937aa,Schwinger:1951nm} 
\mbox{$
(z-z_1)_\mu  {\mathfrak A}^\mu  (z,z_1) = 0 \ .
$} 
It is given by 
\begin{align}
 {\mathfrak A}^\mu  (z; z_1) = (z-z_1)_\nu 
 \int_0^1 \, \dd s \,  s \,  G^{\mu \nu} (z_1+s (z-z_1)) \,  . 
 \label{FSA}
\end{align}
Here,  $z$  denotes   the location of the field,  while  $z_1$
specifies the ``fixed point'' of the FS 
gauge,  and in our case 
 refers  to  an end-point in the Compton amplitude.
Since the field-strength tensor  $G^{\mu \nu} $
has twist equal to (at least)  1, the insertion 
of this field into the free propagator results 
in power $(\Lambda^2/Q^2)^l$ corrections 
to the Compton amplitude. Thus, we can write
\begin{align}
{\cal S}^c (0,z;  A) =  E (0,z; A) { S}_c (z) 
+ {\rm higher \  twists} \  . 
\label{extF}
\end{align}
As a result, at  the leading-twist level, we  deal with   matrix elements   of the 
   \begin{align}
 { M}^\alpha  (z,p) \equiv \langle  p |  \bar \psi (0) \, 
 \Gamma^\alpha \,  { E} (0,z; A) \psi (z) | p \rangle \  
 \label{Malpz}
    \end{align}  
type, where  $ \Gamma^\alpha =  \gamma^\alpha$ or $ \gamma_5\gamma^\alpha$.   
When  $ \Gamma^\alpha =  \gamma^\alpha$, the function $ {\cal M}^\alpha  (z,p) $ may be 
decomposed into $p^\alpha$ and $z^\alpha$ parts
\begin{align}
     { M}^\alpha  (z,p) =2 p^\alpha  {\cal M}  (-(zp), -z^2) + z^\alpha  {\cal M}_z (-(zp),-z^2) \  . 
    \end{align}  
Defining the relevant light-cone PDF, one 
takes  $z=z_-$ (which means   $z_+=0$)   and  $\alpha=+$.  As a result, the  \mbox{$z^\alpha$-part} drops out, and  
PDF    
is  determined by   the $ {\cal M}  (\nu, 0)$  amplitude only.   
On the lattice,   taking $z=z_3$,  
we  choose $\alpha =0$ to    eliminate the     \mbox{$z^\alpha$-contamination\cite{Radyushkin:2016hsy}}  
   and define
 the pseudo-ITD ${\cal M} (\nu, z_3^2) $ by 
 \begin{align}
&  { M}^0   (z_3,p)   
  = 2 p^0 {\cal M} (\nu, z_3^2) 
  \  .
\end{align}

          \begin{figure}[t]
  \centerline{\includegraphics[height=4cm]{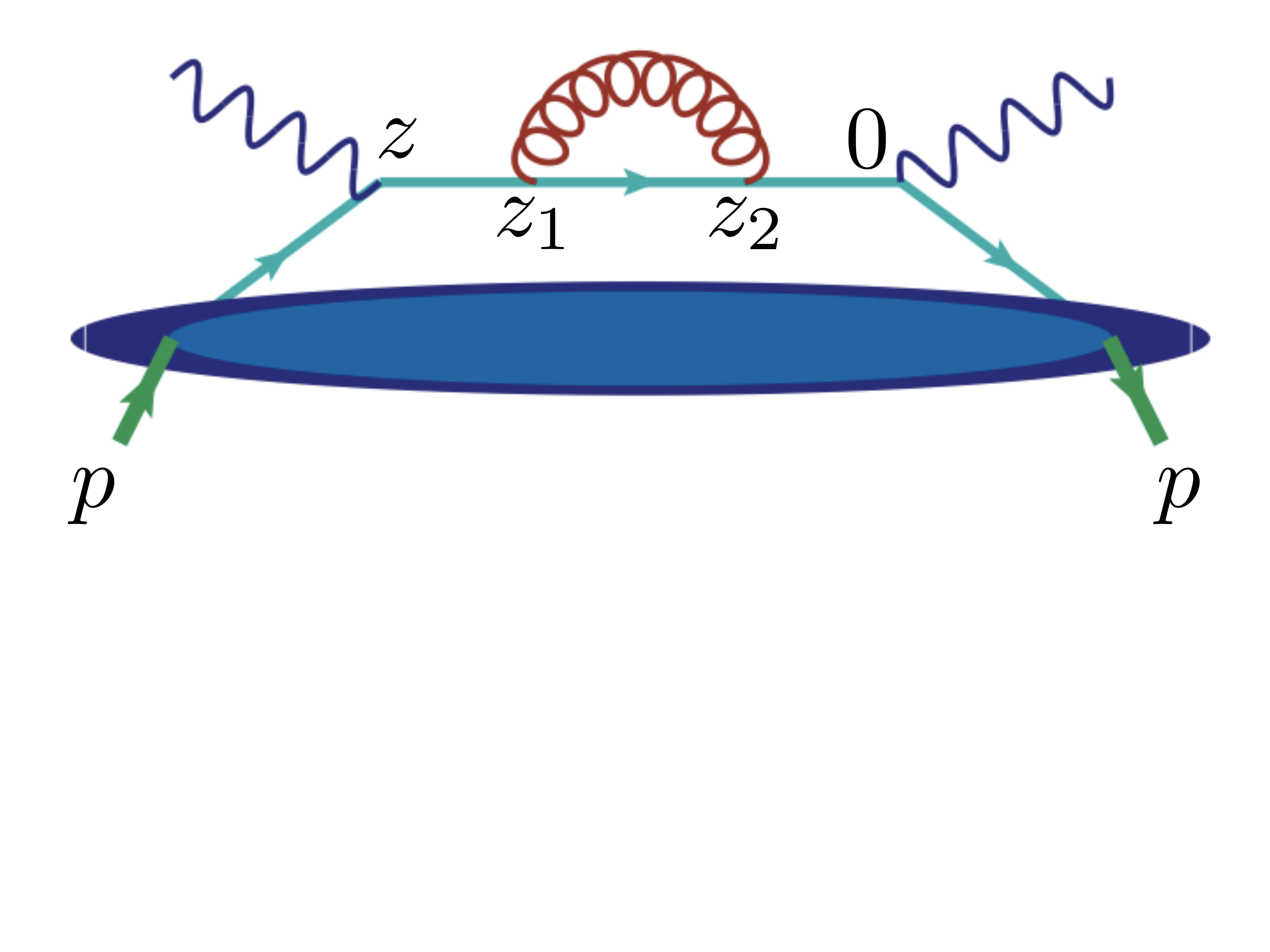}
   }
   \vspace{-1.5cm}
  \caption{
   Self-energy correction to the quark propagator}
  \label{disself.pdf}
\end{figure}

\begin{figure}[h]
   \centerline{ \includegraphics[width=1.5in]{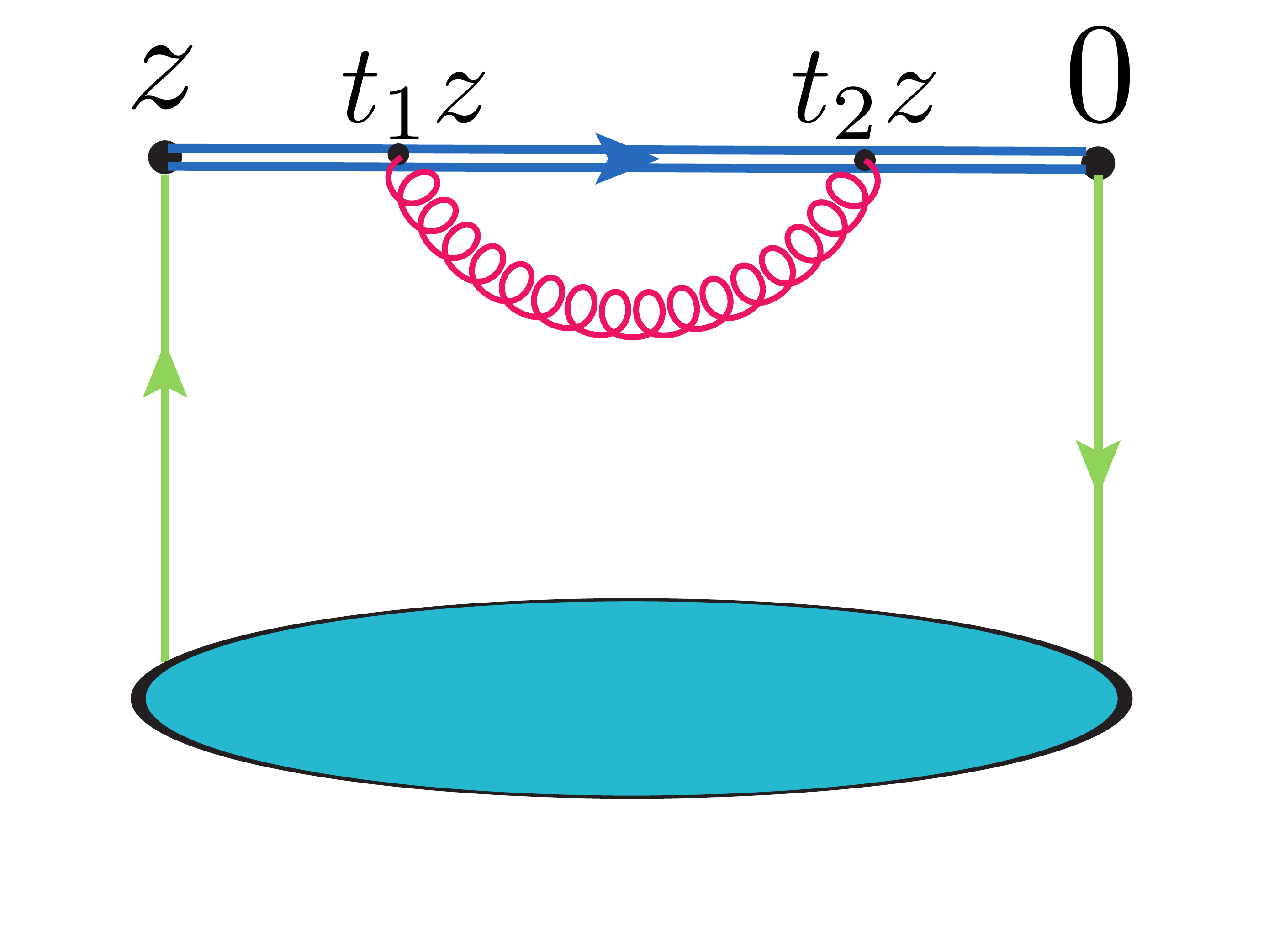}}
   \caption{Self-energy correction to    the gauge link.
   \label{linkself}}
   \end{figure}

It should be noted that the quark self-energy diagram (see  Fig. \,\ref{disself.pdf})
cannot  be factorized into a tree-level coefficient function and 
the matrix element $\langle p | \bar \psi (0)  \ldots  A^{\alpha_1} (z_1) \ldots
  A^{\alpha_1}(z_2)  \ldots  \psi (z) | p \rangle $.
  Its entire contribution belongs to  the one-loop part of the coefficient function.
This means that  the definition (\ref{Malpz}) of $ { M}^\alpha  (z,p)$ 
should  imply that the \mbox{$A^{\alpha_i} (t_iz)$-fields}  in the expansion of the exponential 
(\ref{straightE}) 
are not  contracted with each other. 
In other words, the contributions corresponding to the link self-energy 
corrections (see Fig. \ref{linkself}) should  be excluded.

 However, when the matrix element  (\ref{Malpz})   is   calculated on the lattice,
 such contributions are  included automatically: the lattice ``does not know'' about this restriction. 
Moreover, the link self-energy diagram  produces 
ultraviolet divergences when $z$ is off the light cone.
These   divergences  
require an additional UV  regularization.  
Fortunately, these divergences 
(and also link-vertex UV divergences) are multiplicative\cite{Polyakov:1980ca,Dotsenko:1979wb,Brandt:1981kf,Aoyama:1981ev,Craigie:1980qs,Dorn:1986dt,Bagan:1993zv}  (see also 
 recent papers  
 [\citen{Ishikawa:2017faj,Ji:2017oey,Green:2017xeu}]). They  
form  a  factor $Z(-z^2/a^2)$, where $a$ is a UV cut-off, e.g., the 
lattice spacing. This factor should be included 
in  the right-hand side of the OPE  (\ref{OPE}).  
Thus, to get the PDF $f(x, \mu^2)$ from the pseudo-PDF 
${\cal P } (x, -z^2 ) $ one should ``renormalize'' the latter by dividing 
it by $Z(-z^2/a^2)$.

\subsection{Ioffe-time distributions}

\label{sITD}

\begin{figure}[t]
  \centerline{\includegraphics[height=4cm]{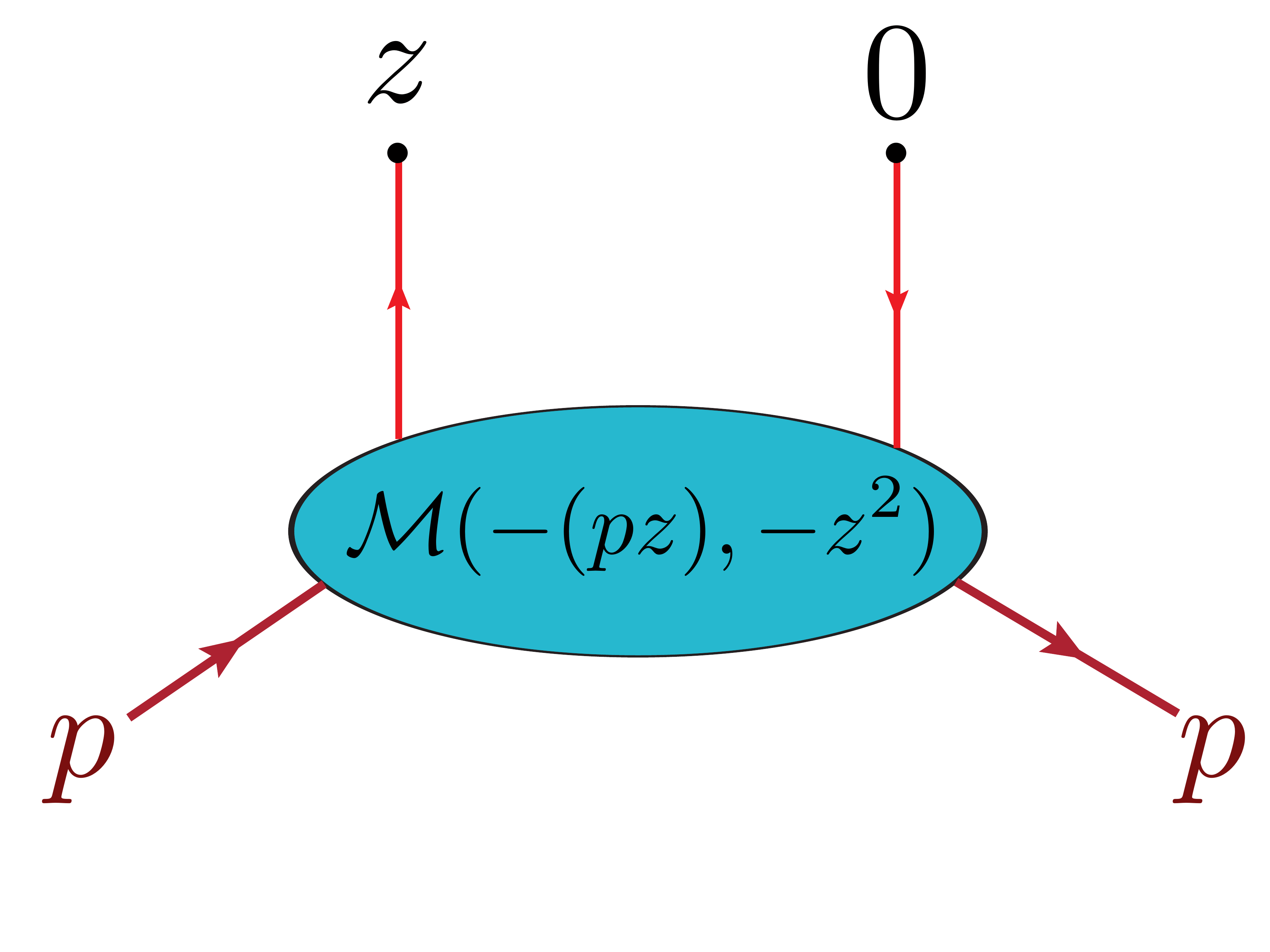}}
  \vspace{-2mm}
  \caption{Ioffe-time distribution.}
  \label{ITDf}
\end{figure}

   The pseudo-PDF representation (\ref{MPD})  separates the dependence 
   $M(z,p)$ on its two $z$-dependent  Lorentz invariants, the Ioffe time $(pz)\equiv - \nu$ 
   and the interval $z^2$ (see Fig. \ref{ITDf}).  Writing $M(z,p)$ as a function  of $\nu$ and
  $z^2$,  we get the {\it Ioffe-time pseudodistribution} ${\cal M}  (\nu, -z^2)$. 
Inverting  Eq. (\ref{MPD}) gives the relation  
   \begin{align}
     {\cal P} (x, -z^2) 
&   = \frac1{2\pi} \, 
 \int_{-\infty}^\infty  \dd \nu 
 \, e^{-i x \nu } \,   {\cal M} (\nu, -z^2) 
 \label{psM}
\end{align}  
that tells us that the pseudo-PDF  is a Fourier transform
of the pseudo-ITD  $ {\cal M} (\nu, -z^2)$   with  respect to $\nu$ for fixed $z^2$.
When $z$ is on the light cone, $z^2=0$, we deal with 
 the  light-cone PDF  $f(x,\mu^2)$ and the light-cone Ioffe-time distribution 
\begin{align}
{\cal I} (\nu , \mu^2)  =
\int_{-1}^1 \dd x \, e^{ix\nu} \,  f(x,\mu^2) \, 
\label{LCITD}
\end{align}
introduced originally in Ref. [\citen{Braun:1994jq}].  
In  terms of  the ITDs, the factorization relation (\ref{OPE})  takes the form 
\begin{align} 
{\cal M } (\nu, -z^2 ) /Z(-z^2/a^2) =&  \int_{-1}^1  {\dd w} \, C (w, z^2 \mu ^2) \, {\cal I} (w \nu, \mu^2)  
 + {\cal O} (z^2) 
  \  .
\label{IOPE}
\end{align}
Combining (\ref{IOPE}) and (\ref{LCITD}), we obtain a {\it kernel relation} 
\begin{align} 
{\cal M } (\nu, -z^2 )  /Z(-z^2/a^2) =&  \int_{-1}^1  {\dd x} \, R (x \nu, -z^2 \mu ^2) \, {f} (x, \mu^2)  
 + {\cal O} (z^2) 
  \ 
\label{ker}
\end{align}
that directly connects the renormalized pseudo-ITD  
with the light-cone PDF  through the kernel
\begin{align} 
R (x \nu, -z^2 \mu ^2)  =&  \int_{-1}^1  {\dd w} \,  e^{i w x\nu}  \, C (w, z^2 \mu ^2) \,  
   \  .
\label{Rker}
\end{align}

The pseudo-PDF strategy is to start with 
the standard lattice choice\cite{Detmold:2005gg,Braun:2007wv,Ji:2013dva}  
  of taking  an equal-time 
 interval  $z=\{0,0,0,z_3\}$ and extract the   ${\cal M } (\nu, -z^2 ) $
 as a function of $\nu$ and $z^2$. 
 As we discussed, it is the $\nu$-dependence of ${\cal M } (\nu, -z^2 ) $
 that governs the $x$-dependence of PDFs. 
When $z=\{0,0,0,z_3\}$,  we have $\nu = p_3 z_3$ and $z^2= -z_3^2$. 

The basic idea of the pseudo-PDF approach  is that it does not matter if $\nu$ is given by $-p_+ z_-$ or by $p_3 z_3$. 
In both cases, one deals with the same functional dependence of    ${\cal M } (\nu, -z^2 )$
on $\nu$.  Using the relations (\ref{IOPE}), (\ref{ker}), we can (at least, in principle) extract light-cone functions $f(x,\mu^2)$
from the ``Euclidean'' pseudo-PDF ${\cal M } (\nu, z_3^2 )$. 

It is worth stressing here that the applicability of
the  basic perturbative relations (\ref{IOPE}), (\ref{ker}) is determined solely
by the size of $z_3^2$. One can take small $p_3$ (even $p_3=0$),
and use perturbative QCD as far as $z_3^2$  is sufficiently small. 
 The size of the momentum $p_3$ changes  the magnitude of $\nu =p_3 z_3$, but it 
 does not  affect the applicability 
of the perturbative  expansion.

Another  key element of the pseudo-PDF approach is
the elimination of  the problematic  UV $Z$-factor by introducing\cite{Radyushkin:2017cyf}  
 the  reduced Ioffe-time  pseudodistribution 
 \begin{align}
{\mathfrak M} (\nu, z_3^2) \equiv \frac{ {\cal M} (\nu, z_3^2)}{{\cal M} (0, z_3^2)} \   . 
 \label{redm}
\end{align}
 Since   $Z(z_3^2/a^2)$ does not depend on $\nu$, the $Z$-factors 
 of the numerator and denominator cancel.    
 The remaining $z_3^2$-dependence of ${\mathfrak M} (\nu, z_3^2)$ for small $z_3^2$ is completely determined by the evolution logarithms, and   may be calculated perturbatively
 using OPE in the form of  Eqs.  (\ref{IOPE}), (\ref{ker}). 
Note also  that the denominator factor 
 \begin{align}
{\cal M} (0 , z_3^2)  =
\int_{-1}^1 \dd x \,   {\cal P}(x, z_3^2) \, 
\label{M0}
\end{align}
 is just  the lowest moment of the pseudo-PDF. Thus, there is nothing singular  here
 in taking $p_3=0$.  Moreover, if
 the local  limit  $z_3 \to 0$ corresponds to a conserved current,
 then ${\cal M} (0 , z_3^2) $ 
 does not have the evolution 
 $z_3^2$-dependence, which provides a
 further simplification.

 The ratio (\ref{redm})   may be also used for elimination of the 
$Z(z_3^2/a^2)$-factor  in the quasi-PDF approach discussed in the next Section. 
However, the quasi-PDF practitioners prefer to use the RI/MOM method (see Ref. [\citen{Zhao:2018fyu}] for a review), 
 in which ${\cal M} (\nu, z_3^2)$ is divided by the  matrix element $m (z_3, p_R)=  \langle p_R | {\cal O}  (z_3) |p_R \rangle$  
 of the same bilocal operator $ {\cal O}  (z_3)$ sandwiched between parton (quark or gluon) states taken 
 at some reference momentum $p_R$.  A   usual choice is to take spacelike $p_R$ with large virtuality,   
 which raises the questions about gauge invariance of such matrix elements.  
 This  difficulty is avoided if ${\cal M} (\nu, z_3^2)$  is divided    by the vacuum matrix element 
 $\langle 0 | {\cal O}  (z_3) |0 \rangle$, as suggested in Ref.  [\citen{Braun:2018brg}].   Still, both of these        
 alternatives have the  disadvantage  that the denominator of the ratio  should be obtained  from a separate calculation, 
 and this increases systematic errors.   

 \subsection{Quasi-PDFs}

  To define the parton {\it quasi-distribution} functions\cite{Ji:2013dva} or  {\it quasi-PDFs} 
  $ Q (y, P) $,
  one takes the separation    
  $z= \{0,0,0,z_3 \}$ and momentum  $p= (E, {\bf 0}_\perp, P)$
  in the same direction. Then   $ Q (y, P) $ is given by 
 the Fourier transform of  the matrix element $M(z_3, P) $  with respect to $z_3$  
\begin{align}
Q (y, P) 
&   = \frac{P}{2\pi} \, 
\int_{-\infty}^\infty  dz_3 
\, e^{-i yPz_3 } \,  M(z_3, P)  
\label{Qdef} \  .
\end{align}   
The inverse representation has the form of a plane-wave decomposition 
\begin{align}
M(z_3, P)
=  & 
\int_{-\infty}^{\infty}   dy \, 
e^{i y  P z_3 } \, 
Q(y, P) \,  
\label{phiQ}
\end{align}  
in which  the function $Q(y,P)$ describes  what fraction $yP$ 
of the  hadron's 
{\it  third momentum component }  is carried by a specified  parton.
 
It is easy  to find  a relation between 
	the quasi-PDF $Q(y,P)$ and the pseudo-PDF ${\cal P} (x,z_3^2)$
	corresponding to the $z=z_3$ separation.
	Indeed,  adjusting     the definition (\ref{MPD}) of the pseudo-PDF 
	to this particular case,
	\begin{align}
	M(z_3,p)  
	&   = 
	\int_{-1}^1 dx 
	\, e^{i x z_3 P  } \,  {\cal P} (x, z_3^2)  \  , 
	\label{MPD3}
	\end{align}   
	and using this result in the definition (\ref{Qdef}) of the quasi-PDF gives 
	\begin{align}
	Q(y,P) =\frac{P}{2 \pi}  \int_{-1}^1 dx \,   \int_{-\infty}^\infty dz_3\,
	e^{-i(y-x) Pz_3}\, 
	{\cal   P} (x, z_3^2)  \ . 
	\label{Py}
	\end{align} 
	This expression  shows  that    the quasi-PDFs $Q(y,P)$   are  
	defined   
on the whole    real  $y$-axis, despite the fact that   the pseudo-PDFs $ {\cal   P} (x, z_3^2)$ have
support on 
	the limited segment $-1 \leq x \leq 1$  only.  

Another  straightforward, but important  observation is that when a  pseudo-PDF  does not depend on 
its second variable, $z_3^2 $, i.e.,	if 
${\cal   P} (x, z_3^2) =f(x)$, then the integral over $z_3$ 
in Eq. (\ref{Py})   gives $\delta (y-x)/P$, so that  the 
resulting quasi-PDF $Q(y,P)$  has no  dependence  on $P$ and coincides with 
the light-cone distribution 
 $f(y)$.  
 
 Alternatively, when ${\cal   P} (x, z_3^2)$  depends on $z_3^2$,
 the integral over $z_3$ gives a nontrivial function 
 \begin{align}
	R(x,y-x, P) \equiv \frac{P}{2 \pi}   \int_{-\infty}^\infty dz_3\,
	e^{-i(y-x) Pz_3}\, 
	{\cal   P} (x, z_3^2)  \ 
	\label{PyR}
	\end{align} 
	that produces $Q(y,P)$ after the subsequent $x$-integration,
	\begin{align}
	Q(y,P) = \int_{-1}^1 dx \,  R(x,y-x, P)    \ . 
	\label{QyR}
	\end{align}

Thus, it is the  $z_3^2$-dependence of ${\cal   P} (x, z_3^2)$ 
(or, equivalently, of  ${\cal   M} (\nu, z_3^2)$)  that  is responsible for  the deviation 
of quasi-PDFs  from lightcone PDFs. In particular, it generates the parts of $Q(y,P)$ outside 
the PDF support region 
$|y|\leq 1$. 

Eq. (\ref{QyR})  has a simple physical interpretation: the fraction $yP$ of the 
third momentum component carried by the parton comes from two sources: (i)  from the  longitudinal 
motion of the hadron as a whole (which gives $xP$), and  (ii)  from 
the part $(y-x)P$ that is generated by a nontrivial dependence 
 of ${\cal   P} (x, z_3^2)$ on $z_3^2$. 
 In its turn, the $z^2$-dependence of ${\cal P}(x, -z^2)$ is related
 to  spatial distribution of partons inside hadrons.



\subsection{Transverse Momentum Dependent  PDFs}

Recall    that ${\cal P} (x, -z^2)$ is a function  defined  in a covariant manner 
 by Eq. (\ref{MPD}).  This means that if we choose  the spacelike part of $z$ 
 in a plane  $z_\perp =  \{z_1,z_2\}$ perpendicular to  the $z_3$ 
 direction, the resulting pseudo-PDF   will be given by  ${\cal P} (x, z_\perp^2)$,
 i.e., by  the  same function, but with $z_3^2$ substituted by $z_\perp^2$. 
As a result, it is possible  to  show  that the pseudo-PDF $ {\cal P} (x, -z^2)$  for space-like $z$ has
a simple interpretation in terms of the transverse momentum-dependent (TMD)  PDFs\footnote{In the case of QCD,
we define   TMD
PDFs  using a straight-line gauge link as in Eq. (\ref{Malpz}) rather 
than   staple-shaped links. 
}.

Take again the frame where $p= (E, {\bf 0}_\perp, P)$ and 
 choose  a separation 
$z$ for which the lightcone component $z_+$ vanishes,
$z_+=0$,  and   only  $z_-$  and 
\mbox{$z_\perp= \{z_1,z_2\}$}   are non-zero.
 With this choice, we have $\nu = - p_+ z_-$ for the Ioffe time, and 
\mbox{$z^2=-z_\perp^2$}  for the interval.    
The  TMD  can be defined in a standard way through a two-dimensional 
Fourier transform with respect to $z_\perp$, which gives 
\begin{align}
{\cal P} (x,  z_\perp^2) 
&   =  
\int  {d^2 k_\perp }      \,  e^{-i( k_\perp  z_\perp)} 
{\cal F} (x, k_\perp^2)  \  
\label{TMD0} 
\end{align}   
when inverted.
This is equivalent to the following representation 
of the original matrix element,
\begin{align}
{\cal M} (\nu,  z_\perp^2) 
&   = 
\int_{-1}^1 dx \,\, e^{i x  \nu  }  
\int  {d^2 k_\perp }      \,  e^{-i( k_\perp  z_\perp)} 
{\cal F} (x, k_\perp^2)  \  . 
\label{TMD} 
\end{align}   
  Due to the   invariance  with respect to  rotations 
in the $z_\perp$ plane,  
  the   TMD  defined in this way depends on $k_\perp^2$  only. 
  
Eq. (\ref{TMD}) corresponds to a plane-wave 
decomposition in which the 
parton   carries a  longitudinal $xp_+$
fraction (with $-1 \leq x \leq 1$), but it also  has a transverse momentum $k_\perp$.
Since $k_\perp$ 
is Fourier-conjugate to $z_\perp$, we may say that the transverse momentum
dependence of TMDs  is governed by the \mbox{$z^2$-dependence}
 of pseudo-ITDs. 
Similarly, 
 the dependence of ${\cal M} (\nu,  z_\perp^2)$ on $\nu$ governs the \mbox{$x$-dependence } 
of ${\cal F} (x, k_\perp^2)$,
 i.e. the longitudinal momentum structure of the hadron.


Though  the definition of the quasi-PDF is based on  a matrix element involving  
a  purely ``longitudinal''  separation  $z=z_3$,  the Lorentz invariance tells us that
the dependence of 
${\cal M} (\nu,  z_3^2)$ on $z_3^2$ is given by the same function that defines 
the TMD by Eq. ({\ref{TMD}).  This observation allows  us 
	to get a relation between 
	quasi-PDFs  and TMDs.
To this end,  we take $z_\perp=\{0,\nu/P \}$ in Eq.  (\ref{TMD})
	and insert  the resulting representation into the definition  (\ref{Qdef}) of  the quasi-PDF. 
	We obtain  \cite{Radyushkin:2016hsy,Radyushkin:2017cyf}
	\begin{align}
	Q(y, P) =  & \,P 
	\int_{-1} ^ {1} d  x \,  \int_{-\infty}^{\infty} d  k_1  {\cal F} (x, k_1^2+(y-x)^2P^2 ) \ . 
	\label{QTMDrel}
	\end{align} 
	This relation  shows again that 
	the quasi-PDF variable $y$ has the  $-\infty < y <\infty$ support, simply 
	because the value   of the transverse momentum  component
	$ k_2$   in \mbox{${\cal F} (x, k_1^2+k_2^2 )$} 
is   not  limited.

We  also see once more  that the  third component $k_3=yP$   of the parton momentum  is composed from 
the part $xP$,
coming from  the motion of the hadron as a whole,
and the remaining  fraction \mbox{$(y-x)P$}  coming from   the same physics 
that  generates the transverse momentum dependence of the TMDs.

 \subsection{TMD  parametrization} 

Since the pseudo-PDF ${\cal P} (x,  z_\perp^2)$ defining the TMD  ${\cal F} (x, k_\perp^2 )$
has the same functional form as the pseudo-PDF ${\cal P} (x,  -z^2)$
for a general spacelike $z$, we can use TMDs to parametrize   the $z^2$-dependence 
of a generic matrix element $M(z,p)$  for an arbitrary  spacelike $z$.  
To this end, let us take  the TMD definition  (\ref{TMD0})    
and integrate over the angle between ${\bf k}_\perp$ and $ {\bf z}_\perp$. This 
 gives 
 \begin{align}
{\cal P} (x,  z_\perp^2) & = 
2 \pi \,   \int_{0}^{\infty} \, d k_\perp\, k_\perp    %
  J_0 \left ( k_\perp z_\perp \right )   \,  {\cal F } (x, k_\perp^2 )  \  , 
 \label{PcalF}
\end{align} 
where $J_0 $ is the Bessel function. 
Use now the fact 
   that ${\cal P} (x, -z^2)$ is a function  defined  by  a covariant  relation 
 (\ref{MPD}).
This implies  that,    for a general spacelike
$z$,   one can write the representation  for the generic matrix element 
$M(z,p)$ in terms of the TMD, 
 \begin{align}
  \langle p | &   \phi(0) \phi (z)|p \rangle  = \,2 \pi \, 
  \int_{-1}^1 dx\,  e^{-i x (pz)}  
  \int_{0}^{\infty} \, d k\, k\,     %
  J_0 \left ( k \sqrt{-z^2}  \right )   \,  {\cal F } (x, k^2 )
   \  . 
 \label{McalF}
\end{align} 
Here, we intentionally  dropped ``$\perp$''  in the notation 
for the momentum variable. By this change, we stress  that $k$ 
is just an integration  parameter of this representation. While $ {\cal F } (x, k^2 )$
is a  function  that coincides with  the TMD, one does not
need   to specify a ``transverse'' plane  and treat $k$ as the magnitude
of a \mbox{2-dimensional }  momentum in that  plane. \ 
In particular,  nothing prevents us from choosing  $z$ 
 in a purely longitudinal direction,  i.e. from taking  \mbox{$z=\{0,0,0,z_3\}$}.
 Then we can  write 
 \begin{align}
  \langle p | &   \phi(0) \phi (z_3)|p \rangle  = \,2 \pi \,  
  \int_{-1}^1 dx\,  e^{i x Pz_3}  
  \int_{0}^{\infty}  d k\, k\,     %
  J_0 \left ( k z_3  \right )   \,  {\cal F } (x, k^2 )
   \  
 \label{McalF3}
\end{align}  
without  asking  (or answering)  the question:  in which plane  ``$k$'' is supposed to be?
As written, ``$k$'' is just a scalar variable in a particular representation (\ref{McalF3})   for the 
matrix element.

\subsection{Taylor expansion} 

\label{structure}

 The {\it TMD parametrization}  (\ref{McalF}) 
is very general. It just reflects the Lorentz invariance
(the fact that matrix element  $M(z,p)$ depends on  $(pz)$ and $z^2$)
and spectral properties of Feynman diagrams (the limits on $x$
are $-1\leq x \leq 1$),   as given by  the underlying pseudo-PDF representation  (\ref{MPD}). 
It holds for any diagram, 
whether it is regular for $z^2=0$, or 
has $\ln (-z^2)$ singularities. 

However, it is instructive to give a derivation 
of Eq. (\ref{MPD}) for the case when  
 the matrix elements $M(z,p)$ is regular 
for $z=0$ to the extent that all the coefficients of  
 a formal Taylor expansion 
 \begin{align}
 \langle  p | \phi (0) \phi(z)  | p \rangle = \sum_{N=0}^{\infty} 
 \frac{1}{N!}z_{\mu_1} \ldots z_{\mu_N} \ 
 \langle  p | \phi (0) \, {\partial}^{\mu_1} \ldots
{\partial}^{\mu_N}   \phi(0)  | p \rangle \ 
\label{Taylor} 
\end{align}
are  finite.  
Now  the information about the hadron  is 
contained in the matrix elements \\ 
$\langle  p | \phi (0)   {\partial}^{\mu_1} \ldots
{\partial}^{\mu_N}  \phi (0)  | p \rangle $ of local operators.
Due to Lorentz invariance, they may be written as
 \begin{align}
\langle  p | \phi (0)   {\partial}^{\mu_1} \ldots & 
{\partial}^{\mu_N}  \phi (0) | p \rangle = (- i)^N \,  p^{\mu_1} \ldots p^{\mu_n}\, A_N^{(0)} 
\nonumber \\ & + {\rm   terms \ containing} \  g^{\mu_i\mu_j} \  .
\label{ln}
\end{align}
Utilizing the fact that the $\mu_k$ indices are symmetrized  in the Taylor expansion  
by the $z_{\mu_1} \ldots z_{\mu_n}$ factor, we may use a more organized 
expression 
 \begin{align}
\langle  p | \phi (0)   (z\partial)^N \phi (0)  | p \rangle =(-i)^N
\sum_{l=0}^{[N/2]}  (-z^2 \Lambda^2)^l (pz)^{N-2l}A_N^{(l)}  \ . 
\label{ln2}
\end{align}
The   information about the hadron structure is now accumulated  in the constants
$A_N^{(l)} $. 
The momentum  scale $\Lambda$ is introduced 
to  secure that all $A_N^{(l)} $'s have the same dimension. 
In general,    one needs $[N/2]$  (the integer part of $N/2$) constants to parametrize 
this matrix element. 
Treating the constants $A_N^{(l)} $ as $x^N$ moments of  some  functions $F_l (x) $
 \begin{align}
A_N^{(l)} =  \frac{N!}{(N-2l)!} \int_{-1}^1 dx \,x^{N-2l}  \,  F_l (x)  \ ,
\label{ln31}
\end{align}
we obtain the desired pseudo-PDF   representation for the matrix element,
 \begin{align}
 \langle  p | \phi (0) \phi(z)   | p \rangle = &  \int_{-1}^1 dx e^{-ix (pz)}  \, \sum_{l=0}^{\infty}  
  (-z^2 \Lambda^2)^l 
 F_l (x)  \nn &
  \equiv  \int_{-1}^1 dx e^{-ix (pz)}  \
    {\cal P} (x, -z^2)  \  .
   \label{Flx}
\end{align}

As we have seen in Section \ref{HB}, the lowest $l=0$ term produces    the twist-2 contribution
to  the  forward Compton amplitude
The function $F_0 (x) \equiv f(x)$ coincides with   the twist-2  light-cone PDF. 
 
  \subsection{Twist decomposition} 
  
  Usually the twist-2 contribution (\ref{sDelta})  for DIS  is obtained using the {\it twist decomposition}, 
  which is 
a  standard way to parametrize the $z^2$-dependence 
of the  generic matrix element. 
It involves expansion of   $z^{\mu_1} \ldots  z^{\mu_n}  $ factor in Eq. (\ref{Taylor}) 
over  traceless tensors $ \{z_{\mu_1} \ldots z_{\mu_n} \} $.  
In the case  of  scalar fields,  it is possible to derive \cite{Radyushkin:1983mj}
 \begin{align}
 \langle p|  \phi (0)   \phi (z) |p  \rangle  =& \sum_{l=0}^{\infty} 
 \left ( \frac{z^2 }{4}  \right )^l  \sum_{N=0}^{\infty} 
 \frac{N+1}{l!(N+l+1)!}  
 \langle p|   \phi(0)
\{z{\partial} \}^{N}
 ({\partial}^2)^l 
 \phi(0)    |p  \rangle \  ,
 \label{twistD}
\end{align}
where we use the notation
  \begin{align}
 \{z \partial \}^n \equiv   \{z_{\mu_1} \ldots z_{\mu_n} \}  \, 
\partial^{\mu_1} \ldots  \partial^{\mu_n}  \  . 
 \label{trless0} 
 \end{align}
  We can parametrize
the matrix elements   entering Eq. (\ref{twistD}) by
 \begin{align}
 &  \langle  p |   \phi(0)
\{z{\partial} \}^{N}
 ({\partial}^2)^l 
 \phi(0)  | p \rangle 
  =\lambda^{2l} (-i )^N \{zp\}^N \,   B_{N } ^{(l)}  \    , 
 \label{twistP}
\end{align}
where the overall scale $\lambda$ with the dimension of mass
is introduced  in order to have  the coefficients $ B_{N } ^{(l)}$  with the same dimension.
This gives  the twist decomposition 
  \begin{align}
 \langle p|  \phi (0)   \phi (z) |p  \rangle  =& \sum_{l=0}^{\infty} \frac1{l!} \, 
 \left ( \frac{\lambda^2 z^2 }{4}  \right )^l  \sum_{N=0}^{\infty} 
 \frac{N+1}{(N+l+1)!}  \, 
 (-i )^N \{zp\}^N \,   B_{N } ^{(l)}
 \  
 \label{twistDp}
\end{align}
involving the powers of $z^2$ and traceless combinations $\{zp\}^N $. 

Alternatively, if one  applies  the Taylor expansion to $ e^{-i x (pz)} $ and the Bessel 
function $ J_0 \left ( k \sqrt{-z^2}  \right ) $ of the TMD representation (\ref{McalF}), one gets 
 a series  which also involves  the powers of $z^2$,  but they  are accompanied by
simple powers  $(zp)^N $, 
 \begin{align}
  \langle p | &   \phi(0) \phi (z)|p \rangle  =  
\sum_{l=0}^{\infty} \frac{1}{(l!)^2}   \left ( \frac{\Lambda^2 z^2 }{4}  \right )^l   \, 
 \sum_{N=0}^{\infty} 
 \frac{1}{N!}  \, 
 (-i )^N (zp)^N \,   C_{N } ^{(l)}
   \  . 
 \label{McalFT}
\end{align} 
The coefficients  $C_{N } ^{(l)} $ here are the combined moments of the TMD 
 \begin{align}
 C_{N } ^{(l)} =   \,2 \pi \, 
  \int_{-1}^1 dx\, x^N   
  \int_{0}^{\infty} \, d k\, k\,     %
 (k^2)^l    \,  {\cal F } (x, k^2 ) 
   \  . 
 \label{AlN}
\end{align} 
 To shorten formulas,  we may   switch here  back \mbox{$k \to k_\perp$}  in the  {\it  notation } 
 for the integration variable,  
and also write the resulting $2\pi k_\perp dk_\perp$ as $d^2 k_\perp$. 
  We can  do this because the TMD ${\cal F} (x, k_\perp^2)$ does not depend on the polar angle. 
  Then 
   \begin{align}
 C_{N } ^{(l)} =   
  \int_{-1}^1 dx\, x^N   
  \int_{0}^{\infty} \, d^2 k_\perp \,      %
 (k_\perp^2)^l    \,  {\cal F } (x, k_\perp^2 ) \equiv \langle  x^{N} 
    k_\perp^{2l} \rangle_{\cal F}
   \  . 
 \label{AlNperp}
\end{align} 
We emphasize  again    that $k$ or  $k_\perp$ should be understood 
  simply as scalar  integration variables.   We do not   need to specify 
  in which plane $k_\perp$ is.

An obvious    advantage of the TMD representation  (\ref{McalF}) 
 is  that, unlike the twist decomposition (\ref{twistDp}), it displays the \mbox{$z$-dependence} 
of $M(z,p)$ by  a closed formula   rather than by a double series.  Furthermore, 
  the $(pz)$-dependence in Eq. (\ref{McalF}) comes 
 through the plane waves $e^{-ix (pz)}$. 
 As a result, most integrals over $z$ (such as the  integral
 (\ref{Qdef}) producing quasi-PDFs) are straightforward. 
 
In contrast,  the twist decomposition (\ref{twistDp})   
is a series expansion over  combinations 
$
 \{pz\}^n $    involving  a  rather  complicated  traceless 
 tensor $ \{z_{\mu_1} \ldots z_{\mu_n} \} \  $.  
 This results in an  involved procedure for  integrations  over $z$.
 Furthermore, the task of summing series in $
 \{pz\}^n $  structures  into a closed form
  is  rather tricky.  
In fact, even when successful, the results of such a summation involve imaginary exponentials
  of $(pz) \pm \sqrt{(pz)^2 -p^2 z^2}$ which are next to impossible to integrate in  a general case.

Another    practical  advantage of the TMD representation  is that it  describes the $z^2$-dependence
  through  Eq.  (\ref{McalF})  involving 
a   ``spectral function'' 
 ${\cal F } (x, k^2 )  $  that has 
a clear physical interpretation of a transverse momentum distribution.  
Based on this interpretation, one may  expect that
${\cal F } (x, k^2 )  $ is a function finite for \mbox{$k=0$}  and monotonically decreasing with $k$ 
when $k$ increases. One should not expect oscillations or other exotics in its 
$k$-dependence. 

Taking some model for ${\cal F } (x, k^2 )$, one can get a model for the pseudo-PDF,
\begin{align}
 {\cal P} (x, -z^2) =      \,2 \pi \, 
  \int_{0}^{\infty} \, d k\, k\,     %
  J_0 \left ( k \sqrt{-z^2}  \right )   \,  {\cal F } (x, k^2 )
   \  . 
 \label{PcalF2}
\end{align} 
The pseudo-PDF representation (\ref{MPD}) gives then a model for the ITD.

 \setcounter{equation}{0}  
 
   \section{Nonperturbative evolution of soft quasi-PDFs} 
   
   \label{soft}

Having a model   for $ {\cal F } (x, k^2 )$  and using the quasi-PDF/TMD relation (\ref{QTMDrel}),
one can  get a model for the quasi-PDF and 
study qualitative features  of   possible \mbox{$P$-dependence}   patterns of the 
 quasi-PDFs $Q(y,P)$.    Such models were proposed  originally in our paper 
 [\citen{Radyushkin:2016hsy}].

  \subsection{Models for  soft quasi-PDFs}

 Since our goal is to study  general features of the 
 \mbox{$P$-dependence, }   it makes sense to take simple, but still realistic  
 models\footnote{A sophisticated model of  power corrections for ITDs  induced by infrared renormalons 
 was recently proposed  in Ref. [\citen{Braun:2018brg}]. However, the discussion of this model 
 is out of the scope of the present paper.}. 
 In this respect, the collinear model ${\cal F} (x, k_\perp^2 ) =f(x) \delta (k_\perp^2)/\pi$
 is not realistic, since it gives $\langle k_\perp^2 \rangle =0$.   
Next in complexity are factorized models, 
 in which 
 ${\cal F}  (x, k_\perp^2)$ is  given   by a   product 
    \begin{align}
{\cal F} (x, k_\perp^2) = f(x) K(k_\perp^2)
\label{fact} 
\end{align} 
 of a 
 collinear PDF  $f(x)$  and a \mbox{$k_\perp^2$-dependent}  factor $K(k_\perp^2)$. 
Such models are used on a daily basis by  TMD practitioners (see, e.g. Ref. 
 [\citen{Anselmino:2013lza}]), with the 
 Gaussian  form  
  \begin{align}
K_G(k_\perp^2) =  & 
\frac{1}{\pi  \Lambda^2}  e^{-k_\perp^2/\Lambda^2}
 \label{Gauss}
\end{align}  
being the most popular choice.  
 For  the Ioffe-time pseudo-distribution  ${\cal M} (\nu,z_3^2)$,  this  corresponds 
to the factorized  Ansatz 
   \begin{align}
   {\cal M}^{\rm soft}  (\nu,z_3^2) = {\cal I}^{\rm soft}  (\nu,0){\cal M} (0,z_3^2) \ 
    \end{align} 
    for its soft part.  In this case, the soft part of the reduced ITD ${\mathfrak M} (\nu, z_3^2)$ 
    does not have $z_3^2$-dependence and coincides with ${\cal I}^{\rm soft}  (\nu,0)$. 
     While there seem to be no first-principle  reasons for 
    such a   factorization property for $ {\cal M}^{\rm soft}  (\nu,z_3^2) $, it   has been  actually observed 
     in a recent lattice study of the 
Ioffe-time pseudo-distributions  in Ref. 
[\citen{Orginos:2017kos}]. 
It is also supported  by  the pioneering study\cite{Musch:2010ka} of the  TMDs 
 in the   lattice QCD performed a decade  ago.

Using the factorized model (\ref{fact})  with the Gaussian shape   (\ref{Gauss}) 
gives the following model  for the quasi-PDF
 \begin{align}
 Q_G(y, P)  = &\frac{P}{\Lambda \sqrt{\pi} }  \,
 \int_{-1}^1 dx\,  %
f(x) \, 
  e^{- (x -y)^2 P^2 / \Lambda^2 }
 \  . 
 \label{QinG}
\end{align} 
It is instructive to   choose   $f(x)$  in 
the  form
 \begin{align}
f(x) =  \frac{315}{32} \, \sqrt{x} (1-x)^3 \theta(x>0)
 \label{qlat}
\end{align} 
obtained in Ref. [\citen{Orginos:2017kos}] 
for the soft part of the $u_v (x) -d_v (x)$  nucleon PDF.
It was  extracted from the lattice  data 
using the {\it pseudo-PDF}-based  method proposed in  our paper [\citen{Radyushkin:2017cyf}].
Our goal is to  investigate, what kind of {\it quasi-PDFs} one would get 
for  such a PDF in the Gaussian factorized model.

 \begin{figure}[h]
 \centerline{\includegraphics[width=2.3in]{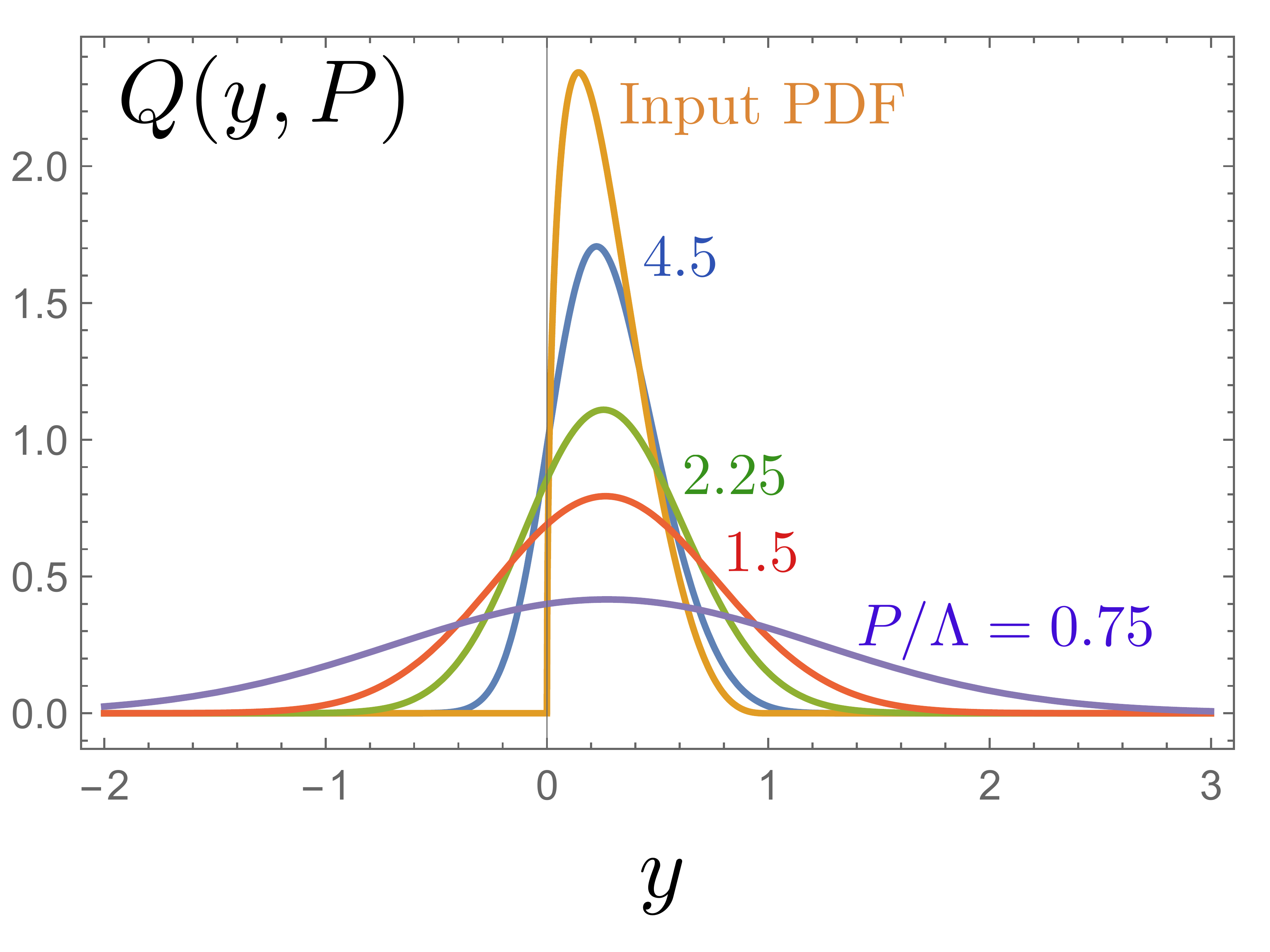}}
    \vspace{-0.2cm}
    \caption{Evolution of $Q(y,P)$  in the Gaussian  model  for $P/\Lambda = 0.75,  1.5,  2.25, 4.5$
 compared to the  limiting PDF 
  \mbox{ $f(y)= \frac{315}{32} \, \sqrt{y} (1-y)^3$.} 
    \label{QgyL}}
    \end{figure}

The curves are shown in Fig. \ref{QgyL}. 
One can see that   the quasi-PDFs have large parts outside the support segment 
$0\leq y \leq 1$ of the input PDFs.  
As we have already emphasized in Sec. IIF, 
 it  is the $z_3^2$ dependence of 
	the pseudo-ITD ${\cal M} (\nu, z_3^2)$ that generates those parts of the quasi-PDFs
	that are outside the $-1\leq y \leq 1$ region.
When  $P$ increases,  the quasi-PDFs in Fig. \ref{QgyL} shrink inside the  $0 \leq y \leq 1$  segment.
In particular, the area under the negative-$y$ part of $Q^{\rm soft} (y,P)$ quickly decreases 
for large $P$, and vanishes in the $P\to \infty$ limit.

The change of  quasi-PDFs with $P$ in  our Gaussian model is very close to that actually 
 observed in the   lattice QCD calculations for the $u_v (x) -d_v (x)$  nucleon PDF
  reported in Ref. [\citen{Lin:2014zya}].
In fact, our curves corresponding to $P/\Lambda$   in multiples of 0.75,
namely for  $P/\Lambda= 0.75 , 1.5, 2.25 $ are  close 
to the curves of Ref. [\citen{Lin:2014zya}] obtained for the momentum values
in the same 1, 2, 3 multiples of 
$p=2\pi/L$ (i.e., for  $p=2\pi/L, 4 \pi/L, 6\pi/L$, correspondingly). 
This is  a  very suggestive   indication that 
the major role  in forming the observed shape of quasi-PDFs 
is played by the non-perturbative physics reflecting the hadron size.

Looking at the curves  shown in  Fig. \ref{QgyL},
one  sees  that the maximal value of the quasi-PDF $Q (y,P)$
for  \mbox{$P=2.25 \Lambda$}  is more than twice lower than the maximal value 
of the input PDF.  A natural   conclusion is that  
the   $P=2.25 \Lambda$ momentum
is  simply too small.  
Namely, to convert  $Q(y, P=2.25 \Lambda)$  into the input PDF, one needs  corrections
of the same size as the  $Q(y, P=2.25 \Lambda)$   quasi-PDF itself. 
It is necessary to at least  double $P$ to get a quasi-PDF that 
is sufficiently close to the limiting PDF form. Only then one may have some hope  that the remaining gap
may be fixed by adding  corrections that are not too large  compared to the starting
approximation.

Larger values of $P$, namely, $P=10 \pi/L$,  were reached in the lattice 
calculation of Ref. [\citen{Alexandrou:2016jqi}]. One may check that the quasi-PDF obtained 
in that paper is  very close to the $P/\Lambda =4.5$ curve of Fig. \ref{QgyL}.
While being much closer to the  input PDF, the $P/\Lambda =4.5$ curve 
still shows strong  artifacts of unfinished nonperturbative evolution,
in particular,  a rather  large signal for negative $y$.

From the value $P=1.29$ GeV indicated in  Ref. [\citen{Lin:2014zya}]  for the highest momentum $p=6\pi/L$,
we can also estimate the magnitude $\Lambda^2 \approx (600$ MeV)$^2$ of the effective Gaussian parameter. 
This  is much larger than  $\langle k_\perp^2 \rangle \approx (300$ MeV)$^2$ that one would expect 
from the transverse momentum distribution. However,  
 $  {\cal M}^{\rm soft}  (\nu,z_3^2)$  in this case 
reflects both the $z_3$-dependence induced by  the  nonperturbative dependence of 
TMDs  and  also the $z_3$ dependence 
of the gauge-link-related  $Z(z_3^2/a^2)$ factor.  The latter 
was not removed in the calculation of  Ref. [\citen{Lin:2014zya}]. 

A more recent lattice calculation 
\cite{Alexandrou:2019lfo} includes renormalization of the link-related UV singularities. 
This procedure should eliminate, to some extent, the nonperturbative 
$z_3^2$-dependence of the   $Z(z_3^2/a^2)$ factor from the renormalized data.
Still, the  $z_3^2$-dependence induced by the transverse-momentum distribution 
may be there.  Indeed, the quasi-PDF shown in Fig. 29 of Ref. [\citen{Alexandrou:2019lfo}] 
has all the features of unfinished  nonperturbative evolution.

   \subsection{Rate of approach} 
  \label{secrate}

 One may  also  be interested in which way 
 the finite-$P$ quasi-PDF curves approach the limiting  PDF  curve.
 To get the answer in a short  analytic form, let us take a very simple  input PDF 
 $f(x)=1-x$ and the same Gaussian Ansatz (\ref{Gauss}) for the $k_\perp$-dependence. 
   In this case, we have 
     \begin{align}
Q(y,P)= & \frac12 (1-y)
   \Big[\text{erf}\left[{(1-y)}P/{\Lambda}
   \right]+\text{erf}\left[{y}P/{\Lambda}
   \right]\Big ]  \nn &
   + \frac{ \Lambda}{2\sqrt{\pi }P}
   \left[e^{-{(1-y)^2}P^2/{\Lambda^2}}-e^{-
   {y^2}P^2/{\Lambda^2}}\right] \  , 
\end{align} 
where  the error function is defined by 
 \begin{align}
\text{ erf} (z)=\frac{2}{\sqrt{\pi}}\int_0^z dt \, e^{-t^2} \  . 
 \end{align} 
 For large $z$, it may be approximated by
  \begin{align}
\text{ erf} (z)=1 -   \frac{ e^{-z^2} }{\sqrt{\pi}z} \left [1- \frac1{2z^2} - \ldots   \right ] \  . 
 \end{align} 
 As a result, the approach to the $P\to \infty$ limit is governed by the exponentials 
 $e^{-(1-y)P^2/\Lambda^2}$  and
 $e^{-y P^2/\Lambda^2}$.  In particular, at    the middle of the 
 $0 \leq y\leq 1$  interval, we have 
     \begin{align}
& Q(1/2,P)= 
 \frac{1}{2} \text{erf}\left(\frac{P}{2
   \Lambda }\right) 
   =  \frac{1}{2}  -   \frac{\Lambda e^{-P^2/4 \Lambda^2} }{\sqrt{\pi} P} 
   \left [1- \frac{2 \Lambda^2}{P^2} - \ldots   \right ]  \ .
     \end{align} 
     Thus, the approach to the limiting value is exponential $\sim  e^{-P^2/4 \Lambda^2} $
     rather than a powerlike.
         \begin{figure}[t]
     	\centerline{\includegraphics[width=2.3in]{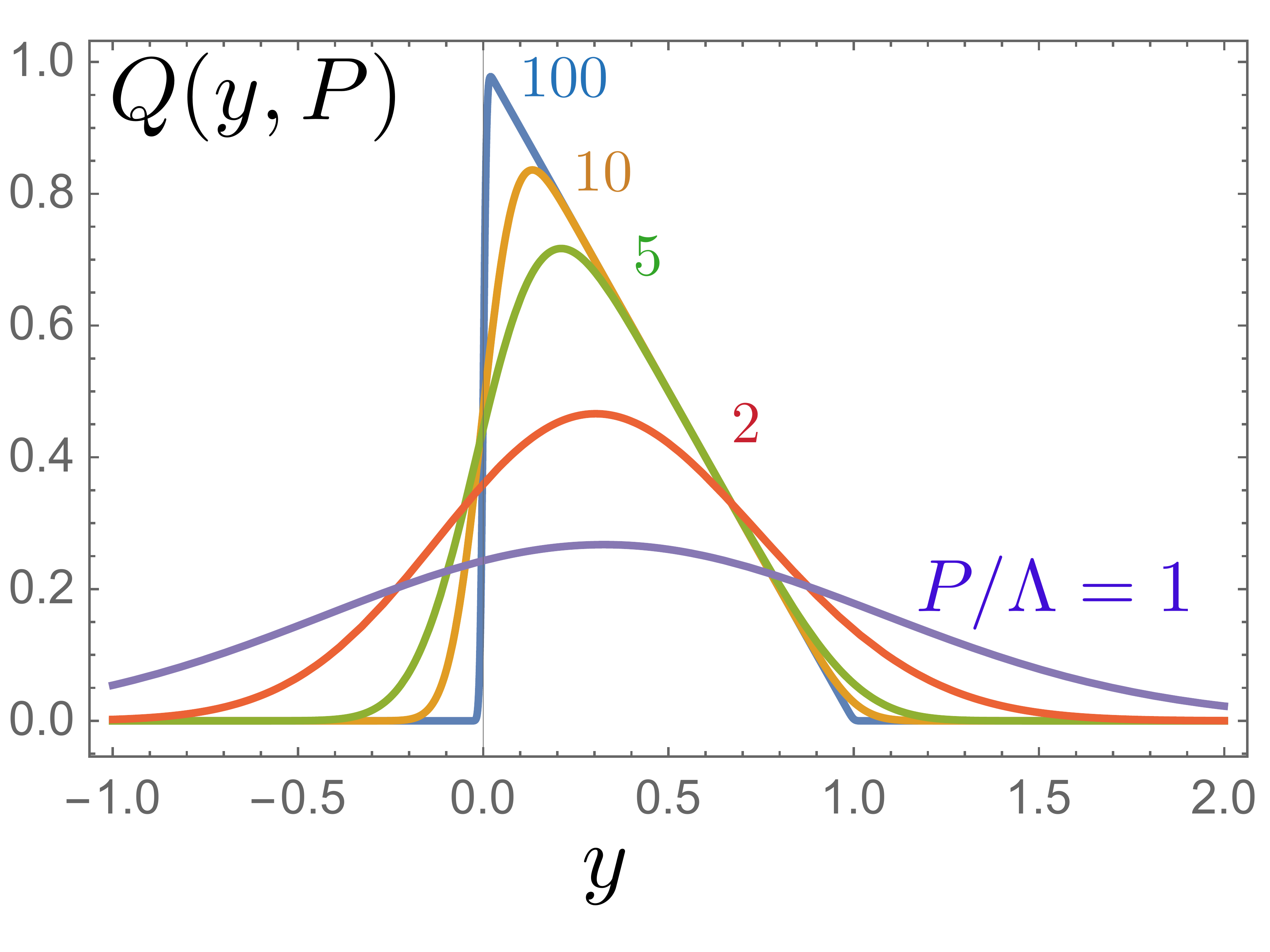}}
     	\caption{Evolution of $Q(y,P)$  in the Gaussian  model  for {$P/\Lambda =1,2,5,10,100$}. 
     		The  limiting PDF  is $f(y) = (1-y)$.
     		\label{Qgy}}
     \end{figure}
     At the end-points,  one  of the exponentials  converts into 1, 
 so these are special cases.  The  input  PDF $f(y)=1-y$ vanishes for $y=1$,
 and the quasi-PDF approaches this limit according to  
       \begin{align}
Q(1,P)= &
     \frac{\Lambda 
   }{2 \sqrt{\pi } P}\left[1-e^{-{P^2}/{\Lambda
   ^2}}\right] \ , 
     \end{align} 
     i.e. like $\sqrt{\Lambda^2/P^2}$ rather than $\Lambda^2/P^2$.
      The non-analytic behavior with respect to $\Lambda^2/P^2$ 
     is  present at another end-point as well 
      \begin{align}
&Q(0,P)= 
 \frac{1}{2}
   \text{erf}\left(\frac{P}{\Lambda
   }\right)+\frac{\Lambda 
   }{2 \sqrt{\pi } P}\left[1-e^{-{P^2}/{\Lambda
   ^2}}\right]
 \nn  & 
 = 
 \frac{1}{2}  
+\frac{\Lambda 
   }{2 \sqrt{\pi } P}\left[1-2 e^{-{P^2}/{\Lambda
   ^2}}   \left (1- \frac{ \Lambda^2}{4 P^2} - \ldots   \right ) \right]  \ . 
    \end{align} 
 As one can see, at $y=0$, the quasi-PDF  approaches 1/2, the average of 
 its $0_+$ and $0_-$ limits  of the input PDF  at that point. 
 The curves for $Q(y,P)$ in this model are shown in Fig. \ref{Qgy}.

 It is also instructive to look at the curves illustrating the $P$-dependence 
 of quasi-PDFs at particular values of $y$ (see Fig. \ref{rate}).
 It is clear that having just three points, at $P/\Lambda=0.75,   1.5$ and 2.25,
 it is rather difficult to make an accurate extrapolation to correct $P=\infty$ values. 
 
 \begin{figure}[h]
        	\centerline{\includegraphics[width=2.3in]{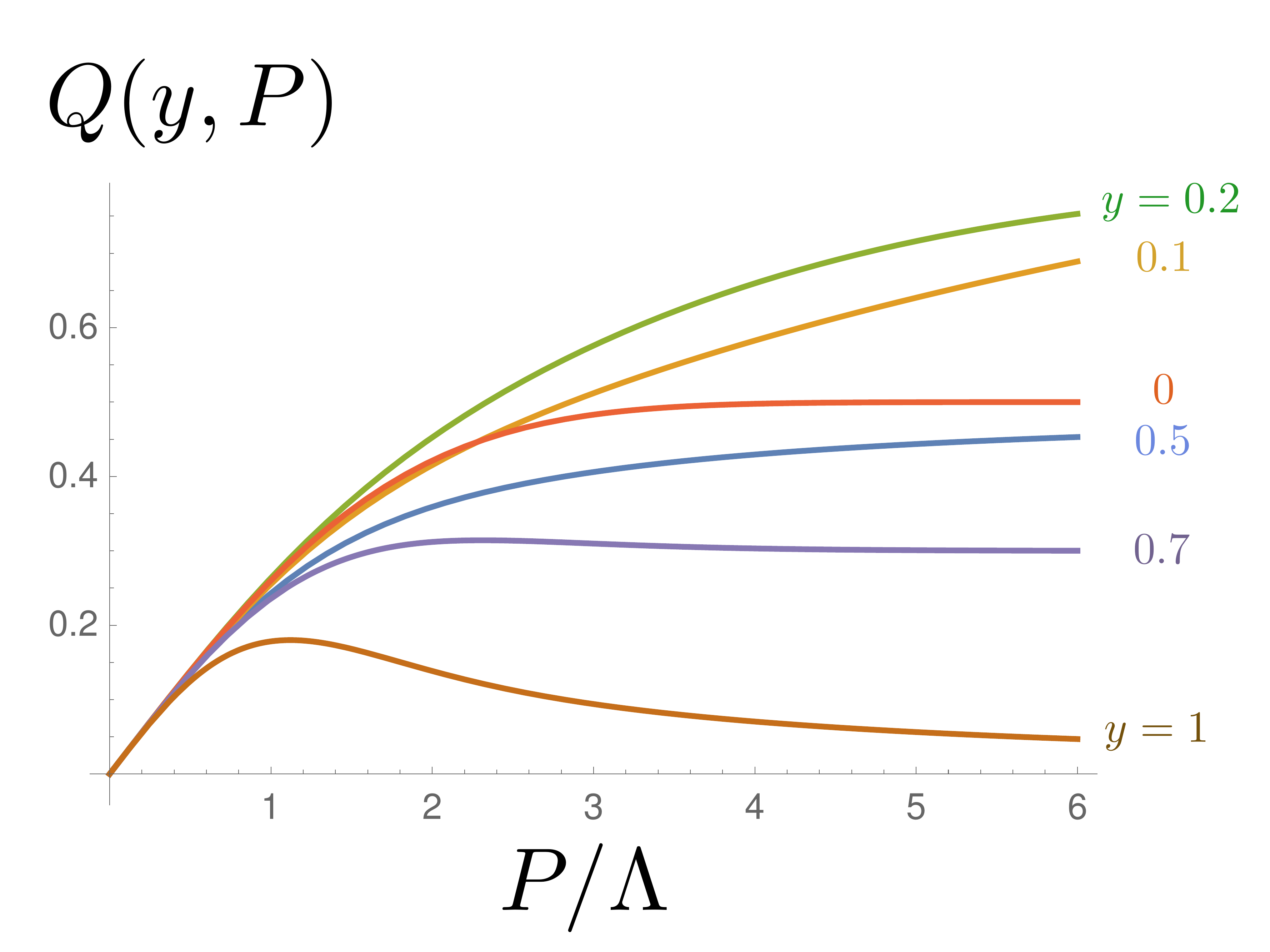}}
        	\caption{$P$-dependence of  $Q(y,P)$  in the Gaussian  model  for indicated values of $y$. 
        		\label{rate}}
        \end{figure}
 
Summarizing, we see  that  the $k_\perp$ effects generate a very nontrivial pattern
of nonperturbative evolution of the quasi-PDFs   $Q(y,P)$.
We also observed that,  in the case of a Gaussian TMD, this evolution
  cannot be described 
by a   ${\cal O} (\Lambda^2/P^2)$ corrections  on the point-by-point   basis  in \mbox{$y$-variable.}


      \subsection{Expansion in $1/P^{2l}$} 
      
      \label{expansion} 
 
Using the TMD parametrization (\ref{McalF3}),
we can write a   formal $1/P^{2l}$ expansion   for  the quasi-PDF $Q(y,P)$   
  \begin{align}
 Q(y, P)  \Rightarrow  &  f (y) +   \sum_{l=1}^\infty  \int  d^2 k_\perp  
  \frac{k_\perp^{2l} } {4^l P^{2l} (l!)^2}     \,  
 \frac{\partial^{2l}}{\partial y^{2l}}
 {\cal F } (y, k_\perp ^2)
  \ . 
 \label{QyPDel}
\end{align} 
  
  Thus,  the  scale  $\Lambda^2$  characterizing  the size of  the higher-twist $1/P^{2l}$ corrections  
 is set by the magnitude of the $k_\perp^{2l}$  moments of the soft TMD  
 ${\cal F }^{\rm soft}  (y, k_\perp ^2)$.  
We attached the subscript ``soft'' here, because it is evident that 
the  $k_\perp^{2l}$ moments diverge for the hard part that has the $\sim 1/ k_\perp^{2}$
behavior for large  $k_\perp$.
Furthermore, Eq. (\ref{QyPDel}), ``as is'',  has a mathematical meaning 
only if the TMD decreases faster than any inverse power of $k_\perp^2$ for large $k_\perp$,
say, like a Gaussian $e^{-k_\perp^2/\Lambda^2}$ or an exponential $e^{-k_\perp/\Lambda}$.
Such distributions may be called   as  {\it ``very soft''. }

 \subsection{Target-mass corrections}


According to the TMD parametrization (\ref{McalF3}),
 the difference between the quasi-PDF $Q(y,P)$ and  the PDF $f(y)$ in
Eq. (\ref{QyPDel}) 
   is  described  by the  $k_\perp^{2l}$ moments of TMDs.
  Thus,   the size of all  the $(\Lambda^2/P^{2})^l$  corrections in the relation
  between $Q(y,P)$ and $f(y)$  is determined 
  by these moments, and  the  scale $\Lambda^2$ 
  is set by the average value $\langle k_\perp^{2} \rangle_{\cal F}$ of the transverse momentum\footnote{Recall  the $ \langle  x^N k_\perp^{2l} \rangle_
 {\cal F} $ notation introduced in Eq. (\ref{AlNperp}). We   also use the notation \mbox{$ 
 \langle   x^N \rangle_
f \equiv \langle  x^N\rangle_ {\cal F}   $}   for averages involving  $l=0$. }.

   Still, a  usual statement  \cite{Ji:2013dva,Lin:2014zya,Alexandrou:2016jqi} is that there  are 
   two types of $(1/P^2)^l$ contributions: 
 {\it target-mass} corrections $(M^2/P^2)^l$ and
   {\it higher-twist}  corrections $(\lambda^2/P^2)^l$. 
   There is no contradiction. Indeed, 
   while there is no explicit source of 
   target mass corrections visible  in the TMD parametrization,  
 such terms do appear if one converts it into 
 the twist decomposition.
The latter 
 can be obtained   by expanding $ (z\partial)^n $ in Eq. (\ref{ln2})  over 
traceless combinations. 
Take the simplest nontrivial case
$n=2$. Then 
 \begin{align}
  (z \partial)^2 = &
\{z \partial \}^2  +\frac14 z^2 \partial^2 
\ , 
\label{twdec2}
\end{align}
where  $\{z\partial\}^{2}$ is the  notation defined in Eq. (\ref{trless0}). 
 Parametrizing  
the matrix element 
 \begin{align}
 \langle p |   \phi (0) \partial^2\phi(0) |p \rangle  = \lambda^2 
 \langle p |   \phi (0) \phi(0) |p \rangle.
 \label{twist4}
\end{align}
we introduce the higher-twist scale $\lambda^2$  generated by $\partial^2$. 
One may interpret $\lambda^2$ as the  average of $(-k^2)$, 
or parton virtuality. By equations of motion, $\partial^2 \phi   = g \psi \phi$,
so one may  also  interpret  $\lambda^2$ as the average strength of the gluon field $g \psi$. 
Since powers of $\partial^2$ are accompanied by powers of
$z^2$, one gets $(\lambda^2/P^2)^l$  corrections for quasi-PDFs. 
From this physical interpretation, one would expect that 
the higher-twist scale $\lambda^2$ is close to the average transverse
momentum scale $\Lambda^2$. 

The target mass correction   $M^2/P^2$ appears when one 
parametrizes the matrix element of the traceless part 
 \begin{align}
 \langle p |   \phi (0) \{z \partial \}^2 \phi(0) |p \rangle  = 
\{z p\}^2  \langle x^2  \rangle_f 
 \label{twist2}
\end{align}
and then 
expands the 
traceless combination  $\{zp\}^2 $ over the powers of the usual scalar product $(pz)$ 
and $z^2$,   
 \begin{align}
\{z p\}^2 =& (z p)^2 -\frac14 z^2 M^2
\label{twTMD} 
\ . 
\end{align}
Hence, applying  the twist decomposition (\ref{twdec2})  to  the  matrix element,  
we  have 
  \begin{align}
  &
 \langle p |   \phi (0) (z\partial)^2\phi(0) |p \rangle 
 =  
 -  \left [ (zp )^{2} - \frac14 z^2 M^2 \right ] \,   \langle x^2  \rangle_f
  +
\frac{z^2 }{4}  
 \,
\lambda^2
 \   . 
  \label{mebil}
\end{align} 
On the other hand, using the TMD parametrization (\ref{QyPDel}),   we get 
  \begin{align}
  &
\langle p |   \phi (0) (z\partial)^2 \phi(0) |p \rangle   =  
 -   \, (zp )^{2} \, \langle x^2  \rangle_f
  +
 \frac{z^2 }{2} \langle  k_\perp^2\rangle_
 {\cal F} 
 \  .
  \label{mebilF0}
\end{align}

This gives a  relation between 
the  parameters of the TMD parametrization and  those of  the twist decomposition
  \begin{align}
  &
 \,
   \lambda^2+ M^2  \langle x^2  \rangle_f= 
2\langle   k_\perp^2\rangle_
 {\cal F} 
 \   . 
  \label{mebilF}
\end{align}
This outcome  may be also obtained by a direct application of 
$ \partial^2$   to the TMD parametrization   (\ref{McalF})
and then taking   $z=0$.  
In the momentum representation, $ \partial^2$ 
results in $(-k^2)$, the parton virtuality.
Thus, we may say that it is given by a kinematical term $(-x^2 M^2)$ 
and a contribution due to the parton's transverse momentum.

When $\partial^2  \phi=0$, or for ``on-shell'' quarks,
the average transverse momentum $\langle   k_\perp^2\rangle_
 {\cal F} $ is completely determined by
 the proton mass and the twist-2 parton   distribution $f(x)$.
 This was    known for a long time\cite{Barbieri:1976rd,Ellis:1982cd}.
Moreover,  it may be  shown\cite{Radyushkin:2017ffo}  that, if   one   
neglects all the higher-twist contributions,
  the TMD  can be expressed in terms 
 of the twist-2 PDF  $f(x)$ 
  \begin{align}
    {\cal F} _{\rm twist-2}(x, k_\perp^2) = - 
\frac1{x\pi M^2} \,   \   
f' (x+  k_\perp^2 /xM^2) 
\  . 
\label{tw2f}
 \end{align} 
Since $f(x)$  has the $0\leq x \leq 1$ support,
this   TMD has a peculiar    restriction \mbox{$k_\perp^2 \leq x (1-x) M^2$,}
  conflicting  with the  expectation that
the values of $k_\perp$ are not limited.

When the higher twists are  nonzero,  the question is essentially  which  basis to choose 
 for $1/P^{2l}$ 
corrections. If one  uses  the twist decomposition, 
then the operators $\phi (\partial^2)^l \phi$ are   accompanied 
by $(\lambda^2/P^2)^{l}$ overall factor, and   there are also further  target mass corrections
 in powers of $M^2/P^2$ generated by  traceless structures.
 
 On the other  hand, if one chooses the TMD parametrization,
 then   there is only one source of $1/P^{2l}$ corrections.
 They are produced by the $k_\perp^{2l}$ moments of TMDs.
 In this sense, the TMD ${\cal F} (x, k_\perp^2)$  serves 
 as a generating function for corrections in powers of $\Lambda^2/P^2$. 
 This is a clear advantage of  the TMD parametrization.
 
Nevertheless,  one can imagine a  scenario  when it would be more 
preferable to use the twist decomposition.
Namely, when the  matrix elements of operators with powers of $\partial^2$ 
are much smaller than the target mass correction terms.
Then, in particular, the moment $\langle  k_\perp^2 \rangle_{\cal F}$ 
will be dominated by $M^2 \langle x^2 \rangle_f /2$ 
and could be calculated from the twist-2 PDF $f(x)$. 
 Thus,  it  is instructive to  make an estimate.  Take 
 a simple model for  valence quark PDF $f(x)= \frac{35}{32} (1-x)^3/\sqrt{x}$, then 
  \begin{align}
&\frac{M^2}{2}   \int_0^1 dx\,  x^2 f (x) = \frac{M^2}{66}  \approx 0.013\,  {\rm GeV} ^2 \ . 
 \label{d21}
\end{align} 
This should be compared to, say,  the value ${\langle k_\perp^2  \rangle} _G = \Lambda^2$, 
that gives a Gaussian TMD (\ref{Gauss}).  
 Even if one takes $\Lambda$ as small as 300 MeV,   $\Lambda^2$
is numerically  
about 0.1\, GeV$^2$, 
So, there are no reasons to expect that the target-mass corrections 
are larger than the $k_\perp$ effects. In fact, all the evidence is that 
they are much smaller.

It should be also  emphasized that the ``target-mass corrections''
appear only within the twist decomposition. 
The latter may   be  obtained 
from the TMD parametrization (\ref{McalF}) 
by an artificial procedure of expanding   the $e^{-ix (pz)}$  factor there 
over the traceless combinations $\{pz \}^k$. 
In this sense, the target-mass corrections are created ``by hand''.
If one chooses to work with the TMD parametrization,
there are {\it no}  kinematical 
 target mass corrections. All the $1/P^{2l}$ corrections are 
described by the $k_\perp^{2l}$ moments of the TMDs.

The same is true for the  twist decomposition of  the pseudo-PDF representation (\ref{MPD}):
there  is no need  to expand  $e^{-ix (pz)}$ there over $\{pz \}^k$. 

 One may wonder, why did we have  the target mass corrections in  
 the expression (\ref{sDelta}) for the handbag structure function?
The answer was given at the end of Sec. \ref{HB}:
 if ${\cal P} (x, -z^2)$ is analytic on the light cone, the scalar handbag diagram 
is given by the twist-2 part alone, because the powers $(-z^2)^l$ from the Taylor expansion 
of ${\cal P} (x, -z^2)$ cancel the $1/z^2$ singularity  of the scalar propagator, 
resulting in   contributions that are treated as zero.  The remaining 
terms are purely 
twist-2 contribution.

 \subsection{Quasi-PDFs  for twist-2 part}
 
 As  discussed  in Ref. [\citen{Radyushkin:2017ffo}],
the quasi-PDFs built from the twist-2 terms may be calculated in explicit form.
  This gives a possibility to    check,  to   which extent the resulting curves agree with 
the curves obtained in actual lattice calculations.  
 
Let us  investigate a  scenario when all the  higher-twist 
operators involving powers of $\partial^2$ vanish, and the
$k_\perp^{2 l} $ 
 moments of the TMD are completely determined  by the \mbox{twist-2}  target mass effects.
 The matrix element $ \langle p |   \phi(0) \phi (z)|p \rangle$  is given  then by 
 its twist-2 part 
 \begin{align}
  \langle p |   \phi(0) \phi (z)|p \rangle |_{\rm twist-2} 
=  &\int_{-1}^1 dx \, f(x)   \sum_{n=0}^\infty (-ix)^n \, \frac{\{z p \} ^n}{n!} \  .
\label{tw2sum}
\end{align} 
In fact, the 
structures 
$
\{zp \}^n 
$
built from the  traceless combinations may be  written 
in  terms of simple powers,
\begin{align} 
\{zp \}^{n}  = (zp)^{n} \frac{ [1+r ]^{n+1} -[1-r ]^{n+1} }{2^{n+1}  r} 
\  ,
\label{zpnR}
\end{align}
where 
$r=\sqrt{1 -z^2 p^2 /(zp)^2} $ (see, e.g., Ref. [\citen{Radyushkin:1983mj}]).
This result  further  simplifies when  \mbox{$z=z_3$}  and  $p=(E, 0_\perp, P)$. Then 
we have  
$  
r=   \sqrt{1+ M^2/P^2  } =E /P
$
and 
\begin{align} 
\{zp \}^{n} =   (-1)^n z_3^n  \frac{ [P + E  ]^{n+1} -[P -  E ]^{n+1} }{2^{n+1}  E } 
\  .
\label{zpnR2}
\end{align} 
For the twist-2 part of the operator $\phi(0) \phi (z)$, this gives 
 \begin{align}
 & \langle p |   \phi(0) \phi (z_3)|p \rangle |_{\rm twist-2} 
=  \int_{-1}^1 dx \, f(x) 
 \left [ \frac{E+P}{2E} e^{ ixz_3(P+E)/2} +
  \frac{E-P}{2E} e^{ ixz_3(P-E)/2} \right ] \ ,
\end{align} 
and we get the ``twist-2 part''  of  the quasi-PDF  in the form 
 \begin{align}
 Q_{\rm twist-2}   (y, P)  = & \,   
  \frac{1}{1+  2 \Delta } \left [ f \left ( y / (1+ {\Delta }\right ) + 
f \left (- y /\Delta  \right )  \right ] 
 \ , 
 \label{newVDFzQin2}
\end{align} 
where 
$$
\Delta = \frac{E-P}{2P} = \frac{M^2}{4 P^2} + \ldots  \ \  . 
$$
This  result   (in  somewhat different way and  notations)  was originally obtained
in Ref. [\citen{Chen:2016utp}].  The derivation  presented  above was given  in our  
paper  [\citen{Radyushkin:2017ffo}].

Let us see what kind of quasi-PDFs one would get for some model PDF.
To begin with, we note that since the quasi-PDFs  $Q  (y, P) $ 
for negative $y$ may come both from the 
$y>0$ and $y<0$ parts of the PDF $f(y)$, it makes sense 
to split $f(y)$ in these two  parts and analyze 
quasi-PDFs  coming from each of them separately. 

We will   take the input PDF that is nonzero for positive $y$ only.
 For illustration, we take again the  model PDF (\ref{qlat})
    obtained in Ref. [\citen{Orginos:2017kos}]  using the approach based on reduced pseudo-ITD.  
The shape  and  the $P$-dependence  of the twist-2 quasi-PDFs shown in 
 Fig. \ref{tw2}  may be compared to that of 
 the  quasi-PDFs   given by the  Gaussian  model 
 and shown in Fig. \ref{QgyL}.
 
  Again, we have a signal for  negative $y$ despite the fact that 
 the input PDF is zero  in that region.
 The area under the negative-$y$ part of the curve decreases
 when $P$ increases and vanishes in the $P\to \infty$ limit. 
 One can see also that 
 the curve for $P=M$     is as close 
    to the input PDF  as the $P=4.5 \Lambda$ (i.e., $P \sim $ 2.5 GeV) 
    curve of the Gaussian model shown in Fig.
      \ref{QgyL}.   
This is a direct illustration of the fact that the $M^2/P^2$ corrections in this case 
are much smaller than the $\Lambda^2/P^2$ corrections of the Gaussian model.

  \begin{figure}[t]
    \centerline{\includegraphics[width=2.2in]{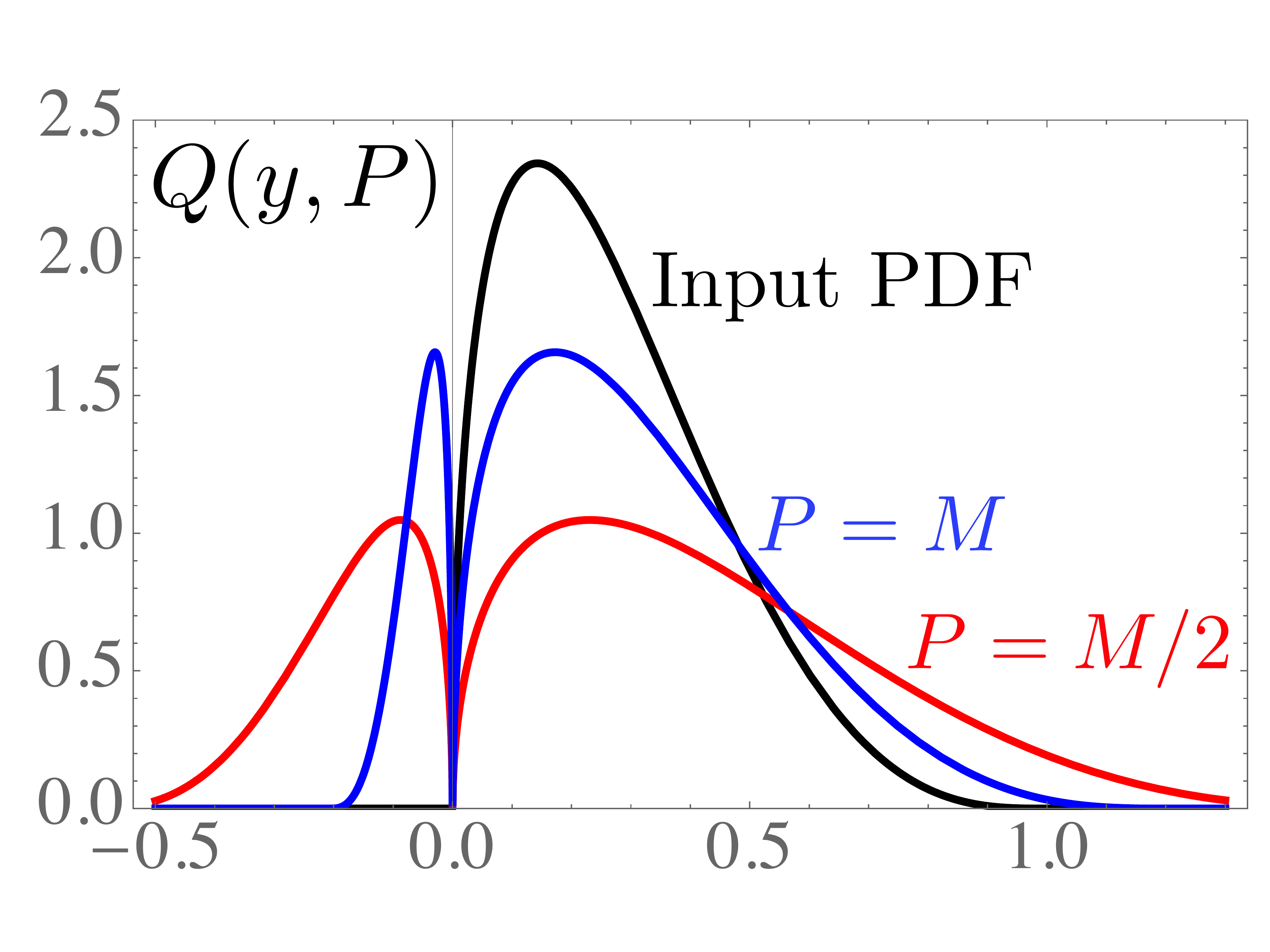}}
    \vspace{-0.4cm}
    \caption{Twist-2 part of  $Q(y,P)$   for {$P=M/2$} and $P=M$
   compared to the  limiting PDF \mbox{$f(y) = \frac{315}{32} \sqrt{y}(1-y)^3 \theta (0<y<1)$.}
    \label{tw2}}
    \end{figure}

Concluding this section,   we repeat again that we  see no reason 
to artificially split the  $(\Lambda^2/P^2)^l$  transverse-momentum 
corrections into the  target-mass and  {higher twist}  terms. 
As we observed, the target-mass part in such a split 
gives a  numerically very small portion. Furthermore,   the only estimate
we can imagine   for the higher-twist terms is that they are given, like in 
 Eq. (\ref{mebilF}),  
by  appropriate $k_\perp$  moments  of the TMD and  kinematical 
$M^2$-dependent  terms.   The latter exactly cancel the target-mass 
 corrections  coming from the lower-twist operators.
 This returns us to the TMD parametrization,
 and the whole idea of spitting looses any sense. 
  
Thus,  the power $1/P^{2l}$  corrections
reflect the transverse momentum effects  only, and in 
 the  (realistic) situation when TMDs are not known,
 these corrections are not calculable 
from  first principles.
Furthermore, as we have seen in Sect. \ref{secrate},
 the formally power-like 
$1/P^{2l}$ terms combine  in  a non-power $P$-dependence
for the  difference  
between the quasi-PDFs $Q(y,P)$ and the PDF $f(y)$, 
the exact form of which is again determined by the TMD. 
The only way to ``scientifically'' get rid of the transverse-momentum corrections  
 is to reach  sufficiently large values of $P$,
for which the {\it nonperturbative}  evolution of quasi-PDFs  may be  neglected.  

At these large $P^2$, one should be able to see the {\it perturbative}
$\ln P^2$ evolution of quasi-PDFs $Q(y,P)$.  It has the same origin as the
DGLAP 
 (for Dokshitzer-Gribov-Lipatov-Altarelli-Parisi 
\cite{Gribov:1972ri,Altarelli:1977zs,Dokshitzer:1977sg})
$\ln \mu^2$-dependence of the light-cone PDFs $f(x,\mu^2)$. 
Strictly speaking, only when this DGLAP-related $\ln P^2$-dependence is observed, 
one may use the perturbative matching relations that
convert the $\ln P^2$ -dependence of quasi-PDFs into the 
$\mu^2$-dependence of the light-cone PDFs. 

However,  the results of all available lattice quasi-PDF calculations, e.g., those of
 Refs. [\citen{Lin:2014zya,Alexandrou:2016jqi,Alexandrou:2019lfo}],  
 show the features of unfinished nonperturbative evolution.
The only reliable way to get rid of it is to use sufficiently  large
 momenta $P$. In practice, this means that one should reach $P\gtrsim 2.5$ GeV.
First, this is not a simple task and, second, the data at the highest 
achievable momentum are the least reliable.

In fact, the perturbative matching between the lattice data for $M(z_3,P)$ 
and the light-cone PDFs   is applicable when $z_3$ is small enough, 
like $z_3\lesssim 0.5$ fm.  The  momentum $P$ may be small, even zero.
In what follows, we discuss the derivation of the perturbative matching 
that is  used in the pseudo-PDF approach.

 \setcounter{equation}{0}  
  
\section{Perturbative QCD corrections at one loop}


     To convert $z_3^2$-dependence of the reduced pseudo-PDFs into
     the $\mu^2$-dependence of the light-cone PDFs, 
     one should know the OPE coefficient function   $C (w, z^2 \mu ^2) $
     (see \mbox{Eq. (\ref{OPE})).} 
An important fact is that the OPE can  be established 
  in the  operator form,  i.e. without 
  specifying the matrix element in which the operators are embedded.
One should just calculate  a modification of the original 
 bilocal operator by 
 gluon corrections.

 \subsection{Link-related UV divergences} 
 
 \label{linkuv}
 
 As  mentioned already, switching off the light cone comes with  a  penalty  in the form of   ultraviolet divergences
 generated by the gauge link. It is convenient and instructive to  analyze them in the Feynman gauge. 

 \subsubsection{Link self-energy} 

The largest  UV-related contribution comes from the 
self-energy correction  to  the  gauge 
 link (see Fig. \ref{linkself}).   At one loop, it is   given by
 \begin{align} 
 \Gamma_ \Sigma (z)  = &  ({i} g)^2\,C_F \,\frac12 \,  \int_0^1 \dd t_1 \,  \int_0^1 \dd t_2 \, 
  z^\mu z^\nu \,  D_{\mu \nu} ^c [ z_3 (t_2-t_1) ] \  , 
    \label{self}
 \end{align}
 where  $D_{\mu \nu} ^c[z_3 (t_2-t_1)]$ is the gluon propagator for the  line connecting  the points 
 $t_1 z_3$ and $t_2 z_3$.  For massless gluons, we have 
  $D_{\mu \nu} ^c (z)  =- g_{\mu \nu}  /4\pi^2 z^2$, and end  up with a divergent expression 
    \begin{align} 
     \int_0^1 \dd t_1 \,  \int_0^1 \frac{\dd t_2}{(t_2-t_1)^2}  \,  .
    \label{t1t2}
 \end{align}
 Though these integrals involve  just dimensionless parameters $t_1,t_2$, the   divergence has an ultraviolet origin. 
 As suggested by Polyakov\cite{Polyakov:1980ca}, it may be regularized for spacelike $z$ by  using   
 the prescription
 \mbox{$1/z_3^2 \to 1/(z_3^2+a^2)$}  for the gluon propagator. 
This regularization 
softens the gluon propagator at distances $z_3\sim $ several $a$,  
and eliminates its singularity at $z_3=0$. In this respect,
it is similar to  the UV regularization  produced by a  finite lattice spacing $a_L$. 
In fact, a comparison with the gluon propagator  in the lattice perturbation  theory 
establishes a simple connection $a=a_L/\pi$ between these two cut-offs\cite{Chen:2016fxx}. 
After the regularization, we have   the expression 
  \begin{align} 
 \Sigma (z_3,a)  = & - g^2\,C_F \,\frac{z_3^2}{8 \pi^2}
   \,  \int_0^1 \dd t_1 \,  \int_0^1  \, \frac{\dd t_2}{ z_3^2 (t_2-t_1)^2 + a^2}  
    \label{selfa}
 \end{align}
that  clearly shows that, for a  fixed $a$ the correction 
  $  \Sigma (z_3,a) $  vanishes at $z_3=0$. 
  The fact that     $\Sigma (z_3=0,a) =0$  means that, at fixed $a$,  $\Sigma$  gives 
 no corrections to the vector current, i.e.  the  number of  the  valence quarks  is not changed.

  Calculating the  integrals gives\cite{Chen:2016fxx} 
  \begin{align} 
  \Sigma (z_3,a)  =   -\,C_F \,\frac{\alpha_s}{2 \pi}
  & 
  \left [ 
   \,
 2  \frac{ |z_3|}{a} \,  \tan
   ^{-1}\left(\frac{|z_3|}{a}\right)   -  \ln 
   \left(1+ \frac{z_3^2}{a^2}\right) \right ] \  .
       \label{selfex}
 \end{align}
If we keep $z_3$ fixed and take the  small-$a$ limit, the result 
   \begin{align} 
\Sigma (z_3,a) |_{a \to 0}  = &  -\,C_F \,\frac{\alpha_s}{2 \pi}
  \left [
   \,
\frac{\pi |z_3|}{a}  - 2 -  \ln \frac{z_3^2}{a^2}  + {\cal O}(a^2/z_3^2) \right ] \   
       \label{selfexexp}
 \end{align}
 (see also Ref. [\citen{Ishikawa:2017faj}]) shows a linear divergence $\sim |z_3|/a$  in the $a \to 0$ limit.
 It  also shows a   logarithmic divergence  $\ln {z_3^2}/{a^2}$. 
 According to the all-order studies \cite{Dotsenko:1979wb,Brandt:1981kf,Aoyama:1981ev} 
  of the   Wilson loops renormalization,
  the one-loop correction (\ref{selfex})  exponentiates. 
  As a result, we get a strong damping factor for large $|z_3|$. In terms of  the  lattice spacing, it reads
     \begin{align} 
Z_{\rm link} (z_3,a_L)\simeq  e^{-A |z_3|/a_L}    \  ,
       \label{Z}
 \end{align}
with $A= C_F \pi \alpha_s/2 \approx 2 \alpha_s$. Taking $\alpha_s =0.2$ for an estimate, we get 
suppression by a factor of 10 starting with $z_3=6a_L$. 
Note also that  the $Z$-factor  is a function of $z_3/a_L$, i.e., it changes when the lattice spacing is changed.
Hence, it is a  lattice artifact,  not  related to actual physical phenomena in the  continuum theory.  
As discussed already,  extracting PDFs, one should divide it out. 
 Still, it is interesting to check if the actual lattice simulations
 are in agreement with its perturbative estimate.

\subsubsection{Vertex contribution} 

    \begin{figure}[b]
   \centerline{\includegraphics[width=2in]{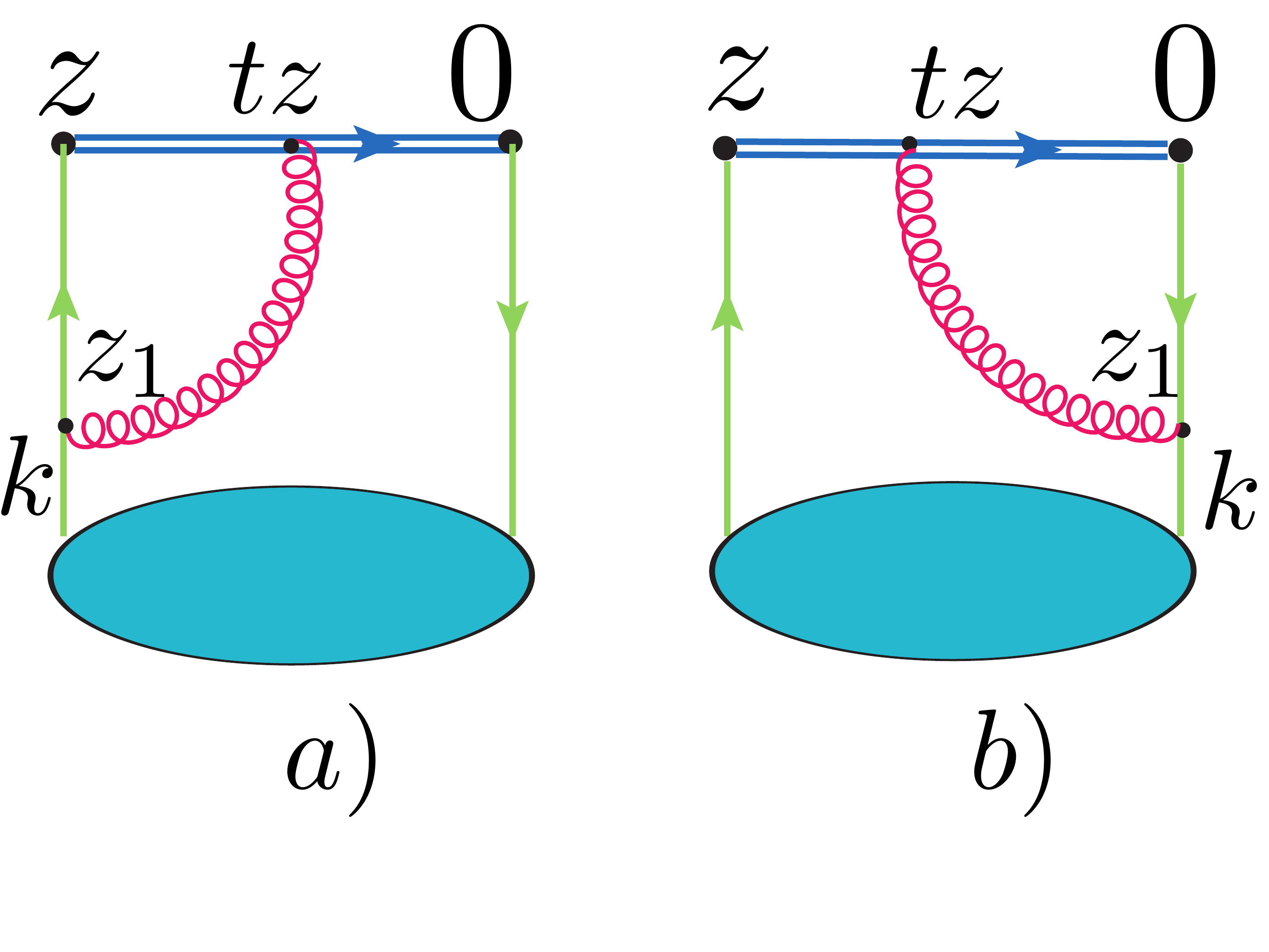}}
        \vspace{-5mm}
   \caption{Insertions of gluons coming out of the gauge link.
   \label{link}}
   \end{figure}

The UV divergent contributions are also present in the  diagrams
involving gluons  that connect the gauge link with the quarks, see  Fig. \ref{link}. 
Regularizing  the gluon propagator by 
 $1/z_3^2 \to 1/(z_3^2+a^2)$, 
 we extract 
the UV-singular term in the form 
  \begin{align} 
  O^\alpha_ {\rm  UV}  (z,a) = &  \frac{ g^2}{4 \pi^2} \, C_F \, \bar \psi (0)  \gamma^\alpha  \psi (0)    
   \int_0^1 d \beta \,   \int_0^1 \dd t\, 
  \frac{t z_3^2 }{ t ^2z_3^2+  a^2/(1- \beta) } \,  . 
    \label{UVaR}
 \end{align}
Taking integrals over $t$ and $\beta$    gives the expression 
    \begin{align} 
 O^\alpha_ {\rm  UV}  (z,a) = &  \frac{ \alpha_s}{2 \pi} \, C_F \, \bar \psi (0)  \gamma^\alpha  \psi (0)  
   \left [  \left (1+\frac{a^2}{z_3^2}  \right  )  \, 
  \ln  \left (1+ \frac{z_3^2}{a^2} \right  )  -1\right ]
       \  
    \label{UVsing}
 \end{align}
that   contains   the same $ \ln  \left (1+ {z_3^2}/{a^2} \right  ) $  logarithmic  term 
as in the self-energy correction (\ref{selfex}).
In the $a\to 0$ limit, this result agrees with that obtained in Ref. [\citen{Ishikawa:2017faj}]. 
The $\ln (1+z_3^2/a^2)$ structure      may be 
  combined   with the UV divergences generated by the link self-energy diagrams.
Again, for a fixed $a$,  the $ O^\alpha_ {\rm UV}  (z_3,a) $ 
contribution vanishes in the $z_3^2 \to 0$ limit.
Just like in the case of the link self-energy  corrections,
the UV divergences coming from vertex diagrams exponentiate in higher orders.

The UV divergent term  comes from the  configuration when the exchanged gluon ends coincide. 
The  study performed in Ref. [\citen{Radyushkin:2017lvu}]    shows that there is also an \mbox{UV-finite}  contribution
coming from the regions where the point $z_1$ is close to some position on the link. 
 The combined   contribution of two  diagrams shown in Fig. \ref{link}  is given by 
    \begin{align} 
  O^\alpha_{\rm reg}  (z_3, a=0) = &    \,     \frac{ \alpha_s}{2\pi} \, C_F
 \,\int_0^1  du \int_0^1  \dd v \, \bar \psi (uz_3) \gamma^\alpha \psi (\bar v z_3) 
  \nn & \times 
   \left \{ \delta (v) \left[ \frac{\bar u }{u }\right ] _+
 + \delta (u) \left[ \frac{\bar v }{v }\right ] _+   \right \}  %
     \ .
    \label{gauge24}
 \end{align}
  We use the notation $\bar v =1-v, \bar u =1-u$, etc. 
 The plus-prescription is defined by
   \begin{align} 
 \,\int_0^1  du \left[ \frac{\bar u }{u }\right ] _+  F(u)  =    \,\int_0^1  du \, \frac{\bar u }{u } \,  [F(u)-F[0] \ ,
    \label{Plus}
 \end{align}
 assuming that $F(0)$ is finite.
Now, it is  the plus-prescription structure  of Eq. (\ref{gauge24}) which 
guarantees that this term   gives no corrections to the local current.

  \subsection{Evolution  terms}

 The contributions considered in the previous section  do not have singularities  
when the quark virtuality $k^2$ vanishes, i.e. they do not need any IR regularization.  In particular,
 the  logarithm  $\ln (1+z_3^2/a^2)$ has $a$ as an UV cut-off, while 
 $z_3^2$ stays on its IR side.   However, vertex diagrams also contain   additional 
 contributions that are      infrared divergent   in the $k^2 \to 0 $ limit. 
 
 Of course, on the lattice everything will  be finite.
 Just like the finite lattice spacing provides a UV cut-off,
 the finite hadron size  provides an  IR cut-off. 
 Unfortunately, the exact form of the IR regularization  imposed  
by the hadron size  is not known.
To get  a feeling,
  let us take an infrared regularization by a  mass term.
 A typical  Schwinger's $\alpha$-parameter integral producing  an IR   singularity  then   has the form
   \begin{align} 
  L_K (z_3^2) = \int_0^\infty \frac{d\alpha}{\alpha} e^{-z_3^2/4\alpha -   \alpha m^2} \ ,
  \label{IK}
     \end{align} 
where  $m$ is the mass
(see, e.g.,  Ref. [\citen{Radyushkin:2017lvu}] for details).  
 One can see that
   \begin{align} 
  L_K (z_3^2) = &2 K_0 (mz_3)
   = - \ln \left (m^2 z_3^2 \frac{e^{2\gamma_E}}{4} \right ) + {\cal O} (z_3^2)  \  ,
  \label{Kexp} 
     \end{align} 
where $K_0(mz_3)$ is the modified Bessel function.  It has a  $\ln z_3^2$ 
singularity for small $z_3$, and  
  exponentially decreases when 
$z_3$  exceeds   $1/m$. 
Since we  want $m$ to mimic  
  the IR cut-off imposed by the hadron size, numerically $m$  
  should   be of an order of 0.5 GeV.
Another type of    the IR regularization is 
 provided by a sharp  cut-off  
    \begin{align} 
  L_G (z_3^2) = \int_0^{\Lambda^2}  \frac{d\alpha}{\alpha}    e^{-z_3^2/4\alpha } = \Gamma [0,z_3^2\Lambda ^2/4]
  =  - \ln  \left (z_3^2 \Lambda^2 \frac{e^{\gamma_E}}{4} \right )  + {\cal O} (z_3^2)  \  
  \label{IDexp}
     \end{align} 
applied to  Eq. (\ref{IK}).   The incomplete 
gamma-function  $ \Gamma (0,z_3^2\Lambda^2/4)$  has a logarithmic singularity
for small $z_3^2$,  
     while for large $z_3^2$ it  has  a Gaussian  $e^{-z_3^2 \Lambda^2/4}$  fall-off.  
  
  As we discussed, the UV link-related $Z$-factor also has a rapid $e^{-A|z|/a}$ decrease 
  for large $|z|$. Thus, one needs to very precisely divide it out 
  from  the  lattice  data  to be able to see  
  the fall-off  reflecting the finite hadron size. 
  
For both cases,  the IR-singular   contribution  from vertex diagrams   is given\cite{Balitsky:1987bk}  by 
    \begin{align} 
  O^\alpha_{\rm  log}  (z_3) & =     \,    L_R  ( z_3^2 ) \,   
   \frac{ \alpha_s}{2\pi} \, C_F
 \,\int_0^1  du \int_0^1  \dd v
  \nn & \times  \, 
 \left \{ \delta (u) \left[ \frac{\bar v }{v }\right ] _+
 + \delta (v) \left[ \frac{\bar u }{u }\right ] _+   \right \} \, \bar \psi (u z_3) \gamma^\alpha \psi (\bar v z_3) 
     \ ,
    \label{Logtot}
 \end{align}
 where $R$   is either $K$ or $G$. One may also 
 use the IR dimensional regularization.  In the $\overline{\rm MS}$ scheme,
 $L_{\overline{\rm MS}}  ( z_3^2 ) = - \ln \left (\mu^2 z_3^2 {e^{2\gamma_E}}/{4} \right ) $. 
 However,  one should realize that the lattice cannot  provide the dimensional 
 IR regularization, and the data will not show the   
 $\ln (z_3^2)$  behavior beyond a few lattice spacings.

Note  that, in contrast to  the UV divergent contribution,
the  $L_ R(z_3^2  \Lambda^2)$ functions are singular in the $z_3^2\to 0$ limit, and 
the parameter 
 $|z_3|$  in the integrals  of Eqs. (\ref{Kexp}), (\ref{IDexp})  works like an  ultraviolet 
 rather than an  infra-red cut-off.


The  integrals   producing the IR-singular terms, also   contain  an IR finite part 
   \begin{align} 
  O^\alpha_{\rm  Fin}  (z)  =&-     \,      \frac{ \alpha_s }{ \pi} \, C_F  \int_0^1 du \int_0^1 \dd v \, 
\bar \psi (u z)    \gamma^\alpha \psi (\bar vz) 
   \left [  \delta (u) \,  s_+(v)+ \delta (v) \,  s_+(u) \right ]
   \ , 
    \label{Ftot}
 \end{align} 
 where $s_+(u)$ is the plus-prescription  version of $s(u)$ given by 
    \begin{align} 
s(u) & \equiv 
  \int_u^1 \dd t \,\frac{ \ln t}{t^2}  = \frac{1-u+\log (u)}{u}  \   . 
    \label{Sv}
 \end{align}

\subsection{Quark-gluon exchange contribution}

There is also an IR-singular  
contribution  given by the diagram  \ref{quarkself}a  
containing  a gluon exchange between
two quark lines. It is given by 
    \begin{align} 
  O^0_{\rm  exch}  (z_3) & =     \,      \frac{ \alpha_s}{2\pi} \, C_F
 \,\int_0^1  du \int_0^{1-u}   \dd v 
  \, \left \{ L_R  ( z_3^2) \,  - 1  \right \} \, \bar \psi (u z_3) \gamma^0 \psi (\bar v z_3) 
     \ 
    \label{exch}
 \end{align}
 for $R= K, G$. For DR in the  $\overline{\rm MS}$ scheme, $L_R-1$ should be substituted by $L_{\overline{\rm MS}} +1$. 
Unlike  the vertex part, the exchange contribution (\ref{exch})
does not have the plus-prescription form.  

One should also include  the 
  quark self-energy diagrams,  one of which is shown in Fig. \ref{quarkself}b.
  As   usual, we should take just a half 
of each,  absorbing  the  other halves into the  soft part.
Since the quark momentum is not changed, these terms have
the $\delta (u) \delta (v)$ structure in the $u,v$-integral.

  \begin{figure}[t]
   \centerline{\includegraphics[width=2in]{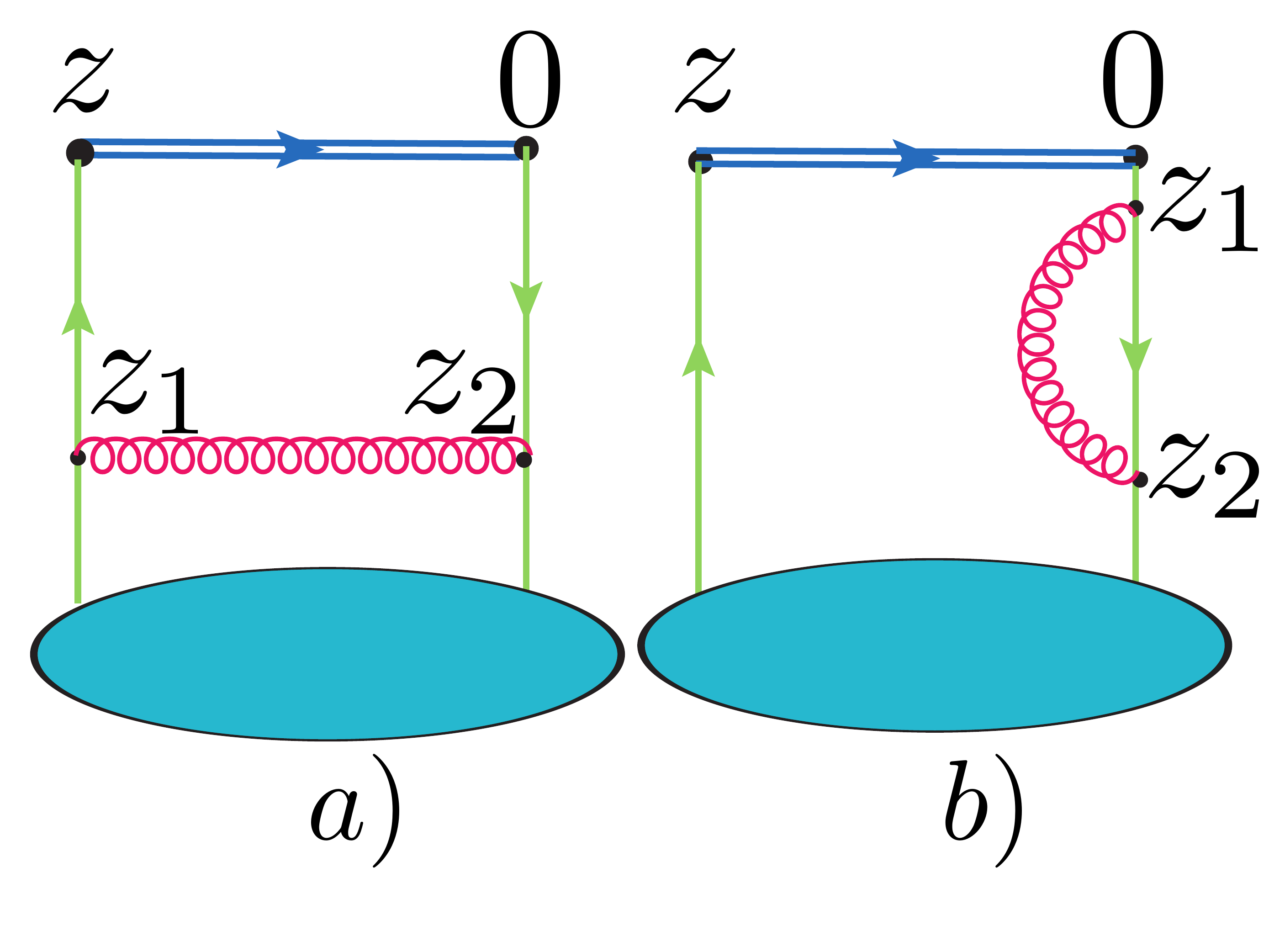}}
        \vspace{-5mm}
   \caption  {a)  Gluon  exchange diagram.  b)  One of quark self-energy correction diagrams.
   \label{quarkself}}
   \end{figure}

\subsection{One-loop correction in the operator form}  

Combining all the  one-loop corrections\cite{Radyushkin:2017lvu}  to the $ { \cal O}^0  (z_3)$ operator gives  
 \begin{align} 
\delta & {\cal O}^0 (z_3)   =  -  \frac{\alpha_s}{2\pi} \, C_F
\,\int_0^1  du \int_0^{1-u}  \dd v \, \bar \psi (uz_3) \gamma^0 \psi (\bar v z_3) 
  \nn & \times  \left \{  \left ( \delta (v) \left[ \frac{\bar u }{u }\right ] _+
 + \delta (u) \left[ \frac{\bar v }{v }\right ] _+  +1  \right ) 
\ln \left [ z_3^2\mu_{\rm IR}^2 \frac{ e^{2\gamma_E}}{4}  \right ] \right. 
 \nn   & \left. \hspace{1cm} 
+    2   \left (\delta (v) \left[ \frac{\ln u }{u }\right ] _+
 + \delta (u) \left[ \frac{\ln v }{v }\right ] _+  -1 \right )
  \right. 
 +Z (z_3)  \delta (u) \delta (v) 
\Bigr \}  \  .
\label{Oha}
 \end{align}
 
 In this  result,    we assume   the dimensional regularization
 and   the $\overline{\rm MS}$ scheme 
 subtraction for the 
 IR singularities,   with  $\mu_{\rm IR}$ serving as the scale  parameter.  
 The function $Z (z_3)$ accumulates information about  corrections 
 associated  with the \mbox{UV-divergent} contributions  like (\ref{selfex}), (\ref{UVsing}). 
 This function 
 in the   $\overline{\rm MS}$ scheme  is  known
 (see  Ref. [\citen{Izubuchi:2018srq}]),
 but we do not need   its explicit form in the pseudo-PDF approach.  As we 
 discussed in Sec. \ref{sITD}, 
     such terms cancel 
 when one forms the reduced Ioffe-time pseudodistributions.

\subsection{Matching for parton distribution functions}

In the PDF case, the one-loop correction to $ { M}^0 (z_3,p)$  is given by 
the forward matrix element 
$\langle  p |\delta {\cal O}^0 (z_3)  | p \rangle $.  
The right-hand-side of Eq. (\ref{Oha})  brings then the matrix element   
 \begin{align}
\langle p |  \bar \psi (uz_3) \Gamma^0 \psi (\bar v z_3)  |p \rangle 
\equiv  {\cal M}_0  (u\nu,\bar v\nu)  \ , 
\label{M0nu}
 \end{align}
 where $\nu =p_3 z_3$ is the Ioffe time  \cite{Ioffe:1969kf}.  
 The structure of Eq. (\ref{Oha})  implies a  scenario in which  the $z_3^2$-dependence
 at   short distances 
 is determined by the ``hard'' logarithms
 $\ln z_3^2$ generated from the  initially ``soft'' distribution ${\cal M}_0(\nu,z_3^2)$
 having  only a polynomial dependence on $z_3^2$ that is negligible
 for small $z_3^2$. For this reason, we skip 
 the \mbox{$z_3^2$-dependence}  in the argument of ${\cal M}_0$-functions,
 leaving just their $\nu$-dependence.
 
The ``vertex'' terms containing $\delta(u)$ or $\delta (v)$ 
are trivially reduced  to one-dimensional integrals in which we change
 $u$ or $v$ to  $1-w$.  Using translation invariance  for the ``box'' terms having a $u,v$-independent coefficient function, we get
   \begin{align} 
 \,  &
 \,\int_0^1  du \int_0^{1-u}   \dd v \, 
 {\cal M}_0 ((1-u-v) \nu) 
  =\int_0^1 \dd w\, (1-w) {\cal M}_0 (w \nu)
     \ .
 \end{align} 
We can represent $(1-w)$ 
 as the sum 
  of the term $(1-w)_+$ 
 that has the  plus-prescription 
 at $w=1$ and  the delta-function term  $ \frac12 \delta(\bar w)$ that we add to  $Z(z_3)$, 
denoting   the  changed $Z$-function  by   $\widetilde Z(z_3)$.
As a result, we have 
 \begin{align} 
{\cal M}(\nu,z_3^2) &  = \Biggl [1-  \frac{\alpha_s}{2\pi} \, C_F    \widetilde Z (z_3) \Biggr ]   {\cal M}_0(\nu)
-\frac{\alpha_s}{2\pi} \, C_F 
\,\int_0^1  \dd w \, {\cal M}_0( w \nu)  \nn &  \times 
\Biggl  \{
   \frac{1+w^2 }{1- w }
\ln \left ( z_3^2\mu_{\rm IR}^2\frac{ e^{2\gamma_E+1}}{4}  \right ) 
+    4    \frac{\ln (1-w) }{1-w }
   -2(1-w)  \Biggr \}_+  
 \  .
\label{deltaM}
 \end{align}
 The combination  
  \begin{align} 
B(w)  =&
       \left [\frac{1+w^2} {1-w}   \right ]_+
  \ 
     \label{V1}
  \end{align}   
  is   the non-singlet 
  Altarelli-Parisi (AP)  evolution kernel \cite{Altarelli:1977zs}.

The next step is to introduce 
the reduced Ioffe-time pseudodistribution  (\ref{redm}) 
of Refs. [\citen{Radyushkin:2017cyf,Radyushkin:2017sfi,Orginos:2017kos}]. 
When the momentum $p$ is also oriented  in the $z_3$ direction,
i.e., $p=\{E, 0_\perp,p_3\}$, the function ${\cal M} (0, z_3^2)$ corresponds to the  ``rest-frame'' $p_3=0$ distribution.
 According to Eq. (\ref{deltaM}), it is given by
  \begin{align} 
{\cal M}(0,z_3^2) &  =  {\cal M}_0(0)    \Biggl  [  1-  \frac{\alpha_s}{2\pi} \, C_F \widetilde Z (z_3) 
  \Biggr ]
 \  .
\label{deltaM0}
 \end{align}
 As a result, the $\widetilde Z(z_3)$ terms disappear from    the ${\cal O}(\alpha_s)$  correction to the ratio
 $ {\cal M} (\nu, z_3^2)/{\cal M} (0, z_3^2)$. 
  Such a cancellation of ultraviolet terms for ${\mathfrak M}(\nu,z_3^2) $ will persist in higher $\alpha_s$ orders,
 reflecting the multiplicative renormalizability of the ultraviolet divergences\cite{Ji:2017oey,Ishikawa:2017faj,Green:2017xeu}
  of ${\cal M}(\nu,z_3^2)$.

 A similar calculation can be performed for the light-cone 
 {\it Ioffe-time distribution}\cite{Braun:1994jq}    ${\cal I} (\nu, \mu^2)$    obtained by taking $z^2=0$ 
 in ${\mathfrak M}(\nu,-z^2)$ and regularizing the resulting UV singularities 
by  dimensional regularization and the $\overline{\rm MS}$ subtraction
specified by a factorization scale $\mu$. The result  may be symbolically written as   
  \begin{align} 
{\cal  I}(\nu,\mu^2)   =  {\mathfrak M}_0(\nu) 
 - & 
\frac{\alpha_s}{2\pi} \, C_F 
\,\int_0^1  \dd w   \, B(w) 
\ln \left ( \mu_{\rm IR}^2/\mu^2 \right ) 
{\mathfrak  M}_0( w \nu) 
 \  .
\label{LCITD2}
 \end{align}
 As a result, we get 
 the matching condition\cite{Ji:2017rah,Radyushkin:2017lvu,Radyushkin:2018cvn,Zhang:2018ggy,Izubuchi:2018srq} 
   \begin{align} 
 {\mathfrak M}(\nu,z_3^2)  
=   {\cal  I}(\nu,& \mu^2) 
-
\frac{\alpha_s}{2\pi} \, C_F 
\,\int_0^1  \dd w   \,   {\cal  I}(w\nu, \mu^2)  \Biggl \{
\Biggl  [ \frac{1+w^2 }{1- w }  \Biggr ]_+ \, 
 \nn & \times
\ln \left ( z_3^2 \mu^2\frac{ e^{2\gamma_E+1}}{4}  \right ) 
+    4    \frac{\ln (1-w) }{1-w }
   -2(1-w)  \Biggr \}_+
 \  
\label{Mtch}
 \end{align}
that  relates  $ {\mathfrak M}(\nu,z_3^2)$ with     the light-cone ITD ${\cal  I}(\nu, \mu^2)$.
Note that this relation works  for small $z_3^2$ only, namely,  in the region 
where the IR sensitive factors $L_R (z_3^2)$ may be approximated 
by  $\ln z_3^2$.   In this region,  $ {\mathfrak M}(\nu,z_3^2)$ satisfies the 
 DGLAP    evolution equation
    \begin{align}
    \frac{d}{d \ln z_3^2} \,  
{\mathfrak M} (\nu, z_3^2)    &=- \frac{\alpha_s}{2\pi} \, C_F
\int_0^1  du \,   B ( u )   {\mathfrak  M} (u \nu, z_3^2)  \ .
 \end{align}

 Eq. (\ref{Mtch}) allows to get $  {\cal  I}(\nu, \mu^2) $ using  lattice data  on 
 ${\mathfrak  M}(  \nu,z_3^2) $. After  that,  inverting the  Fourier transform (\ref{LCITD}) 
one should  be able to get  $f(x,\mu^2)$. 
However, lattice calculations  provide  ${\mathfrak M}(\nu,z_3^2)$ and, hence,  
$  {\cal  I}(\nu, \mu^2) $     in a rather limited range 
of $\nu$,  which   makes taking  this Fourier transform rather tricky (see Ref. [\citen{Karpie:2019eiq}] 
for a detailed discussion). 
  An easier way was proposed in our paper [\citen{Radyushkin:2017cyf}]. The  idea 
  is to assume some parametrization for ${f}(x,\mu^2) $  similar to those  used 
  in global fits (see, e.g., Ref. [\citen{Accardi:2016qay}]), and 
  to fit its parameters using $  {\cal  I}(\nu, \mu^2) $ extracted from the lattice data
through  
 Eq. (\ref{Mtch}).

 An equivalent realization of this idea (similar to that of Ref. [\citen{Cichy:2019ebf}])
 is to use the kernel relation (\ref{ker}), i.e., to 
 substitute    $ {\cal  I}(\nu, \mu^2)$ 
by  its definition (\ref{LCITD}) as a Fourier transform of PDF. This converts (\ref{Mtch}) into 
    \begin{align} 
&{\mathfrak M}(\nu,z_3^2)      = 
  \int_{-1}^1 \dd x \,  \left [e^{ix\nu}  - \frac{\alpha_s}{2\pi} \, C_F 
   R( x \nu, z_3^2 \mu^2) \right] \,  {f}(x,\mu^2)   \ .
 \  
\label{MtchI}
 \end{align}
 The kernel $R(x \nu, z_3^2 \mu^2 )$ is   given by the Fourier transform
 (\ref{Rker}) of the coefficient function, and 
  may be calculated as a  closed-form expression\cite{Izubuchi:2018srq,Radyushkin:2019owq}. 
   
  The PDF  $f(x)$ may be split in its symmetric $f^+(x)$ and antisymmetric  $f^-(x)$ parts.   
 For positive $x$,  they are  related to the quark $f_q (x)$ and antiquark $f_{\bar q} (x)$ distributions
 through  $f^+(x)=f_q (x) - f_{\bar  q} (x)$ and $f^-(x)=f_q (x) + f_{\bar q } (x)$, respectively 
 (see, e.g., Ref. [\citen{Orginos:2017kos}]).
 The  real part of $ R(y, z_3^2 \mu^2)$ generates then the real part of 
 ${\mathfrak M}(\nu,z_3^2) $ from $f^+(x)$, while 
 the  imaginary part of $ R(y, z_3^2 \mu^2)$ connects  the  imaginary  part of 
 ${\mathfrak M}(\nu,z_3^2) $ with  $f^-(x)$. In particular, for the real part we have 
      \begin{align} 
&{\rm Re} \, R(\nu x, z_3^2 \mu^2) =  \Biggl  \{ \frac{1-\cos (\nu  x)}{\nu ^2
   x^2}-\frac{2 \sin (\nu  x)}{\nu 
   x}      + 2\, \sin (\nu 
   x)\, \text{Si}( \nu x )\,  \nn &
 +2\, \cos (\nu  x) \left(\text{Ci}( \nu x)- \log (\nu  x)- \gamma_E +\frac{3}{4}
   \right)  \Biggr \} \ln \left ( z_3^2 \mu^2\frac{ e^{2\gamma_E+1}}{4}  \right ) 
 \nn  &+ 2 {\rm Re} \left[   i \nu  x e^{i \nu 
   x} \, _3F_3(1,1,1;2,2,2;-i  \nu  x
   )  \right] 
 +   \cos (\nu   x)   -2\,  \frac{1-\cos (\nu 
   x)}{\nu ^2 x^2}
\, , 
\label{MtchR}
 \end{align}
 where $\text{Ci}(y)$ and $\text{Si}(y)$  are the  integral cosine and sine functions, and $_3F_3(1,1,1;2,2,2;-i y)$
 is a hypergeometric function.  
Thus, assuming some parametrizations for the $f^{\pm}(x,\mu^2)$ distributions,
 one can  fit  their  parameters and $\alpha_s$ using  Eqs. (\ref{MtchI}), (\ref{MtchR})  and  the  lattice data for ${\mathfrak M}(\nu,z_3^2)$.

 Note that, despite the terms with $\nu x$ factors in their denominators,
 the kernel  $R(\nu x, z_3^2 \mu^2)$ vanishes for $\nu x =0$.
 To this end,   recall that, according to its definition (\ref{Rker}), the kernel 
  $R(0, z_3^2 \mu^2)$ is given by the $w$-integral of the coefficient function
  $C (w, z^2 \mu ^2) $ that has the  plus-prescription form in our case.


 \setcounter{equation}{0}  
 
\section{Exploratory   quenched  lattice study}

\subsection{General features} 
  
 An exploratory  lattice study 
 of the reduced pseudo-ITD  ${\mathfrak M} (\nu, z_3^2)$
 for the valence $u_v(x)-d_v(x)$ parton distribution in the nucleon has  been reported in Ref.
[\citen{Orginos:2017kos}].  
 The calculations were performed in   the quenched approximation
on  
 $32^3\times 64 $
 lattices,  for the lattice spacing $a=0.093$ fm
 at  the pion mass of  $601(1)$ MeV and the nucleon mass of  $1411(4)$MeV. 
Seven  lattice momenta $p\,( 2\pi/L)$, with  $p =0, \ldots 6$ were used.  
The  maximal  momentum  reached is $2.5\, $GeV. 
This simplified setup has allowed to get very precise data 
in a very short time, and its results are a very instructive 
illustration of applications in  the theory of the pseudo-PDFs. 

 \subsection{Rest-frame amplitude}
 
 \label{refraq}

 The  basic idea  of the pseudo-PDF approach  is  to  get information about the reduced
 pseudo-ITD. To this end, one needs  to measure the 
 ratio 
$R(z_3,p_3)=M(z_3,p_3)/M(z_3,p_3=0)$.  
 As we  discussed,  the rest-frame amplitude $M(z_3,0)$
 is basically given by the link UV-factor $Z_{\rm link} (z_3/a_L)$,
 that exponentially decreases for large $z_3$
(see Eq. (\ref{Z})).
 Thus, if $M(z_3,p_3)$ and $M(z_3,p_3=0)$ are obtained from 
 independent measurements, then the errors in the main amplitude
 $M(z_3,p_3)$ are   magnified 
 by the $1/M(z_3,p_3=0)$ factor which is very large for large $z_3$. 
 For this reason, in Ref. [\citen{Orginos:2017kos}], 
the calculations were performed directly for 
 the ratio $R(z_3,p_3)$ itself, rather than for  the  numerator
 and denominator independently.

 However, one can also calculate the rest-frame amplitude separately,
 and analyze its $z_3$-behavior.  The amplitude 
 $M(z_3,p_3)$ has a real and imaginary parts.
Its  real part is an even function of $\nu=p_3 z_3$,  while 
the imaginary part is odd in $\nu$. Hence, 
the imaginary part  should vanish for $p_3=0$. Indeed,  
the results for  the  imaginary part  of $M(z_3,p_3=0)$ obtained in \mbox{Ref. [\citen{Orginos:2017kos}]}
 are compatible with zero.
The real part was found to be    a symmetric function of $z_3$,
as expected.  The results for $z_3 \geq 0$ are displayed  in \mbox{Fig. \ref{Mrest}.}
The curve shown there is the exponentiated version
 \begin{align} 
Z_{\rm pert}  (z_3/a_L)  = \exp \left  \{  -\,C_F \,\frac{\alpha_s}{2 \pi}
  \left [ 
   \,
 2  \frac{ \pi|z_3|}{a_L} \,  \tan
   ^{-1}\left(\frac{\pi |z_3|}{a_L}\right)   - 2  \ln 
   \left(1+ \frac{\pi^2 z_3^2}{a_L^2}\right) \right ] \right \} \  
       \label{Zuv}
 \end{align}
of the UV factors coming from the one-loop  link self-energy (\ref{selfex}) 
and vertex (\ref{UVsing}) corrections, in which we substituted 
the Polyakov regularization parameter $a$ by the lattice spacing $a_L$
using the correspondence $a=a_L/\pi$ found in Ref. [\citen{Chen:2016fxx}]. 
The value of  $\alpha_s$ obtained from the fit  is 0.19. 
Thus,  the ``nonperturbative'' renormalization factor $Z(z_3/a)$  in this particular lattice
simulation 
was found to be  very accurately reproduced by the perturbative formula.
This fact, in our opinion, deserves a further study. 
Still, whatever its form, the UV $Z$-factor completely cancels out in the ratio
${\cal M}(\nu,z_3^2)/{\cal M}(0, z_3^2)$
defining the reduced Ioffe-time pseudodistribution. 

    \begin{figure}[t]
 	\centerline{\includegraphics[width=2.5in]{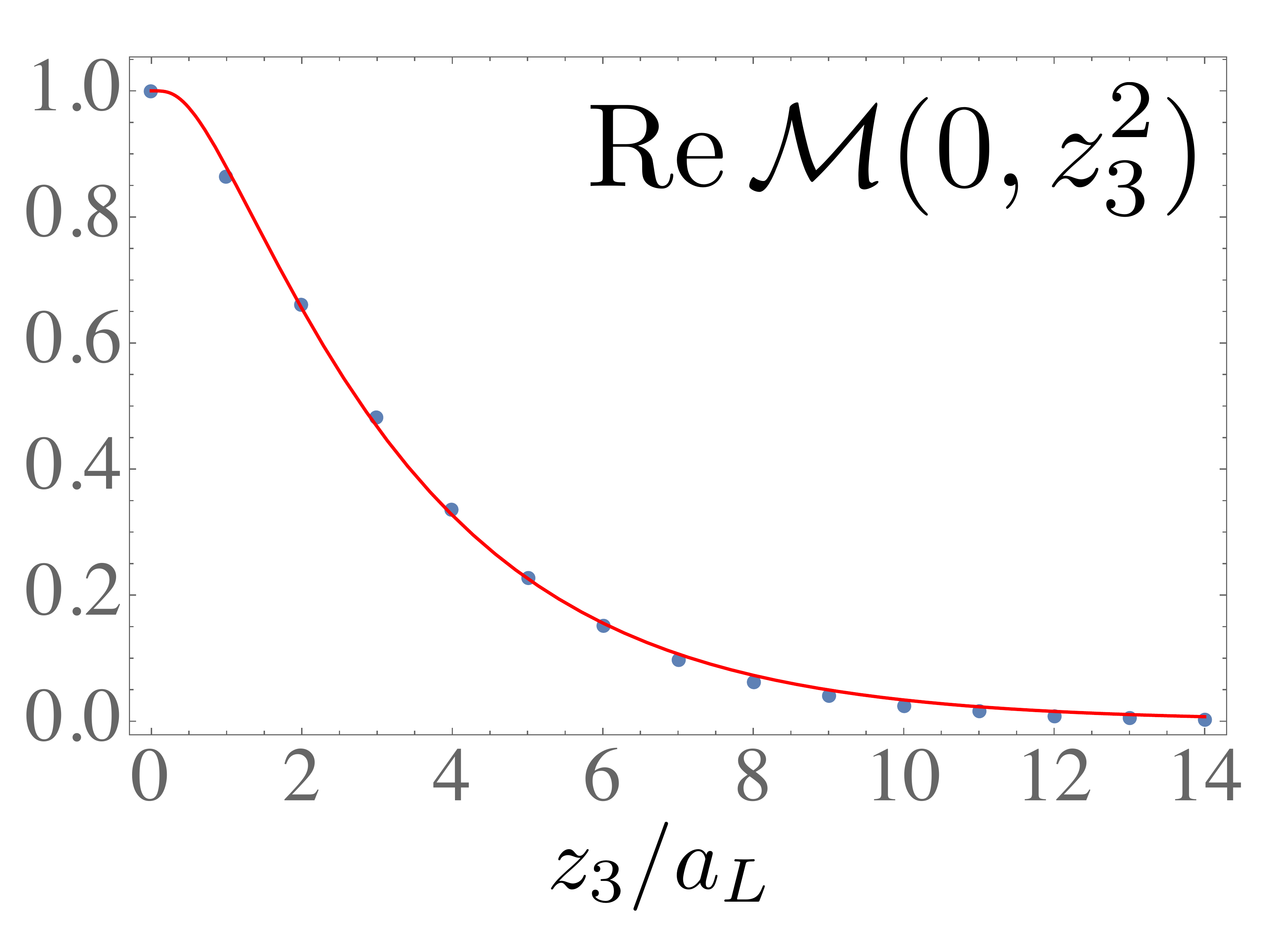}}
 	\caption{Real part of the rest-frame amplitude $M(z_3, p_3=0) ={\cal M}(0,z_3^2)$.
 		\label{Mrest}}
 \end{figure}

\subsection{Reduced Ioffe-time distributions}
\label{redITD1}

 On the left panel  of    Fig. \ref{realz},  we plot the results for 
the real part of the  ratio  ${\cal M} (Pz_3, z_3^2)/{\cal M} (0,z_3^2)$  
taken at six  values of the momentum $P$ and  plotted 
as a function of $z_3$ .  
One can see that   the curves decrease much slower 
with $z_3$ than ${\cal M} (0,z_3^2)$  of Fig. 
\ref{Mrest}. The curves  look  similar  to each other,  all  of them 
having   a  broad  Gaussian-like shape. However, 
 the  width decreases    with $P$. 
 
    \begin{figure}[h]
   \centerline{\includegraphics[width=2.6in]{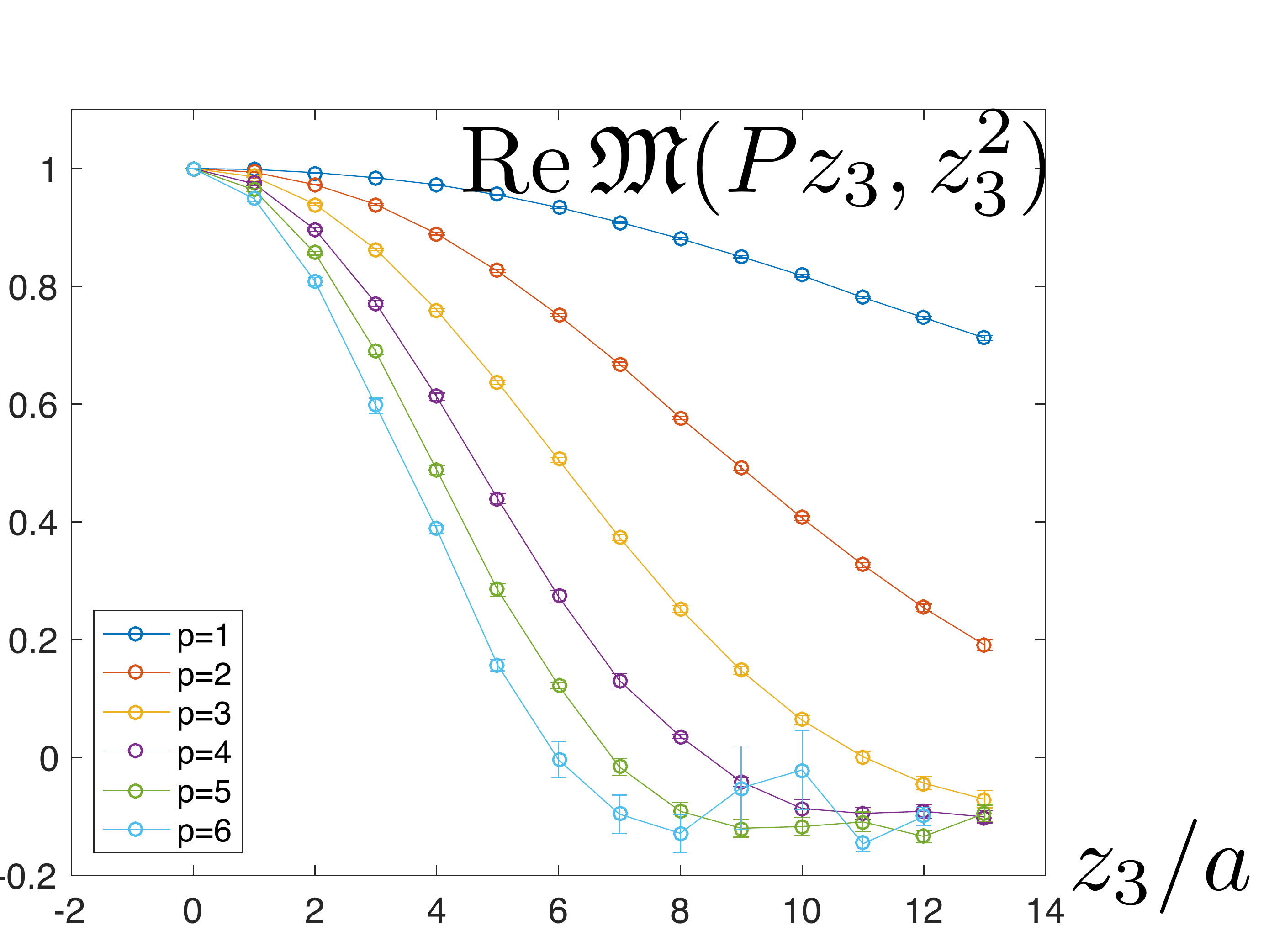} \includegraphics[width=2.38in]{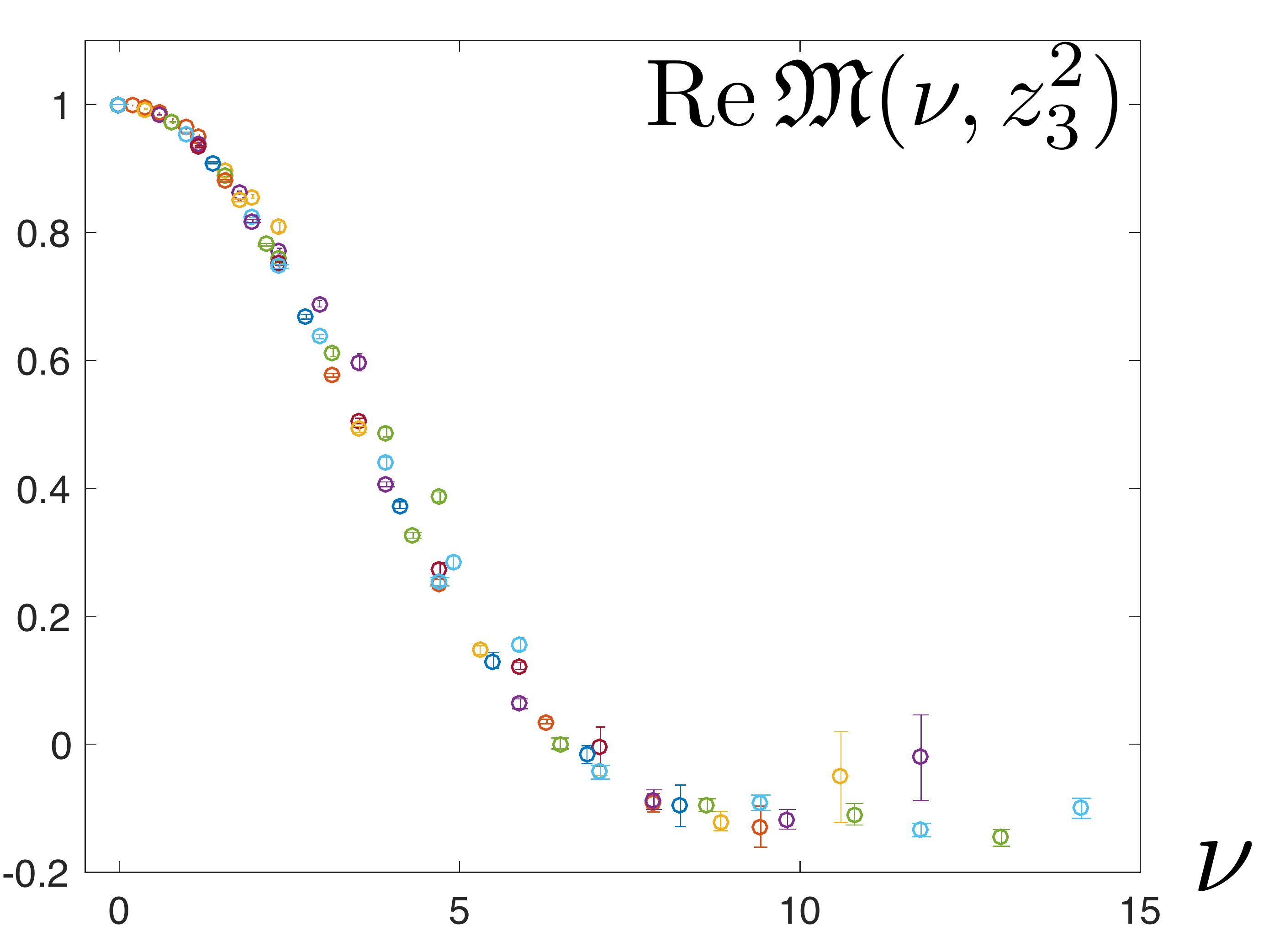}}
   \caption{Left: Real  part of the reduced  distribution ${\mathfrak  M} (Pz_3,z_3^2)$ plotted as
    a  function of
    $z_3$. Here, $P= 2 \pi p/L$. Right: The same  data plotted as a function of $\nu =P z_3$.
   \label{realz}}
   \end{figure}
 
On the right panel  of   \mbox{Fig. \ref{realz} ,} 
 we plot   the  same data, but change the axis to $\nu=Pz_3$.
Now the  data practically fall on the same curve.
 The situation is similar for the  imaginary part.
An evident  interpretation of this outcome is that
the numerator ${\cal M}(\nu, z_3^2)$ and the denominator ${\cal M}(0, z_3^2)$
of the ratio 
defining the reduced pseudo-ITD ${\mathfrak M}(\nu, z_3^2)$
have similar dependence on $z_3^2$.
In other words, the data indicate that the $z_3^2$-dependence of 
${\mathfrak M}(\nu, z_3^2)$  factorizes from its $\nu$-dependence,
${\mathfrak M}(\nu, z_3^2) \approx  {\cal I} (\nu) {\mathfrak M}(0, z_3^2)$. 

Still, one  can also notice some apparently random scatter of the points
corresponding to the same value of $\nu$. 
In fact, there is a regularity in this scatter.
On the left panel of Fig. \ref {M713}, we show the data
corresponding to ``large'' $z_3$-values: from $7a_L$ to $14a_L$. 
As  one can see,  there  is some scatter for the points 
with the largest values of $\nu$ in the region $\nu \gtrsim 10$,
where  the finite-volume effects  become important. 

         \begin{figure}[b]
	\centerline
	{ {\includegraphics[width=2.5in]{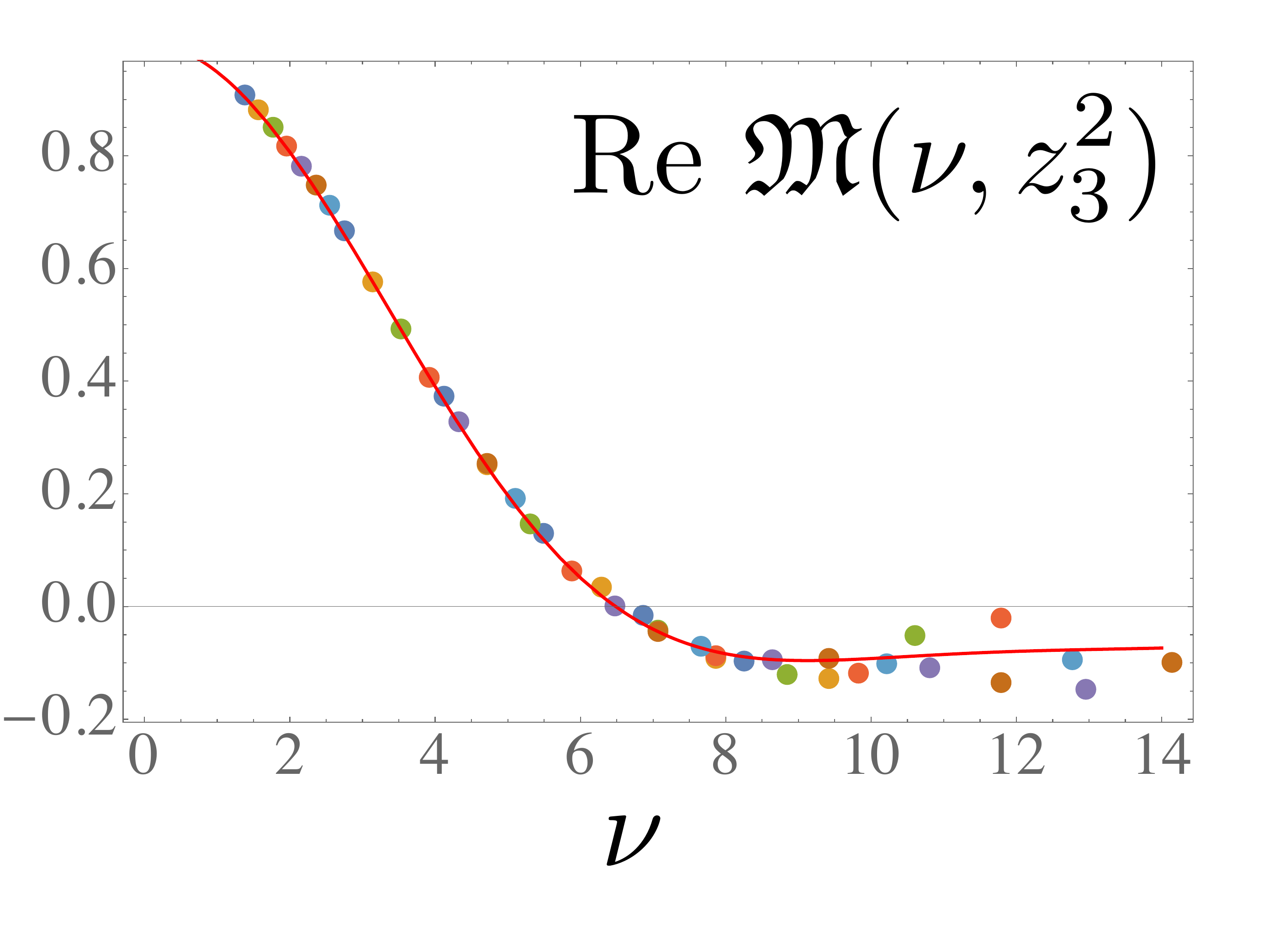} } \includegraphics[width=2.5in]{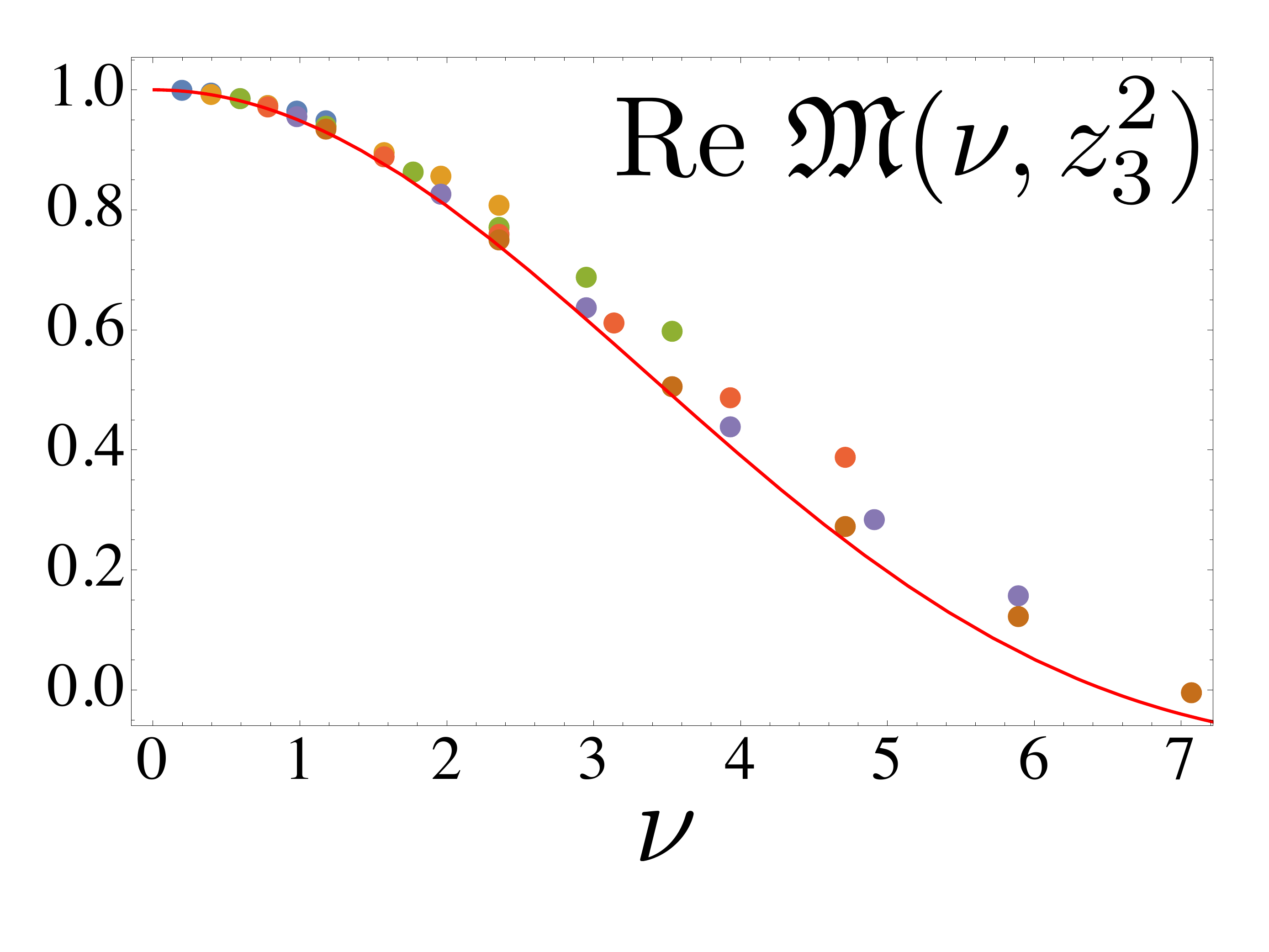}}
	\caption{Real part   of ${\mathfrak  M} (\nu, z_3^2)$  for  $z_3$  ranging  from  $7a_L$ to  $14a_L$ (left) and
		from  $a_l$ to  $6a_L$ (right).
		\label{M713}}
\end{figure}

\noindent Otherwise,  practically all the points lie on the curve 
    \begin{align} 
 {\cal R} (\nu)  =  & \int_0^1 \dd x \,  
 \cos (\nu x)  \, f_v  (x)  
 \label{Recos}
 \end{align}
generated  by    the  function 
     \begin{align} 
    f_v(x) =  \frac{315}{32} \sqrt{x}  (1-x)^{3}\,  . 
    \label{qV}
    \end{align}
Its  shape  was obtained by taking  
normalized $x^a(1-x)^b$-type  functions 
and fixing  the parameters $a,b$  by   fitting    the  data.

Recall   that 
 the real part  of the  light-cone  ITD  ${\cal I} (\nu) $  corresponds  to 
the  cosine Fourier transform of the valence distribution  $q_v(x) = u_v(x) - d_v(x)$
 \begin{align} 
 {\cal I}_R (\nu)  \equiv 
 {\rm Re} \,  {\cal I} (\nu)=  & \int_0^1 \dd x \,  
 \cos (\nu x)  \, q_v  (x)  \ .
\label{Recos2}
 \end{align}

On the right panel of Fig. \ref{M713}, we show the points in the region
of ``small'' $z_3$, ranging in the interval    \mbox{$a_L \leq z_3 \leq 6a_L$}. 
 In this case, all   the points lie  
higher than  the  curve for $R(\nu)$.  Since 
 ${\mathfrak M} (\nu, z_3^2)$,  according to Eq. (\ref{Mtch}), 
contains the evolution logarithm  $\ln z_3^2$  in the region of small $z_3^2$, 
one may conjecture that
the  observed higher values 
of ${\rm  Re}\,{\mathfrak M}$ for smaller-$z_3$ points may be a consequence of the evolution. 

\begin{figure}[h]
  \centerline{\includegraphics[width=2.3in]{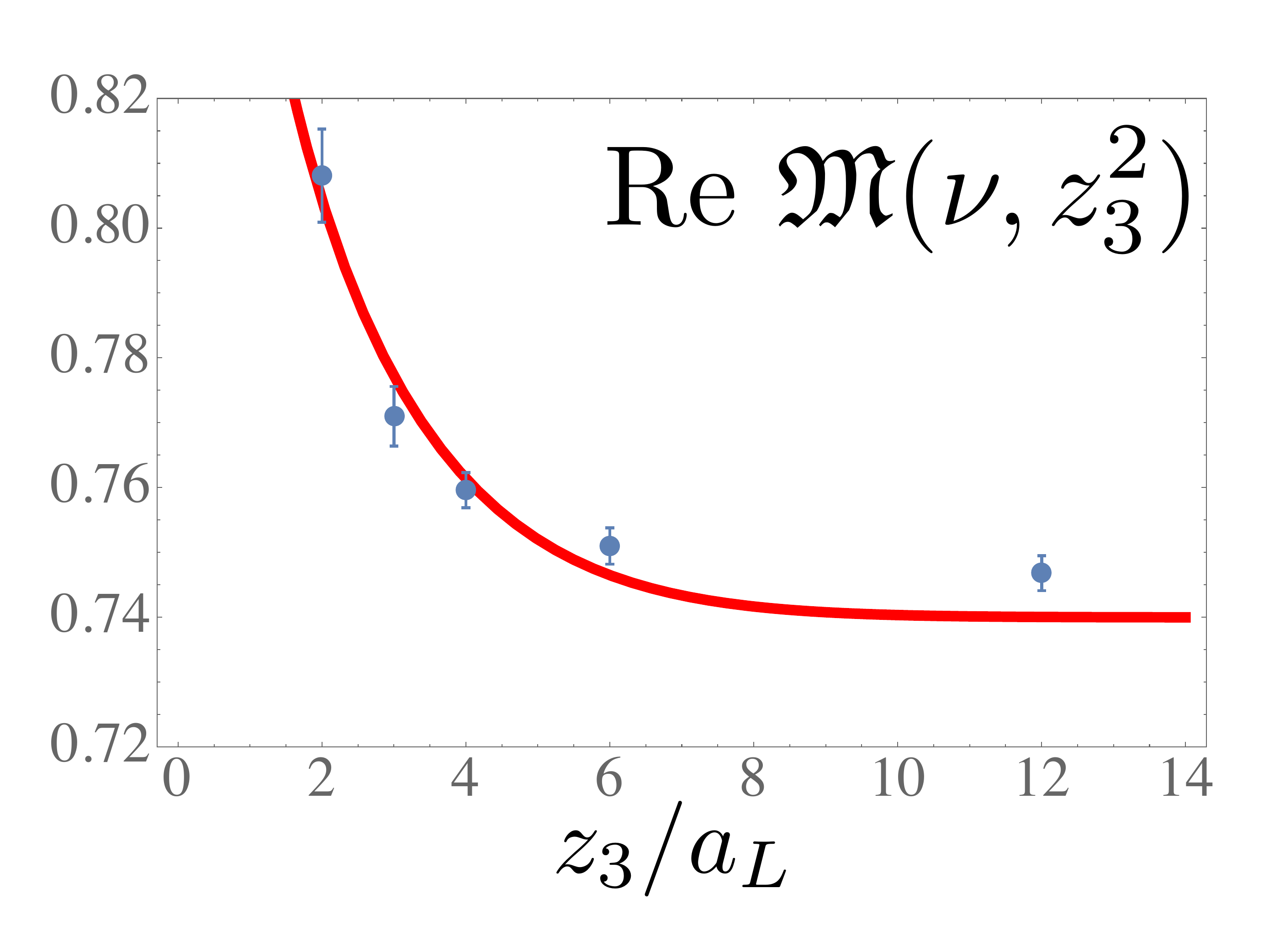}}
    \caption{Dependence  on $z_3$ for $\nu=3\pi/4\approx2.3562$. 
    \label{nu23}}
    \end{figure}

     In \mbox{Fig. \ref{nu23}}   we show 
a  typical pattern of the $z_3$-dependence of the lattice points. We took there  
the  ``magic''  Ioffe-time value $\nu =3\pi/4$ that may be obtained from 
five different combinations of $z_3$ and $P$ values 
used in Ref. [\citen{Orginos:2017kos}].  The shape of the eye-ball fit  line is
given by the incomplete gamma-function $\Gamma (0,z_3^2/30a_L^2)$.
This function  conforms
to our  expectation that the  $z_3$-dependence of the IR-sensitive factors 
$L_R (z_3)$  in (\ref{Logtot}), (\ref{exch}) 
should have    a perturbative  logarithmic $\ln (1/z_3^2)$  behaviour  for small $z_3$,
and 
rapidly vanish  for  $z_3$   larger  than the hadron size  $R_{\rm hadr}$.
We can estimate that  $R_{\rm hadr}$  in this lattice simulation is of an order of  \mbox{$6 a_L \approx 0.55$ fm. }
Looking at Fig. \ref{nu23}, we  may also say  that  perturbative evolution ``stops''  for $z_3 \gtrsim 5 a_L$. 
In this sense, the overall  curve  based on Eq.  (\ref{qV}) corresponds to  a  ``low normalization point'', 
i.e.,  to   the region, where the perturbative evolution is absent.

\subsection{Building $\overline{\rm MS}$  ITD}

 Thus, we see   that the data of \mbox{Fig. \ref{nu23}}  
show a logarithmic evolution behavior  in  the small  $z_3$ region.  
Still,    the 
\mbox{$z_3$-behavior }  starts to visibly deviate 
from a pure logarithmic $\ln z_3^2$  pattern for $z_3\gtrsim 5a$.     
Thus,   $z_3 \leq 4a$  is  the ``logarithmic region'' where 
one may   use Eq. (\ref{Mtch})  to construct 
the  light-cone $\overline{\rm MS}$ ITD. 
To this end, it is convenient to invert it and write  
   \begin{align} 
{\cal  I}(\nu,& \mu^2)  =  {\mathfrak M}(\nu,z_3^2)  
+
\frac{\alpha_s}{2\pi} \, C_F 
\,\int_0^1  \dd w   \,   {\mathfrak M}(w \nu,z_3^2) 
 \nn & \times  \Biggl \{
\Biggl  [ \frac{1+w^2 }{1- w }  \Biggr ]_+ \, 
\ln \left ( z_3^2 \mu^2\frac{ e^{2\gamma_E+1}}{4}  \right ) 
+    4    \frac{\ln (1-w) }{1-w }
   -2(1-w)  \Biggr \}_+ \  .
 \  
\label{MtchI2}
 \end{align}

Let us start with  the real part of this relation. At the leading order in 
$\alpha_s$,   
we have $  {\cal  I}_R(\nu, \mu^2) ={\rm Re} \ {\mathfrak M}(\nu,z_3^2)   $. 
In its turn, ${\rm Re} \ {\mathfrak M}(\nu,z_3^2)$ is given by $ {\cal R} (\nu) $ of Eq. (\ref{Recos})
plus scatter, which we intend to describe by the $\ln z_3^2$ part of the  
${\cal O} (\alpha_s)$ 
correction.
This means that  we  should approximate  ${\rm Re} \,  {\mathfrak M}(w \nu,z_3^2) $  
by ${\cal R} (w\nu) $ in the ${\cal O} (\alpha_s)$ term. 
Using further the definition (\ref{Recos}) of $R(\nu)$ in terms of $f_v(x)$ given by  (\ref{qV}) we get 
   \begin{align} 
{\cal  I}_R(\nu,& \mu^2)  =  {\rm Re} \, {\mathfrak M}(\nu,z_3^2)  
+
\frac{\alpha_s}{2\pi} \, C_F 
\,\int_0^1  \dd x    \,  f_v(x) \, {\rm Re} \, R(x\nu,z_3^2) \ , 
 \  
\label{IRker}
 \end{align}
where $ {\rm Re}  \,  R(x\nu,z_3^2)$ is the kernel specified by   Eq. (\ref{MtchR}).

The next step is to check if 
   the actual $z_3^2$-dependence of the data on 
  ${\mathfrak M} (\nu, z_3^2)$ plus 
   the \mbox{$ \ln  z_3^2$-dependence}   
  of the one-loop correction produce together the result that has 
   no (or little)   $ z_3^2$-dependence.
  In the worst case scenario, this will  not happen for any value  of $\alpha_s$,
   the only free parameter that we have. 
   This will mean that our  data are simply  inconsistent with the DGLAP evolution equation. 
   
   Fortunately,   as  it was found  
    in the original paper  [\citen{Orginos:2017kos}],  
 the  \mbox{$z_3^2$-dependence}  of the data  
 matches  \mbox{$ \ln  z_3^2$-dependence}   of the one-loop correction
   if one takes $\alpha_s/\pi =0.1$.  
Using   this   value
   in  Eq.  (\ref{IRker}) and   the  data  on  ${\rm Re} \ {\mathfrak M}(\nu,z_3^2)    $,  
one  can generate the ``data points''  for   ${\cal I}_R  (\nu, \mu^2) $.  
  This was done in  Ref. [\citen{Radyushkin:2018cvn}] for $\mu =1/a_L$ that 
  corresponds to $\mu = 2.15$ GeV.   The results are shown in  the left panel of \mbox{Fig. \ref{Msbar16}.}

      \begin{figure}[h]
  \centerline{\includegraphics[width=2.05in]{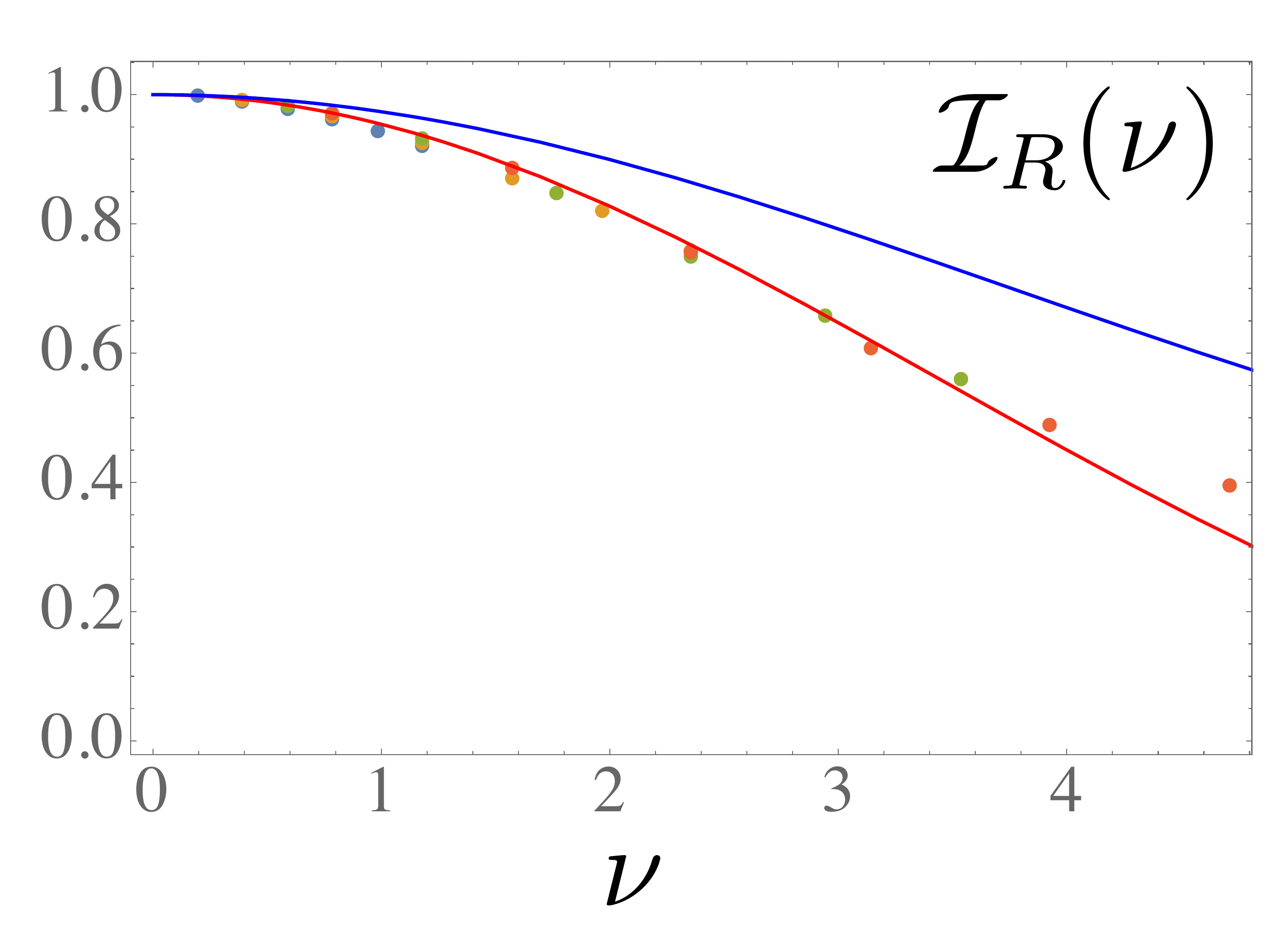}\hspace{5mm} \includegraphics[width=2in]{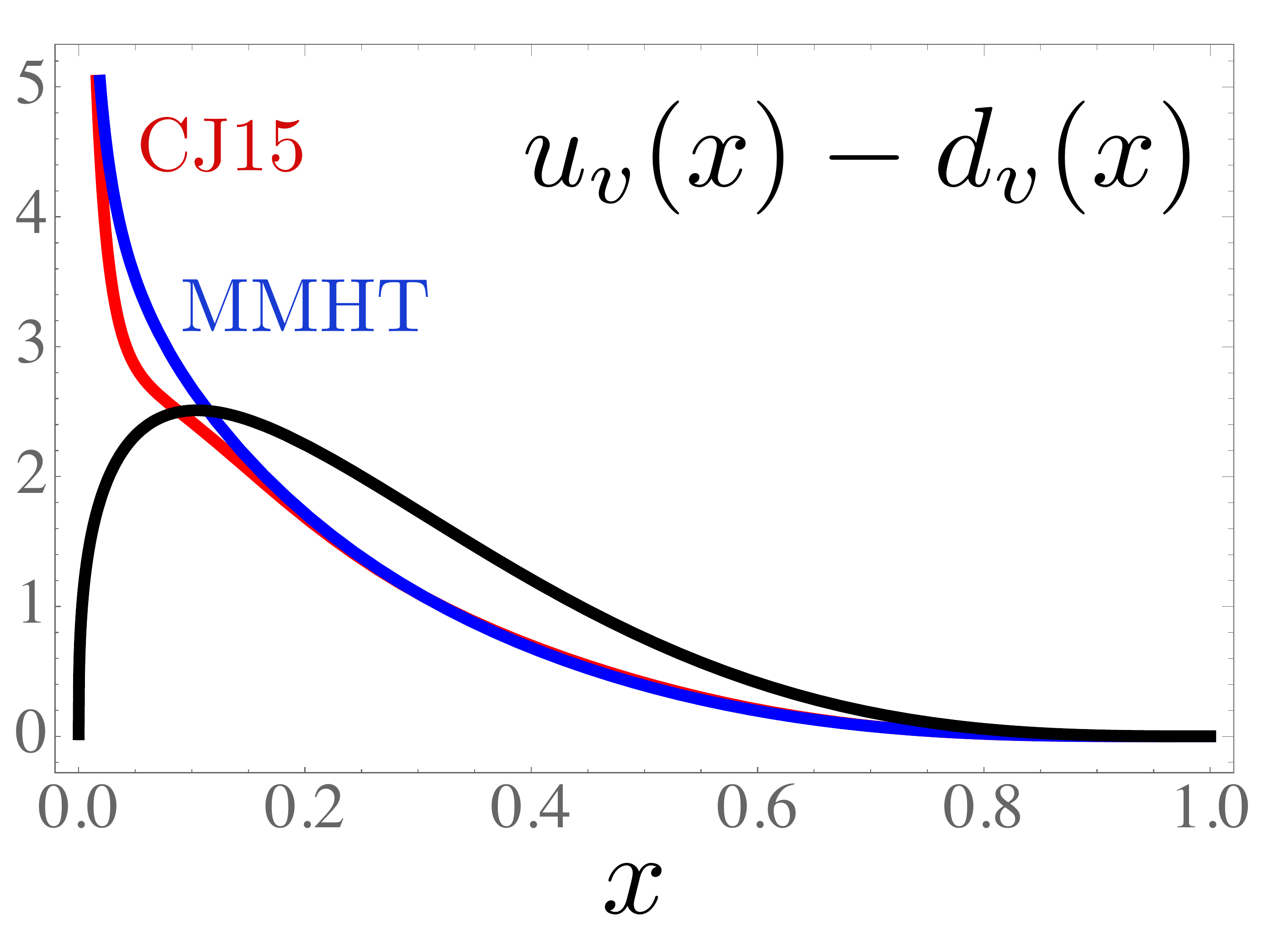} }
    \caption{Left: Function  ${\cal I}_R  (\nu, \mu^2) $  for
    $\mu = 1/a_L$  calculated using  the data with   $z_3$   from  $a_L$ to  $4a_L$.
        The upper curve corresponds to the ITD of the CJ15 global fit PDF. Right: Curve for $u_v(x)-d_v(x)$  at  $\mu=2.15$ GeV  built from the  data shown on the left 
  compared to CJ15 and MMHT global fits.
    \label{Msbar16}}
    \end{figure}

  One can see that 
all the  points for  ${\cal I}_R  (\nu, \mu^2) $ are  close to  some  universal    curve
   with a rather small scatter.  
     The  curve itself  
     was obtained by fitting the points by 
 the cosine transform  of a normalized $N x^a (1-x)^b$ distribution, 
which gave   
   $a=0.35$ and $b=3$.
   The magnitude of the scatter illustrates  the error of the fit for the 
   ITD in the $\nu \leq 4$ region. 
   For comparison, we show 
 the ITD  
 obtained from the global fit PDFs   
 corresponding to the CJ15     global fit.\cite{Accardi:2016qay}  
 One can see that our ITD is systematically below the curve based on the global fit PDFs.  

The ``mathematical''  reason for the discrepancy may be understood from  
the right panel of  \mbox{Fig. \ref{Msbar16},}  
where 
 we compare 
 the normalized  \mbox{$N x^{0.35} (1-x)^3 $} \mbox{$\equiv q_v (x, \mu=2.15$ GeV)}  
 distribution to CJ15  [\citen{Accardi:2016qay}] and MMHT 2014 
   [\citen{Harland-Lang:2014zoa}]  global fit PDFs, 
 taken at the  scale $\mu=2.15$ GeV. 
 Unlike the $\sim  x^{0.35} $ function, these PDFs are singular for small $x$,  
 which leads to the enhancement of  ITDs for large and moderate values of $\nu$.  
 
 The singular small-$x$ behavior of the global fit PDFs reflects the Regge dynamics,
 in particular, the parameters of the $\rho$-trajectory. 
 Since the $\rho$-meson may be treated as a resonance  in the two-pion system,
 a possible 
 ``physical'' reason for the discrepancy lies in the simplified  features of the lattice simulation 
used in \mbox{Ref.  [\citen{Orginos:2017kos}]:} the  quenched approximation and very large pion mass.

\subsection{Imaginary part}

Imaginary part of the pseudo-ITD may be considered in a similar way.
 It  corresponds  to 
  the sine Fourier transform 
  \begin{align} 
{\rm Im} \,  {\mathfrak  M} (\nu)=  & \int_0^1 \dd x \,  
\sin  (\nu x)  \, [q (x)  +  \bar q (x)]
\label{MS}
     \end{align}
of the function   given by the sum   $q(x) + \bar q (x) $    of 
quark and antiquark   distributions.  This function 
differs from   the valence combination  $q_v(x) = q(x) -\bar q(x)$   by $2 \bar q (x) = 2[ \bar u (x) - \bar d(x)]$.
In the left panel  of Fig. \ref{imor713}, we show the data for large $z_3$ values $z_3 \geq 7a$. 
Just like in the case of the real part (see Fig. \ref{M713}), the points with $\nu \lesssim 10$
are close to a universal curve. Representing  $q (x)  +  \bar q (x) =q_v(x) +2 \bar q (x)$  
and taking $f(x)$ of Eq. (\ref{qV})   as $q_v (x)$,   the difference is fitted to be given by   
  \begin{align} 
    \bar q (x) \approx  0.1 \,[20  x \, (1-x)^{3}]\,  . 
    \label{barq}
     \end{align}

   \begin{figure}[h]
  \centerline{\includegraphics[width=1.6in]{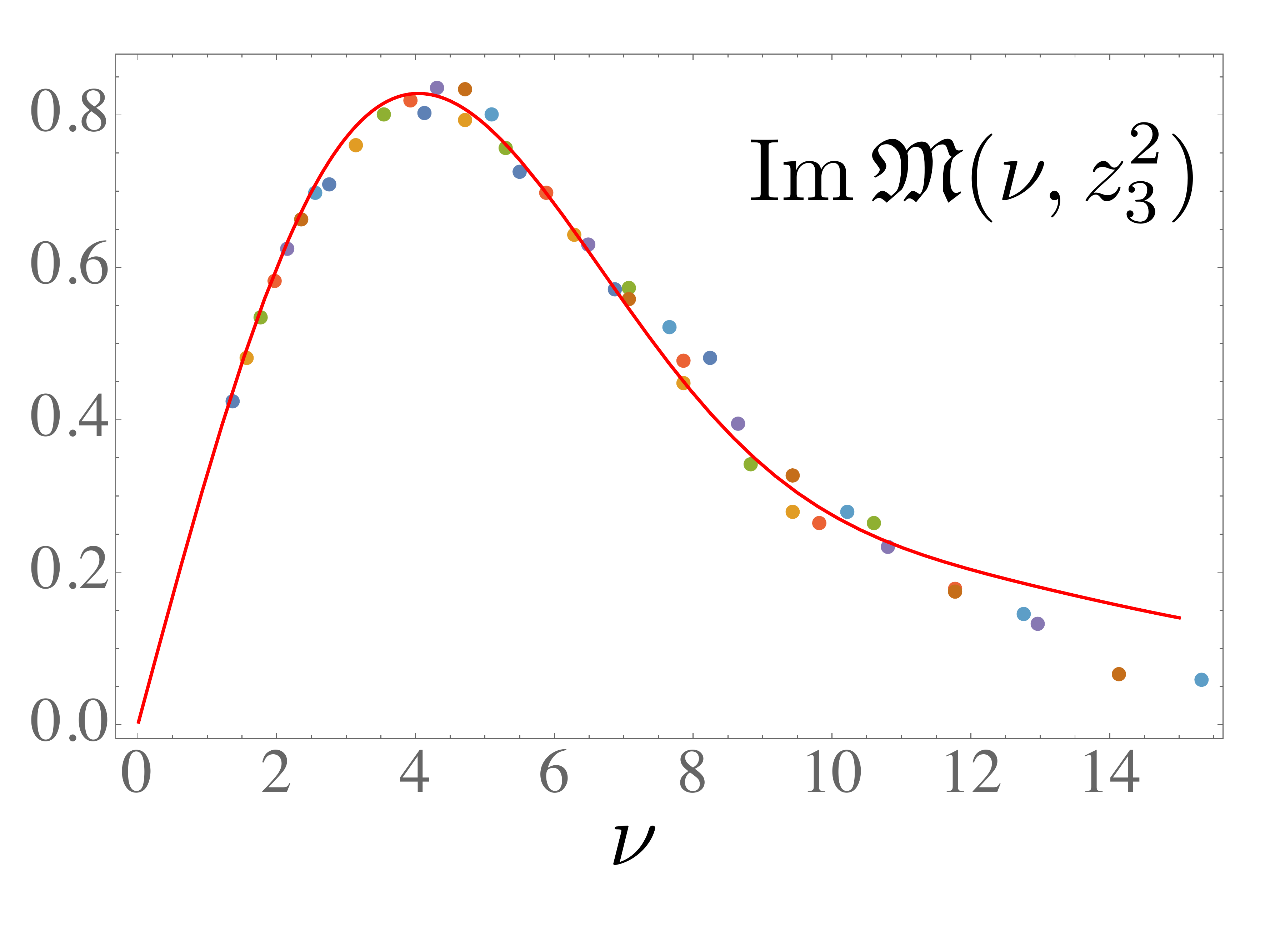} \includegraphics[width=1.6in]{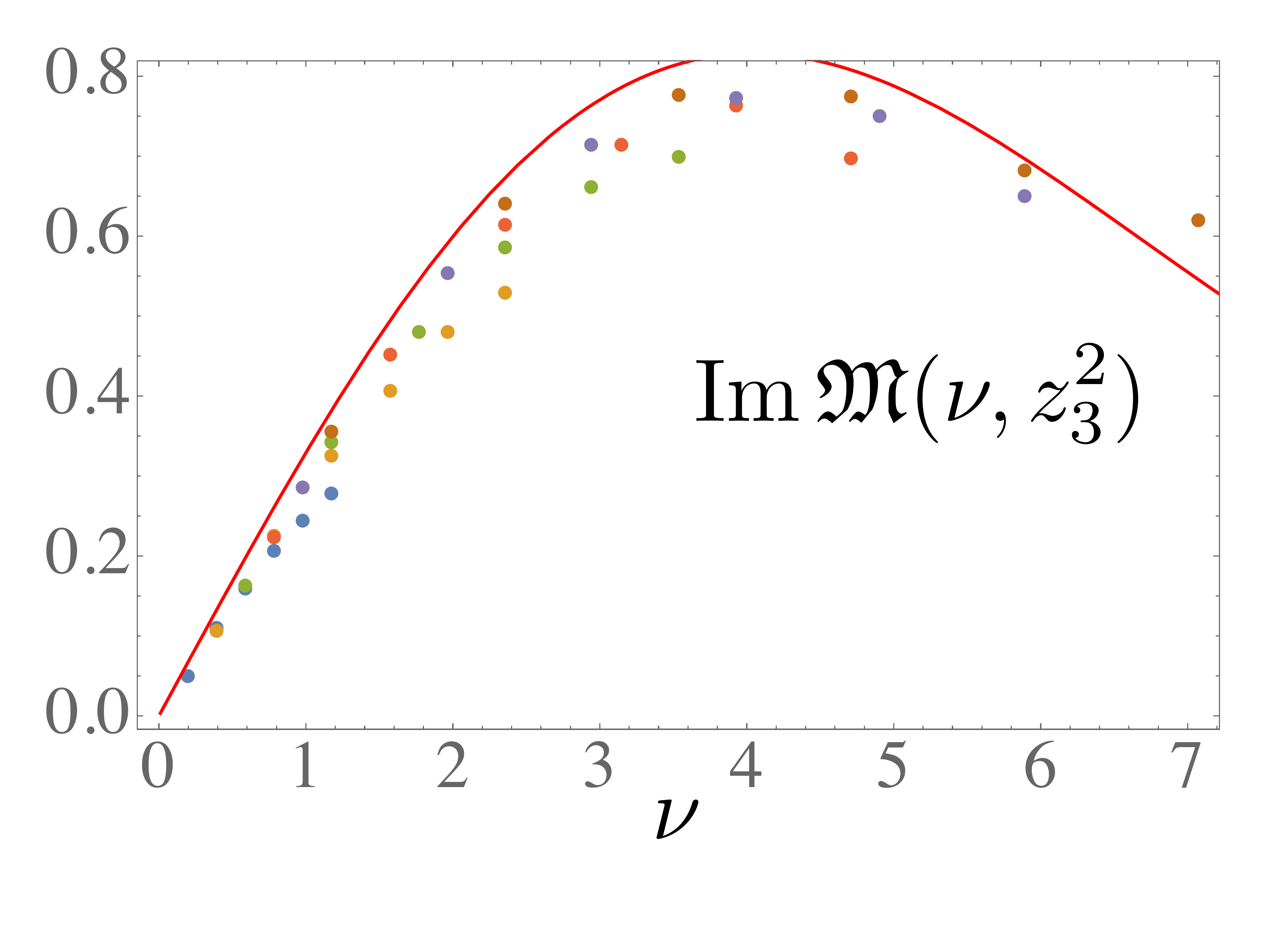}\includegraphics[width=1.6in]{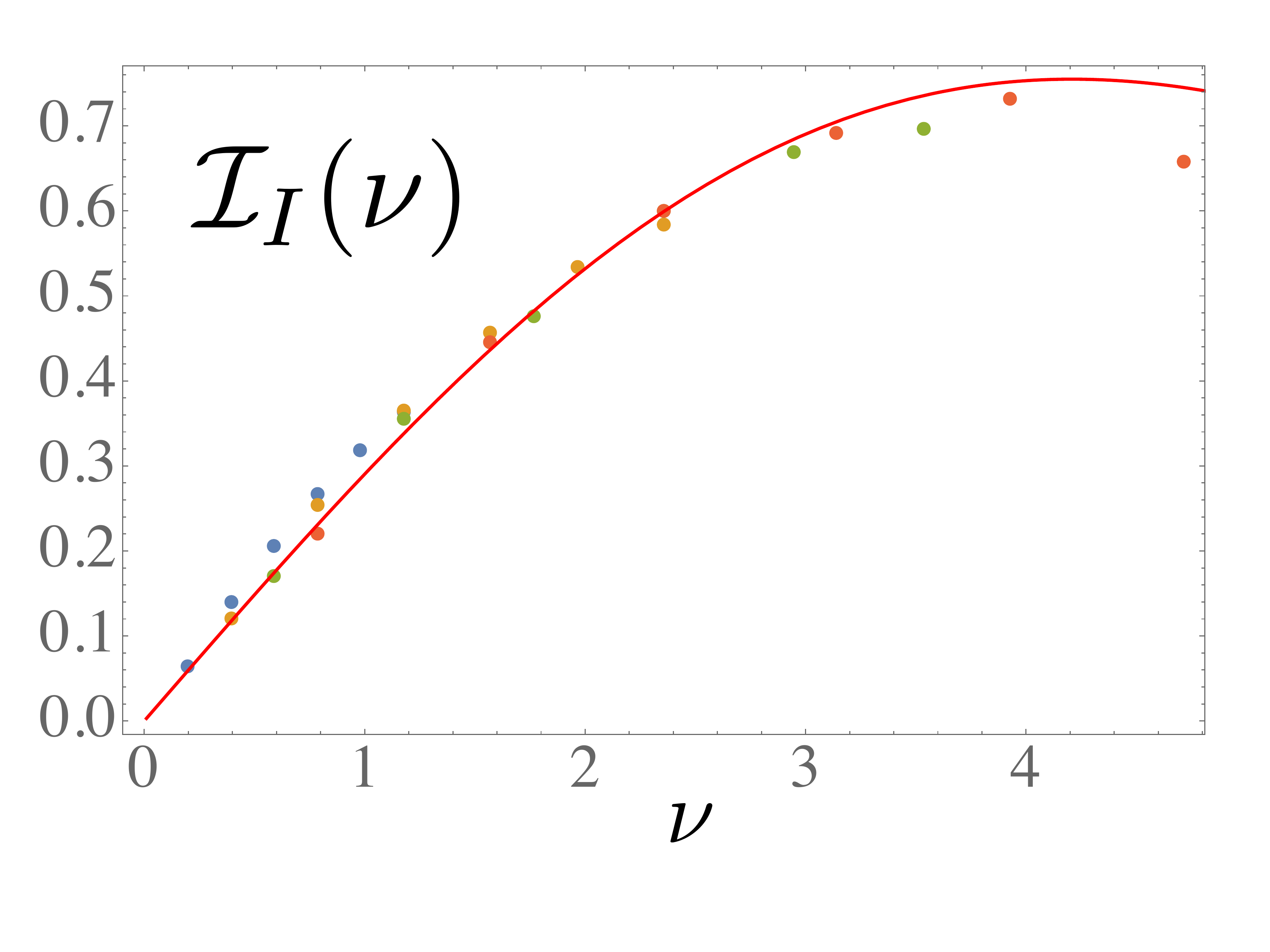}}
    \caption{ Left: Imaginary  part   of ${\mathfrak  M} (\nu, z_3^2)$  for  $z_3$  ranging  from  $7a_L$ to  $13a_L$. 
     Middle: The same for $z_3$ from  $a_L$ to  $4a_L$. In both cases, 
    the curve corresponds to $q (x)  +  \bar q (x)= f_v(x) + 2 \bar q (x)$, with $f_v(x)$ given by Eq. (\ref{qV}) 
    and $\bar q (x)$ given by Eq. (\ref{barq}). Right: Function  ${\cal I}_I  (\nu, \mu^2) $  for
    $\mu = 1/a_L$  calculated using  the data with   $z_3\leq 4a$.
    The curve is described in the text. 
    \label{imor713}}
    \end{figure} 

In the middle  panel of  Fig. \ref{imor713}, we show data with $z_3 \leq 4a$. 
All these  points are below the curve  obtained by fitting the $z_3 \geq 7a$ data. 
This is in agreement with the fact 
that, in the region $\nu \lesssim 6$, the perturbative evolution  
decreases   the imaginary part of the pseudo-ITD
when $z_3$ decreases. 
The construction of 
the $\overline {\rm MS}$  function ${\rm Im}\,  {\cal I} (\nu, \mu^2) \equiv {\cal I}_I (\nu, \mu^2)$
proceeds in the same way  as for the real part. 

 The results are shown in the right panel of \mbox{Fig. \ref{imor713}.}
Again, all the  points are rather close to a   universal    curve
   with a rather small scatter.  The curve shown corresponds to the sine Fourier transform 
   of the sum of the  valence distribution  $q_v (x, \mu =1/a_L)$  obtained 
  from the study of the 
   real part,  and the antiquark contribution $2 \bar q (x, \mu =1/a_L$).  The latter was found from the fit  to be 
   given by $ \bar q (x, \mu =1/a_L=2.15$ GeV$)= 0.07 [20 x (1-x)^3]$.

Note that  the result for $ \bar q (x)$ is a   positive function of $x$,
which means that $\bar u (x) > \bar d(x)$ in the lattice simulation of 
Ref. [\citen{Orginos:2017kos}]. For  the quenched approximation,
this is a natural outcome: in the absence of quark loops, the ratio $\bar u/\bar d$ 
reflects the number of the $u$- and $d$-quarks in the proton.

     \setcounter{equation}{0}

 \section{Calculation with dynamical fermions}

 A calculation with  dynamical fermions was reported in Ref. [\citen{Karpie:2018zaz}].
The analysis was performed using three lattice ensembles for a pion mass of about 400 MeV.
Two lattice spacings have been used. For the 
 lattice spacings $a=$ 0.127 fm, the calculations have been performed 
 on $24^3\times 64$ and $32^3 \times 96$ lattices.
 For a smaller lattice spacing of 0.94 fm, a  $32^3 \times 64$ lattice was used. 
All three ensembles have produced similar results, perfectly compatible  between themselves. 

The dynamical calculations are  more  time-consuming and noisy 
compared to the quenched calculations, so the results have bigger statistical errors than 
those of Ref.  [\citen{Orginos:2017kos}].  Still,  the structure 
of the pseudo-ITDs in both calculations is very similar, and their analysis 
follows the same steps. 
  
  \subsection{Rest-frame amplitude}

As discussed earlier in Sec. \ref{sITD}, \ref{linkuv}, \ref{refraq},   
the rest-frame amplitude ${\cal M} (0, z_3^2)$ within the pseudo-PDF approach plays 
the role of the UV-renormalization $Z$-factor. 
In Fig. \ref{Mrestdyn}, we show the results for two explored lattice spacings of  0.094 fm
and 0.127 fm. In the latter case, we show the points 
for a bigger $32^2 \times 96$ lattice. 
The results obtained on a smaller $24^3\times 64$  lattice 
practically  coincide with them. 
Just like in the quenched calculation,
these  points  are well described by the perturbative formula (\ref{Zuv})
(shown by a curve in Fig. \ref{Mrestdyn}),
but now with the value of 0.26 for the $\alpha_s$.

    \begin{figure}[h]
 	\centerline{\includegraphics[width=2.5in]{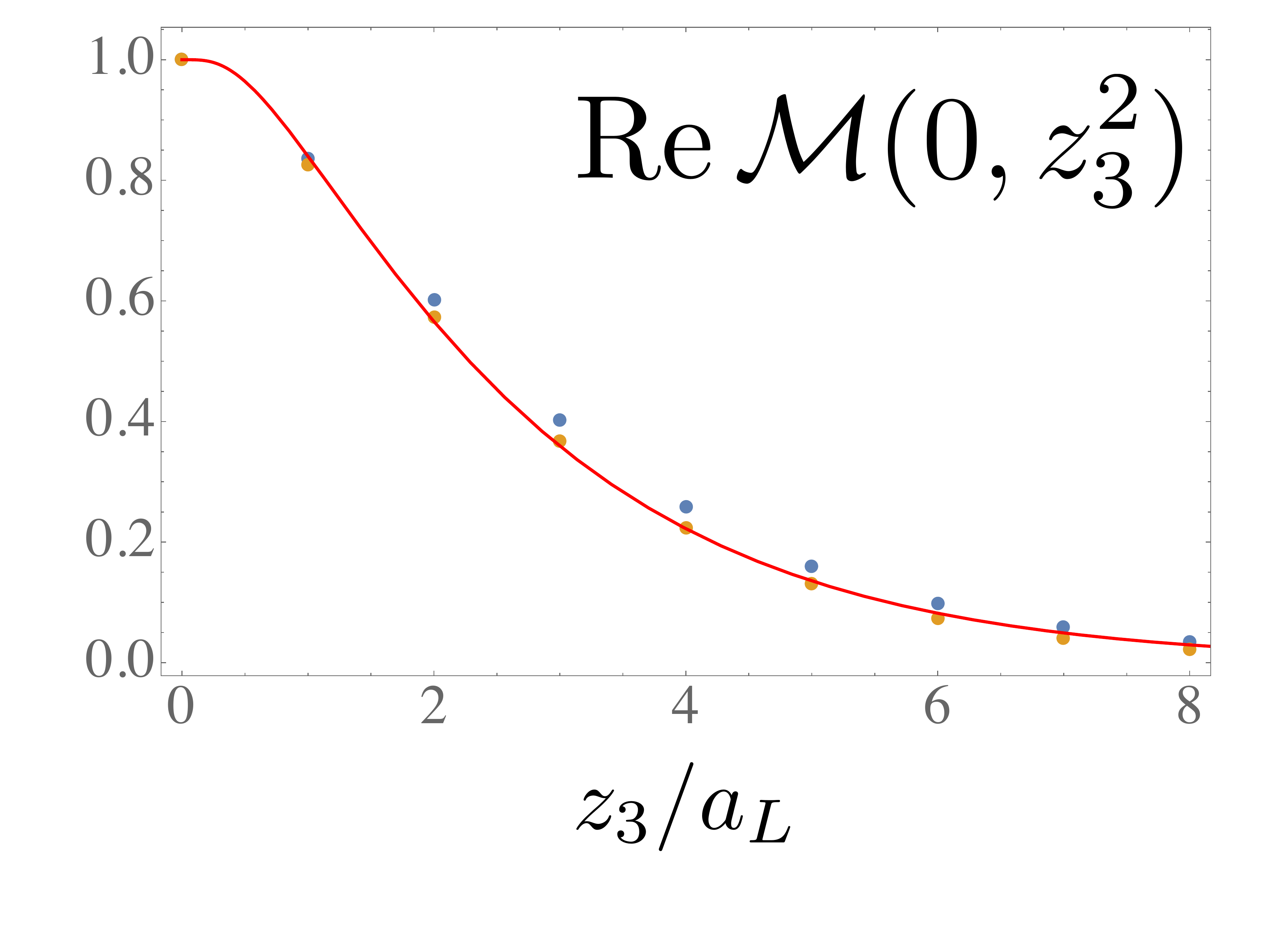}}
 	\caption{Real part of the rest-frame amplitude ${\cal M}(0,z_3^2)$
	for lattice spacings 0.094 fm (higher points) and 0.127 fm.
 		\label{Mrestdyn}}
 \end{figure}

Note that the points for the  two  different lattice spacings  are plotted as functions 
of the ratio $z_3/a_L$ rather than versus the physical distance $z_3$.
Such a choice is suggested by the perturbative calculation  that 
shows that the $Z$-factor should be a function of $z/a_L$. 
Indeed, one can see that the two sets of points in Fig. \ref{Mrestdyn}
are very close to each other. The points corresponding to the 
0.094 fm lattice spacing 
are just slightly above the curve in Fig. \ref{Mrestdyn} describing the 0.127 fm points. 
In fact, the 0.094 fm points  are also 
well described 
by the perturbative formula (\ref{Zuv}), if one uses a  smaller value  $\alpha_s=0.24$.

The fact that the $Z$-factor was found to be given by a function 
of $z_3/a_L$ (modulo a natural change of $\alpha_s$ to a smaller value
in the case of a smaller lattice spacing) is a clear demonstration 
that it is an artifact of the lattice calculation rather than a 
function describing physical effects. 
    
\subsection{Reduced Ioffe-time distributions}
\label{redITD2}

The data on the reduced  pseudo-ITD are shown in Fig. \ref{Dynda} 
for the lattice spacing 0.094 fm (left) and for 0.127 fm on the large  $32^3\times 96$ lattice (right).
 The  curves in both cases correspond to $e^{-0.05 \nu^2}$,
and were drawn to demonstrate  that the results  in both cases are rather similar. 
The data on ${\mathfrak  M} (\nu, z_3^2)$ have been used to obtain the light-cone ITD
${\cal  I} (\nu, \mu^2)$ at the  scale $\mu =2$ GeV using a technique similar
to that described in Sec. \ref{redITD1}. 

  \begin{figure}[h]
	\centerline
	{ {\includegraphics[width=2.5in]{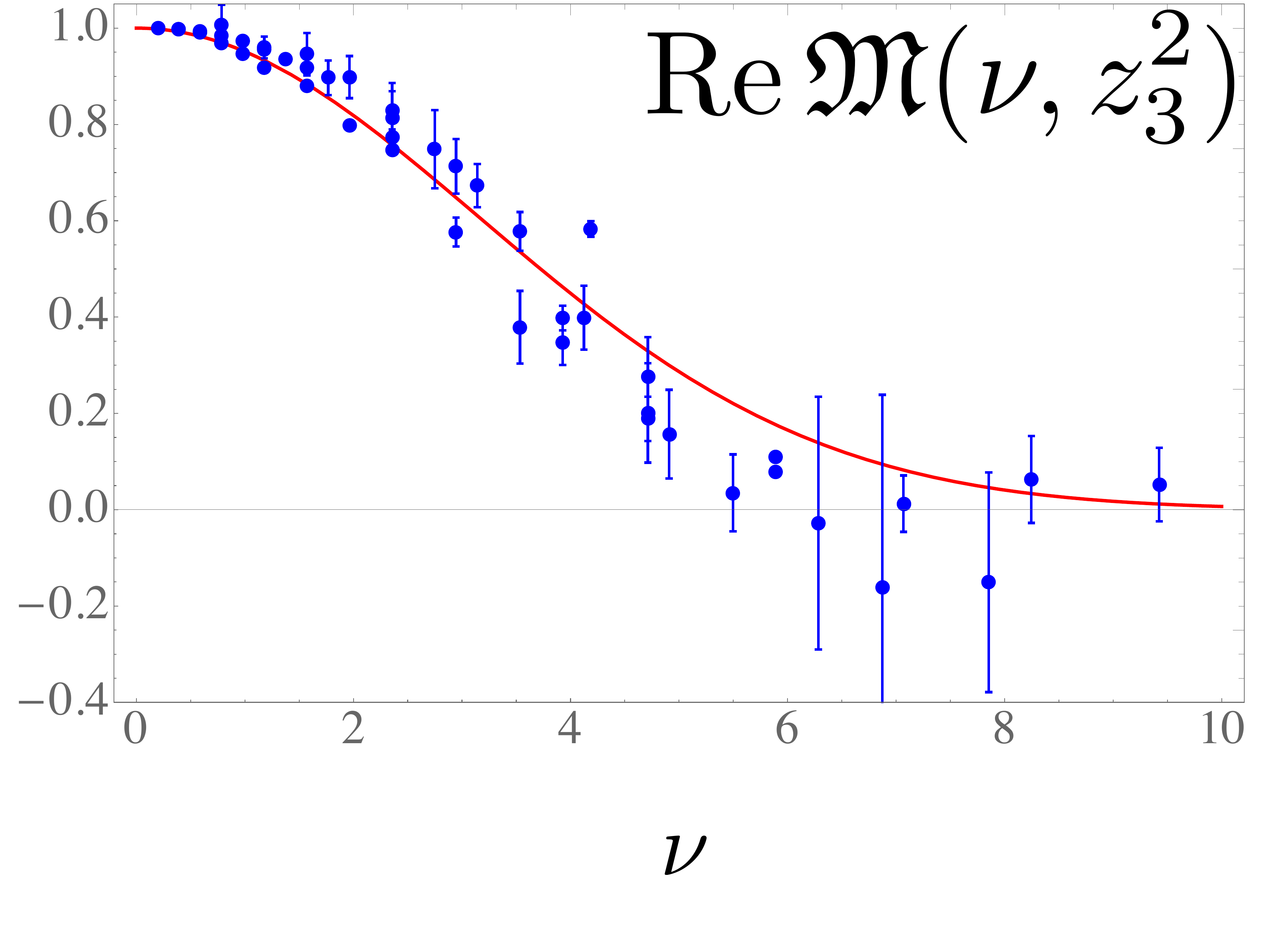} } \includegraphics[width=2.5in]{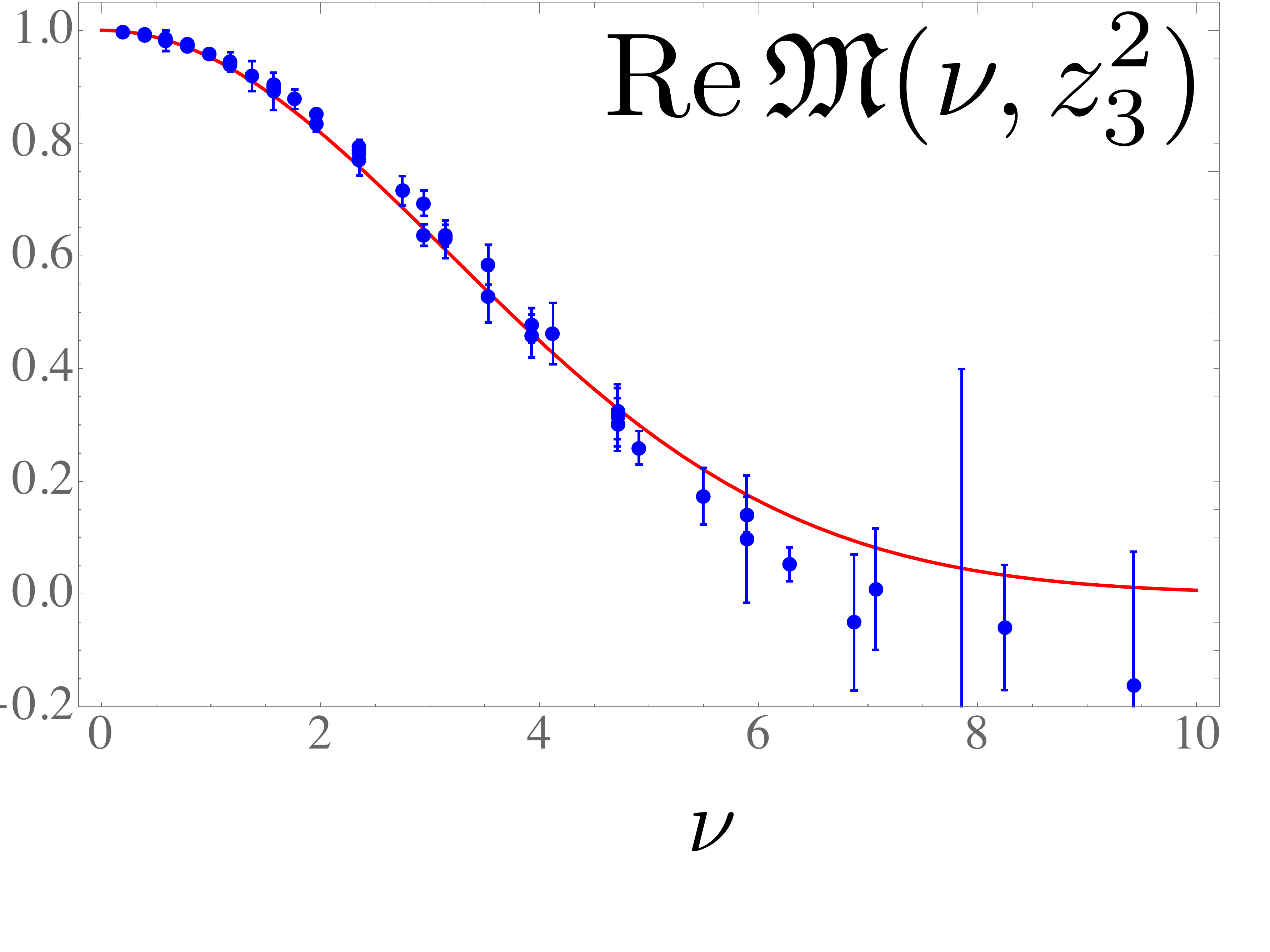}}
	\caption{Real part   of ${\mathfrak  M} (\nu, z_3^2)$  for lattice spacing 0.094 fm (left) and
		0.127 fm (right).
		\label{Dynda}}
\end{figure}

 The function ${\cal  I} (\nu, \mu^2)$ is plotted  on the left panel of Fig. \ref{Dynda2}.
 The light-cone PDF $q_v (x) = u_v(x)-d_v(x)$  extracted from this ${\cal  I} (\nu, \mu^2)$  is shown on the 
right panel.  The central line for the  result of this calculation with  dynamical 
fermions is in a much better agreement with phenomenological curves.  Still, the error band 
is very wide, which calls for a simulation having  a better statistics. 

    \begin{figure}[t]
	\centerline
	{ {\includegraphics[width=2.5in]{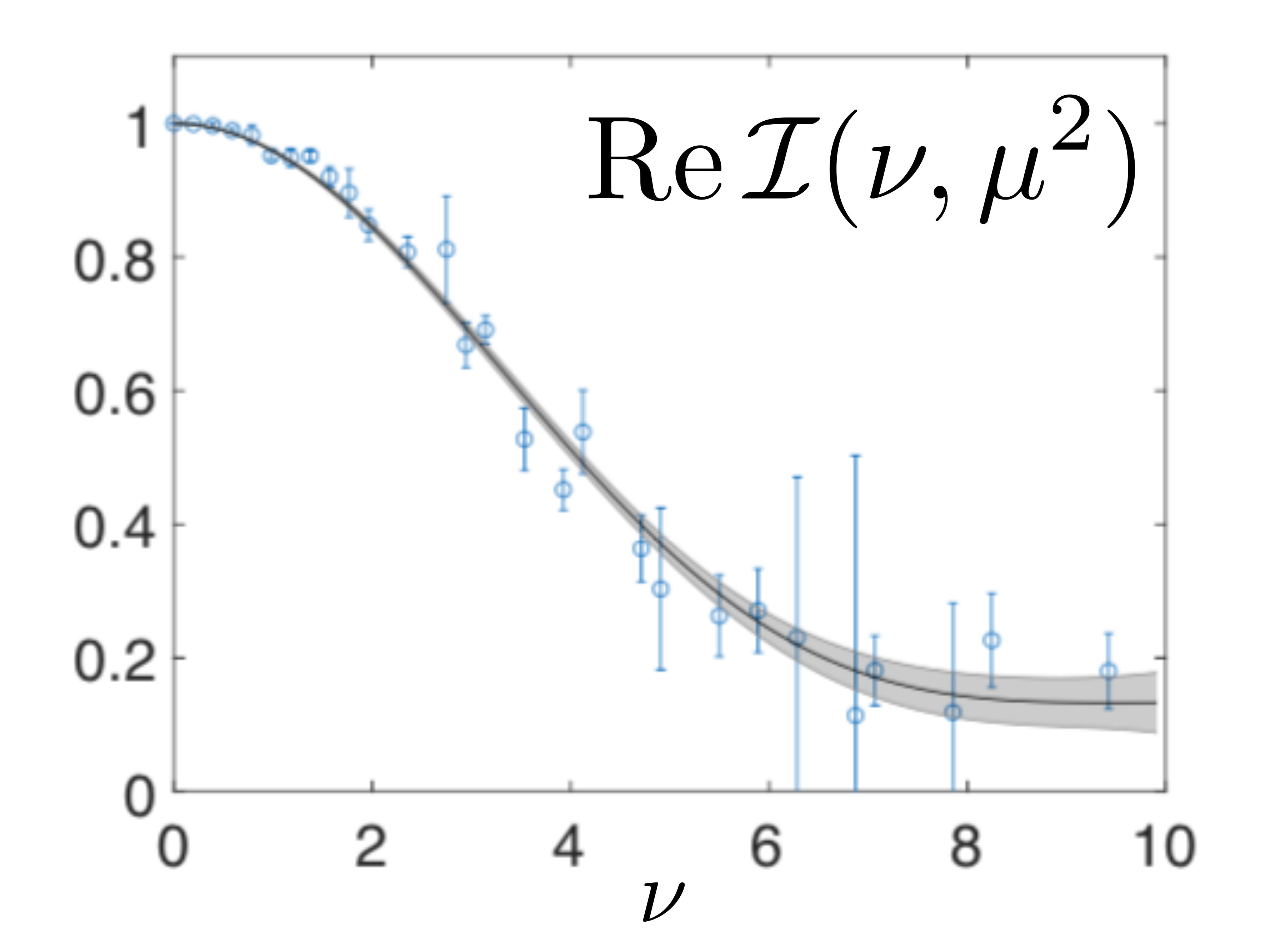} } \includegraphics[width=2.5in]{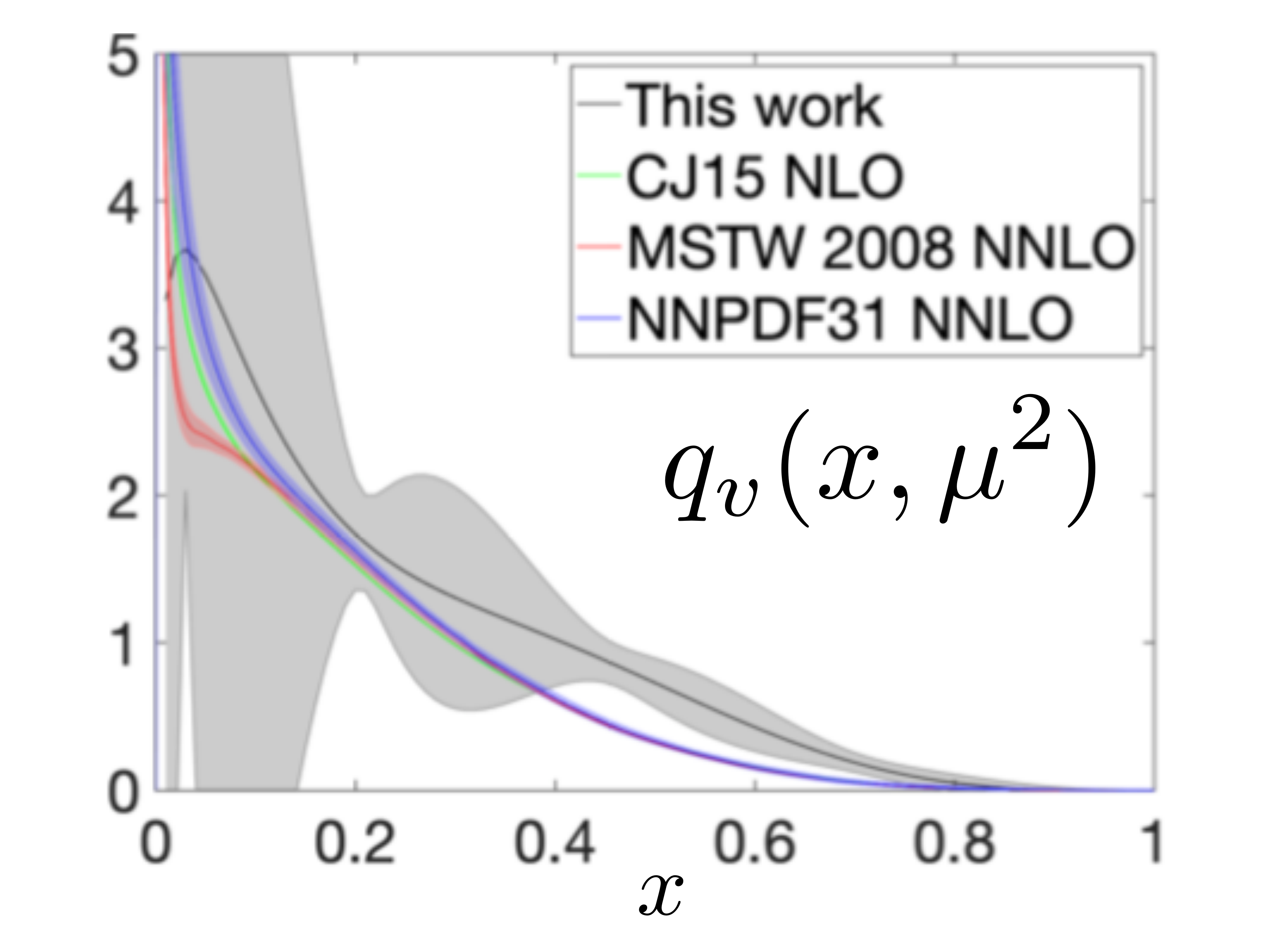}}
	\caption{Real part   of the light-cone ITD  ${\cal I } (\nu, \mu^2)$  (left)  
	and the valence PDF  $q_v (x,\mu^2)$ (right)  for $\mu = 2$ GeV extracted from the data for lattice spacing 0.094 fm. 
			\label{Dynda2}}
\end{figure}

 \subsection{Moments}
 
 The basic matching relation (\ref{Mtch})  has a $w$-convolution structure  in  its
 ${\cal O} (\alpha_s)$ part. However, it  may be converted into a
 product form 
if one considers  the 
 $x^n$ 
 moments 
 \begin{align}
  b_n(z_3^2) \equiv  \int_0^1 \dd x x^{n} {\mathfrak P}(x,z_3^2) =
\left.   i^n \frac{\partial^n\,{\mathfrak M}(\nu,z_3^2)}{\partial \nu^n} \right|_{\nu=0}\
  \label{bn}
 \end{align}
  of the renormalized pseudo-PDF ${\mathfrak P}(x,z_3^2)  \equiv {\cal P}(x,z_3^2)/{\cal M} (0,z_3^2)$.
 This gives 
 \begin{align}
  b_n(z_3^2) = K_n(z_3^2 \mu^2 ) a_n(\mu^2) + {\cal  O} (z_3^2\Lambda^2_{\rm QCD},\alpha_s^2) \ , 
    \label{momrel}
 \end{align}
 a connection  between $  b_n(z_3^2)$ 
and the  $x^n$  moments 
 \begin{align}
  a_n(\mu^2) = \int_0^1 \dd x x^{n} {f}(x,\mu^2) \
  \label{an}
 \end{align}
 of the light-cone PDF $f(x,\mu^2)$.  
The kernel $K_n(z_3^2 \mu^2 ) $ is given by\cite{Karpie:2018zaz}
 \begin{equation}\label{eq:pmom_match}
K_n(z^2\mu^2,\alpha_s) =  1  -  \frac{\alpha_s}{2\pi} C_F \left[\gamma_n \ln\left(z^2\mu^2\frac{e^{2\gamma_E +1}}{4}\right) + l_n\right]\,,
 \end{equation}
where the anomalous dimensions 
\begin{equation} 
\gamma_n = \int_0^1 \dd u\, B(u) u^n= \frac{1}{(n+1) (n+2) } - \frac{1}{2}  
- 2 \sum_{k=2}^{n+1}\frac{1}{k} 
\end{equation}
are the moments of the Altarelli-Parisi kernel $B(u)$, and the coefficients 
\begin{equation}
l_n =2\left[ \left(\sum_{k=1}^n \frac{1}{k}\right)^2 + \sum_{k=1}^n \frac{1}{k^2}
+\frac12 - \frac{1}{(n+1)(n+2)} \right]
\end{equation}
are the moments of the remaining terms in the second line of Eq. (\ref{Mtch}). 
Thus, one can now obtain the $\overline{\rm MS}$ moments 
directly from the reduced ITD  ${\mathfrak M}(\nu,z^2)$ by using
\begin{equation}
a_{n}(\mu^2) = (-i)^n\frac{1}{K_n(z_3^2\mu^2,\alpha_s)}\left.\frac{\partial^n\,{\mathfrak M}(\nu,z^2)}{\partial \nu^n} \right|_{\nu=0}
 + O (z_3^2\Lambda^2_{\rm QCD},\alpha_s^2)\,.
 \label{eq:msbarMOM}
\end{equation}
   The first moment $a_1 \equiv \langle x \rangle$ is obtained from the slope of the imaginary 
   part of ${\mathfrak M}(\nu,z^2)$, while $a_2 \equiv \langle x^2 \rangle$
   from the $\nu^2$-fit  of the real part.  The results for  $\langle x \rangle$ and 
   $ \langle x^2 \rangle$  obtained from all  three ensembles are presented in Ref. 
   [\citen{Joo:2019jct}].   In Fig. \ref{fig:mom}, we show the results for 
    $ \langle x^2 \rangle$  from the 0.094 fm ensemble. 
 
\begin{figure}[h]
\centering
\includegraphics[width=0.395\textwidth]{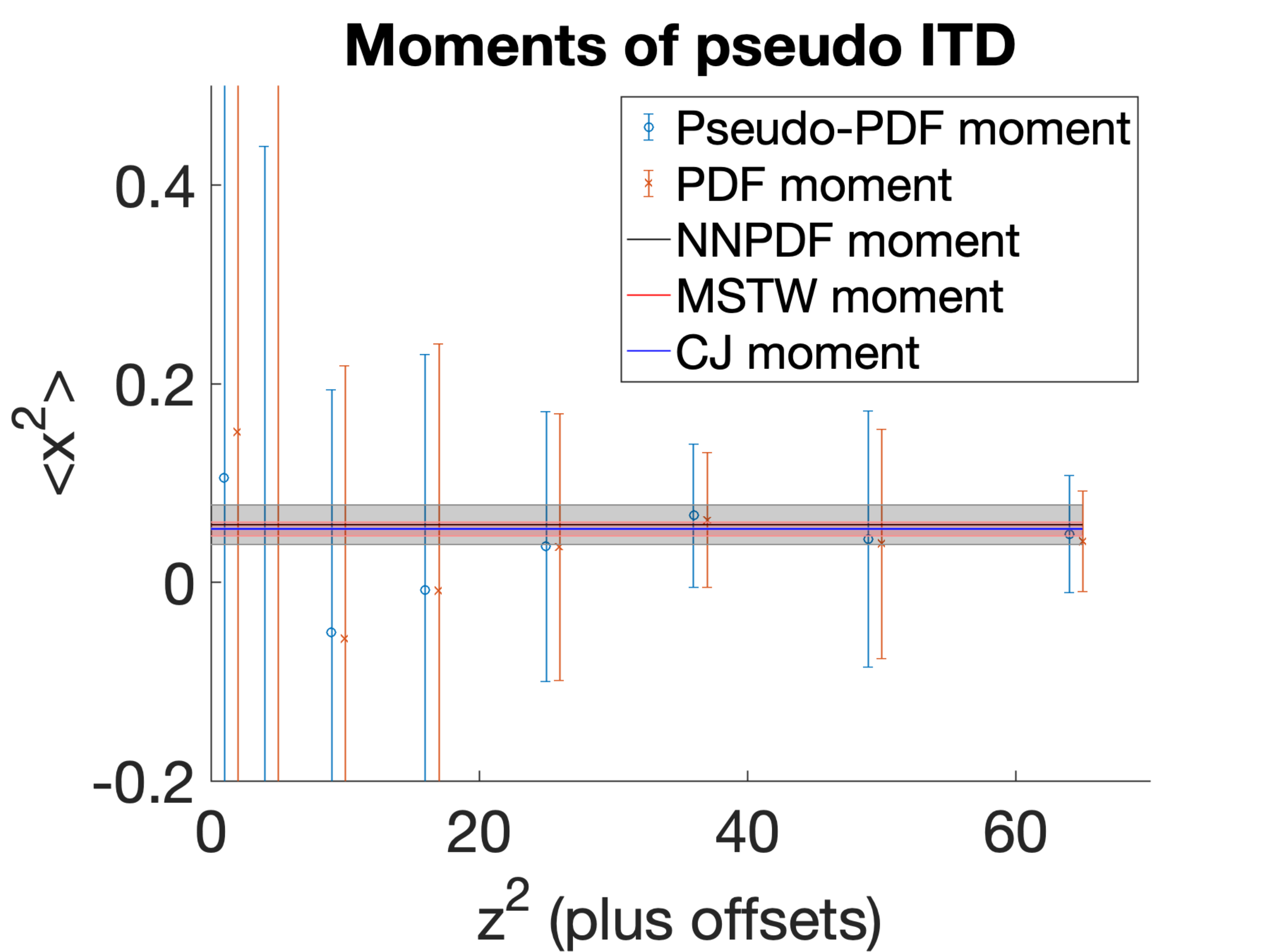}
\caption{The    $ \langle x^2 \rangle$  moment of the pseudo-PDF obtained 
 from the 0.094 fm ensemble and  compared to phenomenologically determined PDF moments from the NLO global fit CJ15nlo [\citen{Accardi:2016qay}], and the NNLO global fits MSTW2008nnlo68cl\_nf4 [\citen{Martin:2009iq}]  and NNPDF31\_nnlo\_pch\_as\_0118\_mc\_164 [\citen{Ball:2017nwa}],  all evolved to 2 GeV. }
\label{fig:mom}
\end{figure}  
 
   \setcounter{equation}{0}
 
 \section{Matching in nonforward kinematics }

 The matching relations (\ref{Mtch}) for PDFs were  derived from the 
 operator  expression  (\ref{Oha}) for  the one-loop correction 
  by inserting it into a forward matrix element
 $\langle p| \ldots |p\rangle$. 
The same  expression (\ref{Oha})  
 may be used to  deal with  nonforward matrix elements\cite{Radyushkin:2019owq}.   
 In the simplest case, we have the $\langle 0| \ldots |p\rangle$ matrix element
 corresponding   to  the pion distribution amplitude.
 A more complicated case is the matrix element 
  $\langle p_2| \ldots |p_1\rangle$ corresponding to a
  non-singlet  generalized parton distribution (GPD).

 \subsection{Matching relation for  the  pion distribution amplitude}

    Within a  framework
    of  covariant quantum field theory,  the  pion  distribution amplitude 
     was    introduced in our 1977 
    paper (see  \mbox{Ref. [\citen{Radyushkin:1977gp}]}). 
    The  starting point of the definition is 
   the  matrix element  
 \begin{align}
M^\alpha (z,p)=  \langle  0 | \bar \psi (0) \,
 \gamma^\alpha \, \gamma_5 { \hat E} (0,z; A) \psi (z)    | p \rangle \ , 
\label{DA}
 \end{align}
with $z$  taken   {\it on the light cone}.  
Here,   $ | p \rangle $  is a pion  state with momentum $p$.
    In Ref. [\citen{Lepage:1980fj}],
    a  similar object was introduced within 
    the {\it  light-front}  quantization formalism
    (see  Ref. [\citen{Radyushkin:2017gjd}] for comparison of the two definitions).
    
  For  lattice applications, we  take $z=z_3$ and the 
   \mbox{$\alpha=0$}  component to eliminate the $z^\alpha$ 
   contamination from the  decomposition of $M^\alpha (z,p)$ over   Lorentz structures
  and extract  the $p^\alpha {\cal M}(\nu, -z^2)$ 
  part.  The reduced Ioffe-time distribution is built through ${\mathfrak M}(\nu, z_3^2)=
  {\cal M}(\nu,z_3^2)/  {\cal M}(0,z_3^2)$. 
  
  As shown in Ref. [\citen{Radyushkin:1983wh}],      for all contributing  Feynman diagrams  
we have 
   \begin{align}
 {\cal M} (\nu, z_3^2) 
&   = 
 \int_{0}^1 \dd x 
 \, e^{i x \nu  } \,  {\cal F}  (x, z_3^2)  \   . 
  \label{MPDA}
\end{align}   
The function  $  {\cal F} (x, z_3^2) $ is the pion {\it pseudodistribution amplitude} 
(pseudo-DA).   Similarly to pseudo-PDFs, we get  a covariantly defined 
variable $x$, having  in this case  the $0\leq x \leq 1$ support. 
To   exploit the symmetry properties of $  {\cal F} (x, z_3^2) $ 
with respect to the $x \to 1-x$ interchange, 
 it is convenient to use the $(-z/2,z/2)$ endpoints instead of $(0,z)$.
The relation between the two cases is provided by   translation invariance, 
   \begin{align}
  \widetilde {\cal M} (\nu, z_3^2)   \equiv 
   \langle  0 | \bar \psi (-z_3/2) \,  \ldots 
\psi ( z_3/2)    | p \rangle  &=
e^{-i  \nu/2 }   {\cal M} (  \nu, z_3^2)  \ . 
\label{Mtilde}
\end{align}

 Using  explicit form (\ref{Oha}) of the one-loop correction, 
(\ref{Oha}) and   parametrizing  
   \begin{align}
   \langle  0 | \bar \psi (uz_3) \,&  \ldots 
\psi (\bar v z_3)    | p \rangle  = 
e^{i u \nu}   {\cal M}_0 [(1-u-v )  \nu]  \ , 
\  
  \label{ubarv}
\end{align} 
one may derive  the matching condition for  the pion 
DA\cite{Radyushkin:2019owq}    
    \begin{align} 
&   \widetilde {\mathfrak M} (\nu, z_3^2)= \widetilde  {\mathcal I}(\nu,\mu^2)   
  -
\frac{\alpha_s}{2\pi} \, C_F 
\,\int_0^1  \dd w   \, \widetilde   {\mathcal I}(w\nu,\mu^2)   \nn & \times 
\Biggl  \{ \ln \left [ z_3^2 \mu^2\frac{ e^{2\gamma_E+1}}{4}  \right ]
\Biggr (
\Biggl  [\frac{2w }{1- w } \Biggr ]_+ \, \cos ( \bar w \nu/2  ) 
+  \frac{ \sin (\bar{ w} \nu/2)}{\nu/2} -\frac12 \delta( \bar w)\Biggr )\nn & + 
   4    \Biggl [\frac{\ln (1-w) }{1-w } \Biggr ]_+ \cos ( \bar w \nu/2  )  
   -2  \frac{ \sin (\bar{ w} \nu/2)}{\nu/2}+ \delta( \bar w) \Biggr \} \  .
\label{ITDmDAf}
 \end{align}
 We use here the ``tilded''  light-cone ITD $ \widetilde  {\cal I } (\nu, \mu^2)$  
 corresponding  to  the $(-z/2,z/2)$ endpoints. It     is related 
 to the light-cone pion DA $\Phi (x, \mu^2) $ by 
      \begin{align}
\widetilde  {\cal I } (\nu, \mu^2) 
&   = 
 \int_{0}^1 \dd x 
 \, e^{i (x-1/2) \nu  } \,  \Phi (x, \mu^2)  \   .
  \label{MPDa}
\end{align}   
Thus, if $ \Phi (x, \mu^2) $ is even (odd) with respect to the $x \to 1-x$ interchange,
then $\widetilde  {\cal I } (\nu, \mu^2) $ is even (odd) function of $\nu$. 
   
 To extract $ \Phi (x, \mu^2)$,
   we recommend, just like in the PDF case,   to assume 
   some parametrization for it, say,  $N (x \bar x)^a$ times some polynomial of $x$, and 
   then to  fit the parameters of the model by $\widetilde  {\cal I } (\nu, \mu^2) $ extracted from 
 the lattice data.
    Another way is to use a kernel relation,
     analogous to   Eq. (\ref{MtchI}),  which expresses  $ \widetilde {\mathfrak M} (\nu, z_3^2)$ 
   in terms of  $  \Phi (x, \mu^2)$.   It is straightforward to 
   calculate the analog of the  $R(x \nu, z_3^2 \mu^2)$ 
   in  a  closed form. The  further procedure is to 
    fit $\alpha_s$ and the parameters of 
   the model for the  light-cone DA $\Phi (x, \mu^2)$  using the lattice data for 
   the reduced pseudo-DA $ \widetilde {\mathfrak M} (\nu, z_3^2)$.

\subsection{Definitions and kinematics of GPDs}

In the case of GPDs, we should consider a nonforward matrix element 
  $\langle p_2| \ldots |p_1\rangle$ involving hadronic states with
  two different momenta. The 
 simplest case is  the pion.  It has just one    light-cone GPD  $H(x,\xi,t;\mu^2)$  that 
  may be defined\cite{Ji:1996ek} 
 by 
  \begin{align}
  \langle p_2 |&  \bar   \psi(-z/2) \gamma^\alpha \hat E(-z/2,z/2; A) \psi (z/2)|p_1 \rangle 
  \nn &
   = 2 {\cal P}^\alpha
  \int_{-1}^1 \dd x \, e^{-i x ({\cal P}z)} \, H(x,\xi,t;\mu^2) \ , 
\,  
 \
 \label{Hxi}
\end{align} 
(see also Refs. [\citen{Mueller:1998fv,Radyushkin:1997ki}]), where 
 the coordinate  $z$ has only the $z_-$ light-cone component
 and the choice $\alpha=+$ is made to eliminate the $z^\alpha$ part.  
 As usual,  $\mu$ is the  factorization scale.
 Note that this definition involves the $(-z/2,z/2)$ endpoints, which 
simplifies the  analysis of the $x \to -x$ symmetry properties of $H(x,\xi,t;\mu^2) $.

The momentum 
${\cal P}= (p_1+p_2)/2$ here  is the average of the hadron momenta.
The   {\it skewness}  variable $\xi$  is related to the plus-component of their difference $p_1-p_2 \equiv r$.
Namely, $\xi= r^+ /2{\cal P}^+$. 
One more variable is given by the invariant momentum transfer  $t=(p_1-p_2)^2$.
In principle, the right-hand side of Eq. (\ref{Hxi}) may 
have  also the $r^\alpha$ term.   
However, when we take $\alpha=+$, such a term is redundant, since   $r^+=2\xi {\cal P}^+$. 
 
 A similar definition  holds for the spin non-flip GPD 
$H(x,\xi,t;z^2)$ of the nucleon. One should just  
substitute  $2{\cal P}^+$ by $\bar u(p_2)\gamma^+ u(p_1)$. 

 For a general case,  the skewness   $\xi$ may be  defined as 
  \begin{align}
  \xi = \frac{(p_1z)-(p_2z)}{(p_1z)+(p_2z)}
  \ . 
   \label{xi}
\end{align} 
Thus, we deal with  two Ioffe-time invariants
\mbox{$\nu_1 \equiv -(p_1z) $}  and \mbox{$\nu_2 \equiv -(p_2z) $. }
 For lattice applications,   we 
choose  $z=z_3$.  
Decomposing
$p_1=\{E_1,\Delta_{1,\perp},P_1\}$ and 
\mbox{$p_2=\{E_2,  \Delta_{2,\perp},P_2\}$,}  
we have $\nu_1= P_1 z_3$ and $\nu_2 =P_2 z_3$. 
The skewness variable  is  given by 
  \begin{align}
  \xi =& \frac{\nu_1-\nu_2}{\nu_1+\nu_2} = \frac{P_1-P_2}{P_1+P_2}  \,
\,  .
 \
 \label{Zz1z2}
\end{align} 
Using the $\xi$-definition (\ref{Zz1z2}), we may write \mbox{$P_1= (1+\xi)P$}  and $P_2= (1-\xi)P$,
 where  
$P\equiv {\cal P}_3$. 

Again, we choose $\alpha=0$ to eliminate the $z^\alpha$ part from the parametrization of 
$  \langle p_2 |  \bar   \psi(-z/2) \gamma^\alpha \hat E(-z/2,z/2; A) \psi (z/2)|p_1 \rangle 
$
for $z=z_3$. 
 Note that  the  $\Delta_\perp^\alpha$ contributions will be  also 
absent in the parametrization.  Hence,  we can define the double Ioffe-time pseudodistribution 
$  \widetilde{M}(\nu_1,\nu_2,t;z_3^2) $
  \begin{align}
  \langle p_2 | \bar   \psi(-z_3/2) \gamma^0 \ldots  \psi (z_3/2)|p_1 \rangle = 2 {\cal P}^{0}     \widetilde{M} (\nu_1,\nu_2,t;z_3^2)
\,   . 
 \
 \label{Mnn}
\end{align} 
We use here the ``tilde'' notation   indicating that $  \widetilde{M} (\nu_1,\nu_2,t;z_3^2)$ 
parametrizes  the  operator with the $(-z_3/2, z_3/2)$ endpoints. 
Denoting 
   $\nu=(\nu_1+\nu_2)/2
   $, 
 we  define the {\it generalized Ioffe-time pseudodistribution} (pseudo-GITD) by  
\begin{align}
  & \widetilde{M} (\nu_1,\nu_2,t;z_3^2) =  \widetilde{\cal M} (\nu, \xi ,t;z_3^2)\ . 
  \end{align} 
 It  is related to    the {\it  pseudo-GPD} by 
  \begin{align}
  & \widetilde {\cal M} (\nu, \xi,t;z_3^2)
 = e^{i \xi \nu } \,  \int_{-1}^1 \dd x \, e^{i x \nu } \, 
    {\cal H} \left (x,\xi,t;z_3^2\right ) 
\,  .
 \
 \label{MH}
\end{align}

Using  the 
operator   expression (\ref{Oha})  for the one-loop contribution  gives
the matching relation for GPDs
\begin{align} 
& \widetilde {\mathfrak M} (\nu, \xi, t, z_3^2) =  \widetilde  {\mathcal I}(\nu,\xi,t,\mu^2)    
  -
\frac{\alpha_s}{2\pi} \, C_F 
\,\int_0^1  \dd w   \, \widetilde  {\mathcal I}(w \nu,\xi,t,\mu^2)  \nn & \times
\Biggl  \{ \ln \left [ z_3^2 \mu^2\frac{ e^{2\gamma_E+1}}{4}  \right ]
 \Biggr (
\Biggl  [\frac{2w }{1- w } \Biggr ]_+ \, \cos ( \bar w \xi \nu  ) 
+  \frac{ \sin (\bar{ w} \xi \nu)}{\xi \nu} -\frac12 \delta( \bar w)\Biggr )\nn & + 
   4    \Biggl [\frac{\ln (1-w) }{1-w } \Biggr ]_+ \cos ( \bar w \xi \nu  )  
   -2  \frac{ \sin (\bar{ w} \xi  \nu)}{\xi \nu}+ \delta( \bar w) \Biggr \} \  \ . 
\label{ITDmGPD}
 \end{align}
 Its structure is  similar to the matching relation (\ref{ITDmDAf})
  for the pion DA. Eq. (\ref{ITDmGPD}) 
relates 
 the reduced pseudo-GITD
   \begin{align}
\widetilde {\mathfrak M} (\nu, \xi,t, z_3^2) \equiv \frac{\widetilde 
{\cal M} (\nu, \xi, t, z_3^2)}{\widetilde {\cal M} (0,0,0, z_3^2)} \  .
 \label{redITDGPD}
\end{align}
with 
 the light-cone  GITD 
  \begin{align} 
&  \widetilde  {\mathcal I}(\nu,\xi,t,\mu^2)   =   \int_{-1}^1 \dd x\, e^{i x \nu} H(x,\xi,t;\mu^2) \ .
\label{LCITDGPD}
 \end{align}

 We  propose again to extract $H(x,\xi,t;\mu^2)$  by  taking  some parametrization 
for it, and then to  fit its  parameters by using the lattice data on $\widetilde {\mathfrak M} (\nu, \xi,t, z_3^2) $.
Building the model, one should  take into account the {\it polynomiality} property \cite{Ji:1996ek,Mueller:1998fv,Radyushkin:1997ki} of 
   GPDs, i.e.,  the requirement  
that, in the non-singlet case,  the  $x^N$ moment of $H(x,\xi,t;\mu^2)$ 
should be a polynomial of the  $N$th degree in $\xi$. 
An efficient    way to satisfy this requirement is to use the {\it double distribution Ansatz} \cite{Radyushkin:1998es}. 

Another (but  equivalent)   strategy is to convert (\ref{ITDmGPD})  into a  kernel relation. 
It is obtained by  writing 
  the light-cone  GITD 
 $ \widetilde  {\mathcal I}(\nu,\xi,t,\mu^2) $  in terms of $H(x,\xi,t;\mu^2)$ using  Eq. (\ref{LCITDGPD}).
The  kernel relation allows  then to  fit the parameters of  $H(x,\xi,t;\mu^2)$  from the lattice data on $\widetilde {\mathfrak M} (\nu, \xi,t, z_3^2)$. 

\subsection{Lattice implementation}

Lattice measurements involve 
 a discrete  set of  values both  for coordinates $z_3 = n_z a$ and 
 for longitudinal momenta $P_1= 2\pi N_1 /L$,   
$P_2= 2\pi N_2 /L$, where $L=n a$ is the lattice size in the $z_3$ direction. 
Hence,  possible values of the Ioffe-time parameters  are given by 
discrete sets $\nu_1=2\pi n_z N_1/n$ and  $\nu_2=2\pi n_z N_2/n$.
As a result,  possible values for skewness are given by 
rational numbers 
   \begin{align}
\xi = \frac{P_1-P_2}{P_1+P_2} =\frac{N_1-N_2}{N_1+N_2} \ . 
\label{xipn}
\end{align} 
Changing $N_1$ and $N_2$ from 0 to 6, one ends up with 
  13 possible values for $\xi$. They  range from 0 to 1 and 
represent rather well   the whole $0\leq \xi \leq 1$ segment.
A complication is that,   varying the  skewness  $\xi$,
 one also    changes the value of the momentum \mbox{ transfer $t$.}
For  given $\xi$, the momentum transfer has its minimal value $t_0$
that is achieved for  purely longitudinal initial and final momenta, 
\begin{align}
{t_0 }= & - \frac{8\xi^2 {M^2}  }{   1-\xi^2+{M^2}/{P^2}  +
 \sqrt{  ( 1-\xi^2+{M^2}/{P^2})^2  + 4\xi^2 {M^2}/{P^2}}}
 \  .
 \label{tmin}
\end{align} 
 To relax  the correlation between the   values of $t$ and  $\xi$,
 one may 
 add a transverse component $\Delta_\perp$ to the momentum 
transfer. In particular, taking  \mbox{$p_1=\{E_1,\Delta_\perp,P_1\}$}  and $p_2=\{E_2,0_\perp,P_2\}$, 
 gives 
 \begin{align}
&t = 
2M^2+ 2 P_1 P_2-\Delta_\perp^2  
- 2 \sqrt{ M^2+  P_1^2+\Delta_\perp^2}  \sqrt{ M^2+  P_2^2} 
\,  .
 \
 \label{tDel}
\end{align}
 A possible further  strategy is to choose first some particular values of $P_1$ and $P_2$.
This  fixes the value of $\xi$ and $\nu$.  The next step is to  take  several different values of $\Delta_\perp$
 to change $t$. That will give the \mbox{$t$-dependence}  for fixed $\xi$ and $\nu$.
 After this, changing $z_3$, we will change $\nu$ leaving $\xi$ and $t$ unchanged.
 Finally, using  the matching conditions to convert the \mbox{$\nu$-dependence}  into the $x$-dependence,
 we will end up with  $H(x,\xi,t;\mu^2)$ for a fixed $\xi$ as a function  of $x$ and $t$.


  \section{Summary}
  \label{summary}
  
In this paper, we reviewed   the basic  ideas   of the  pseudo-PDF  approach
to extraction of parton densities from lattice calculations, and also discussed 
the results of  practical implementations of these ideas. 

The main object of this approach, the Ioffe-time
pseudodistribution  ${\cal M} (\nu, -z^2)$, is just  
the matrix element 
  $M(z,p)$ of the correlator of parton fields,  written in terms 
 of two Lorentz invariants, the Ioffe time $\nu =-(pz)$ and $z^2$.
 We have emphasized that it is exactly this matrix element  
 that enters into  the handbag contribution 
 for the forward Compton amplitude in the DIS analysis.
 And it is this matrix element  that is the starting object for  a lattice extraction of PDFs
 both in the quasi-PDF and  pseudo-PDF  approaches. 
 
 The crucial idea  of the pseudo-PDF approach is the realization  that  
it does not matter how the product  $(pz)$ is composed.  One can build 
it using   a light-front separation $z=\{z_+=0,z_-, z_\perp \}$  or a Euclidean separation $z=\{0,0,0,z_3\}$.
In both cases, the function 
  ${\cal M} (\nu, -z^2)$  is  the same. 
This observation allows to calculate  
${\cal M} (\nu, -z^2)$ on the lattice.
 
A distinct feature  of the  pseudo-PDF approach is to study 
${\cal M} (\nu, -z^2)$ ``as is'', without converting it into an auxiliary  
object, such as a  quasi-PDF.  Since the OPE provides 
a direct relation (\ref{ker}) between the renormalized ${\cal M} (\nu, -z^2)$ and the light-cone 
PDF $f(x,\mu^2)$, no such intermediaries are  necessary.

The ``renormalization'' of  ${\cal M} (\nu, -z^2)$ is needed because it contains 
artificial 
ultraviolet divergences generated
by
the gauge link for space-like intervals. 
In the present paper, we discussed these  divergences   in some detail. 
We argued that they may be eliminated 
by just dividing  ${\cal M} (\nu, -z^2)$  with  the  rest-frame function  ${\cal M} (0, -z^2)$.
This  procedure is very  simple  and  transparent. It allows to avoid the  use of more complicated
tricks such as  
the RI/MOM  scheme  method 
(see Refs. [\citen{Cichy:2018mum,Zhao:2018fyu}] for  its recent reviews and references).

The remaining $z^2$-dependence of  $ {\cal M} (\nu, -z^2)/{\cal M} (0, -z^2)$
corresponds to perturbative evolution,
and   can be converted into the scale-dependence of the light-cone PDFs 
$f(x,\mu^2)$ using matching relations.  We gave  such relations
for nonsinglet PDFs, for the pion DA, and for nonsinglet GPDs.
All of them have been obtained from one and the same  operator expression (\ref{Oha}) 
for the one-loop corrections. 

Matching  conditions rely on perturbation theory, so they  are valid for small
$z_3^2$ only. 
Furthermore,    the applicability of
the  OPE  is determined solely
by the size of $z_3^2$. The size of the momentum $p_3$ changes  the magnitude of $\nu =p_3 z_3$, but it 
 does not  affect the applicability of the perturbative QCD expansion. 
We have emphasized that one  can take small $p_3$ (even $p_3=0$),
and use perturbative QCD as far as $z_3^2$  is sufficiently small.

The perturbative evolution was successfully observed 
in the exploratory quenched lattice calculation\cite{Orginos:2017kos}.
The analysis of its very precise data provides a methodological framework for 
extraction of parton densities using the pseudodistribution
approach. This framework  has been used in   recent  calculations\cite{Joo:2019jct,Joo:2019bzr} of the nucleon 
and pion valence quark distributions. It is also used in the ongoing 
calculations of  the pion distribution amplitude and generalized parton distributions.

\section*{Acknowledgments}

I  thank my collaborators  B. Jo{\'o}, J. Karpie, K. Orginos, D. Richards, 
R. Sufian and S. Zafeiropoulos, 
 who performed the 
 lattice simulations,   the  results of which  were  discussed  in the present  paper.
 I am also grateful to  I. Balitsky, 
V. Braun, \mbox{M. Constantinou,} R. Edwards, X. Ji, W. Melnitchouk, J. Qiu, N. Sato  and  Y. Zhao  for stimulating  discussions and correspondence. 
This work is supported by Jefferson Science Associates,
 LLC under  U.S. DOE Contract \mbox{ \#DE-AC05-06OR23177} 
 and by U.S. DOE Grant \#DE-FG02-97ER41028.

\appendix


\section{Spectral properties of pseudo-PDFs}

Pseudo-PDFs correspond   to the generic matrix element (see Fig.  \ref{chipk})
\begin{align}
 \langle  p | \phi (0)  \phi(z)  | p \rangle = \frac{1}{\pi^2}  \int { d^4 k}  \, e^{- ikz} \,   \chi (k,p)   \ , 
 \label{twist2parz}
\end{align}
where    the   momentum space  function  $\chi (k,p)$  is an analog of 
the Bethe-Salpeter  amplitude \cite{Salpeter:1951sz}.  
The complications related to spin
 do not affect  
  the  spectral properties, so  we  use 
 simplified  scalar  notations.

      \begin{figure}[h]
  \centerline{\includegraphics[width=2in]{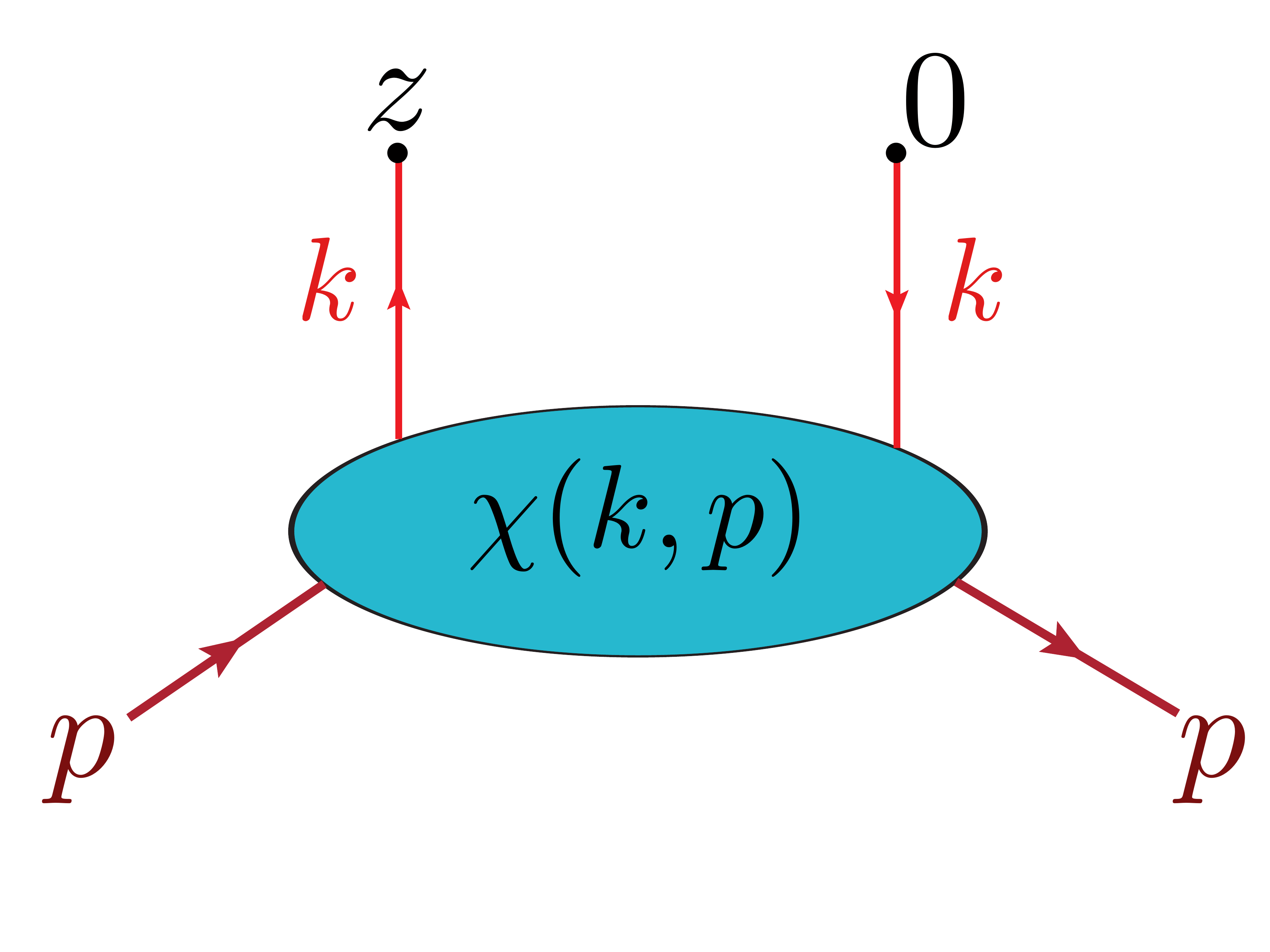}}
    \caption{Structure of generic  matrix element.
    \label{chipk}}
    \end{figure}
  
  The function $ \chi  (k,p)$ depends on the momenta $k$ and  $p$.
The analysis of Feynman diagrams  
in the Schwinger $\alpha$-representation (see, e.g., Ref. [\citen{Nakanishi:1971graph}])  tells that, a general scalar 
handbag diagram  $d_i$ of the Fig. \ref{chipk} type may be written  as 
\begin{align}
  &i  \chi_{d_i}  (k,p) =  i^{l} \, \frac{P({\rm c.c.})}{(4\pi i)^{2L}}
\int_0^{\infty} \prod_{j=1}^l   d\alpha_{j} [D(\alpha)]^{-2}
\nonumber \\ & \times 
\exp \left \{ i k^2  \frac{A (\alpha)}{D(\alpha) } +i 
  \frac{ (p-k)^2 B_s (\alpha)+   (p+k)^2 B_u (\alpha) }{D(\alpha) }
 \right \} 
\nonumber \\ & \times 
\exp \left \{ 
i  p^2  \frac{ C(\alpha) }{D (\alpha) }
- i  \sum_{j} \alpha_{j} (m_{j}^2- i\epsilon) \right \}  \ , 
\label{alphap}
\end{align}
where ${P({\rm c.c.})}$ is the relevant
 product of  coupling constants, 
 $L$ is the number of loops of the diagram,  $l$ is the number  of its
 lines,
 and the argument $(\alpha)$ of the $A,B,C,D$ functions  should be understood as
 $(\{\alpha_{d_i} \})$. Using  this representation, we get 
 \begin{align} 
& i \chi (k,p)  = \int_{0}^{\infty} d\lambda  \,  \int_{-1}^{1} d x \,
e^{i  \lambda[  k^2 - 2x (kp) +i\epsilon]}  \, F (x, \lambda ;p^2)  
\label{chisca} 
\end{align}
where the function $F (x, \lambda ;p^2) $ is given by the sum over all  contributing diagrams, 
 \begin{align} 
&  F (x, \lambda ;p^2)  =  \sum_{d_i}  \int_{0}^{\infty} d\lambda_{d_i}   \delta (\lambda - \lambda_{d_i} )
 \int_{-1}^1  \dd x_{d_i}  \delta (x-x_{d_i} ) \ \, F_{d_i} (x_{d_i} , \lambda_{d_i} ;p^2)   \ , 
\label{chisca2} 
\end{align}
with the   functions $F_{d_i} (x_{d_i} , \lambda_{d_i} ;M^2)$  specific 
for each diagram, and 
\begin{align}
\lambda_{d_i}= & \frac{A_{d_i}(\alpha)+ {B_s}_{ \, d_i}(\alpha) +  {B_u}_{\, d_i}(\alpha)}{D_{d_i}(\alpha) } \  \  ,  \label{lambda}  \\ 
 x_{d_i}= & \frac{{B_s}_{\, d_i}(\alpha) - { B_u}_{\,d_i}(\alpha)}{A_{\, d_i}(\alpha)+ {B_s}_{\, d_i}(\alpha) +  {B_u}_{\, d_i}(\alpha)} \ . 
 \label{xal}
  \end{align}
  
Eq. (\ref{chisca}) expresses an  evident   fact that  
 the function $\chi (k,p)$ 
depends on $k$ through the Lorentz invariants 
 $(kp)$ and $k^2$. 
A  nontrivial  property    is that  
$A(\alpha)$, $B_s(\alpha)$,  $B_u(\alpha)$,  $C(\alpha)$ and $D(\alpha)$ 
are  non-negative functions, namely,  sums of products 
of  non-negative $\alpha_j$-parameters 
of a  diagram. 
This immediately gives 
$0\leq \lambda \leq\infty$. 
The limits for $x$ in  general case are  obviously \mbox{$-1\leq x\leq 1$. } 
The negative $x$ values appear when $B_u (\alpha) \neq 0$,  
which happens for some   nonplanar diagrams. 
Integrating over $\lambda$ in Eq.~(\ref{chisca})  gives  
      a Nakanishi-type  representation\cite{Nakanishi:1969ph} for this amplitude.

 Note that {\it no} restrictions 
are imposed on $k$ and $p$ in \mbox{Eq. (\ref{chisca}).}
In particular, $p$ is the actual external momentum with $p^2=M^2$. 
Transforming  Eq.  (\ref{chisca})   to the coordinate 
representation and changing $\lambda =1/\sigma$  gives
 \begin{align}
  \langle p |   \phi(0) \phi (z)|p \rangle 
=  & 
\int_{0}^{\infty} d \sigma \int_{-1}^1 \dd x\,  %
 \Phi (x,\sigma; M^2) \,
  \,  e^{-i x (pz) -i \sigma {(z^2-i \epsilon )}/{4}} \,  
 \
 \label{newVDFx}
\end{align} 
where 
 \begin{align}
\Phi  (x, \sigma; M^2) = 
\exp [-ix^2  M^2/\sigma] F  (x, 1/\sigma; M^2)
\end{align} 
is the  {\it Virtuality Distribution Function}\cite{Radyushkin:2014vla,Radyushkin:2015gpa}  (VDF) and
Eq. (\ref{newVDFx}) is  the  {\it VDF representation}.  
It  basically reflects the fact that the matrix element 
$\langle p |   \phi(0) \phi (z)|p \rangle$  depends on $z$ through 
$(pz)$ and $z^2$.  

The main 
     non-trivial   feature of the representations (\ref{chisca}),   (\ref{newVDFx})
is  in their  specific  limits of integration over  $x$ and $\lambda$ (or $\sigma$).  
These are dictated by the properties of the  contributing Feynman  diagrams,
 in particular, by positivity of the functions
 $A,B,D$  determining $x$ and $\lambda$.
 It should be emphasized that  these functions are determined purely 
 by   denominators of propagators, and are not affected by their
 numerators present in non-scalar theories.
 
 Thus, the VDF representation (\ref{newVDFx})  is valid  for any diagram and 
 reflects very general  features of quantum field theory.  On these  grounds, we 
assume that it holds nonperturbatively.
Integrating over $\sigma$, we get the pseudo-PDF representation
 \begin{align}
  \langle p |   \phi(0) \phi (z)|p \rangle 
=  & 
\int_{-1}^1 \dd x\,  %
 {\cal P} (x, -z^2) \,
  \,  e^{-i x (pz) } \,  
 \ . 
 \label{newPDFx}
\end{align}

Eq. (\ref{newPDFx})   gives a covariant definition of $x$ as 
a variable that is Fourier-conjugate to the Ioffe time $(pz)$. To define $x$,  we do not need to assume that 
$p^2=0$ or that $z^2=0$. We also do not  need  to  base the definition of $x$ on   the ideas of the light-front quantization, the analysis
in the  infinite momentum frame,  Sudakov variables, etc. 



\end{document}